\documentclass[bibyear]{aa} 

\usepackage{graphicx}
\usepackage{txfonts}
\begin{document} 
\def\hi {H\,{\sc i}}
\def\hii {H\,{\sc ii}}
\def\water {H$_2$O}
\def\meth {CH$_{3}$OH}
\def\dg{$^{\circ}$}
\def\kms{km\,s$^{-1}$}
\def\ms{m\,s$^{-1}$}
\def\jyb{Jy\,beam$^{-1}$}
\def\mjyb{mJy\,beam$^{-1}$}
\def\solmass {\hbox{M$_{\odot}$}}
\def\solum {\hbox{L$_{\odot}$}} 
\def\d {$^{\circ}$}
\def\n {$n_{\rm{H_{2}}}$}
\def\kmsg{km\,s$^{-1}$\,G$^{-1}$}
\def\tbo {$T_{\rm{b}}\Delta\Omega$}
\def\tb {$T_{\rm{b}}$}
\def\om{$\Delta\Omega$}
\def\dvi {$\Delta V_{\rm{i}}$}
\def\dvz {$\Delta V_{\rm{Z}}$}
\def\code {FRTM code}
\def\NW {Nedoluha \& Watson}
\title{EVN observations of 6.7~GHz methanol maser polarization in massive star-forming regions IV. 
Magnetic field strength limits and structure for 7 additional sources\thanks{Tables from A.1 to A.7 
are only available in electronic form at the CDS via anonymous ftp to cdsarc.u-strasbg.fr (130.79.128.5) 
or via http://cdsweb.u-strasbg.fr/cgi-bin/qcat?J/A+A/}}
\author{G.\ Surcis  \inst{1}
  \and 
  W.H.T. \ Vlemmings \inst{2}
 \and
  H.J.~van Langevelde \inst{3,4}
  \and
  B. \ Hutawarakorn Kramer \inst{5,6}
  \and
  A. Bartkiewicz \inst{7}
  }

\institute{INAF - Osservatorio Astronomico di Cagliari, Via della Scienza 5, I-09047, Selargius, Italy\\
 \email{gabriele.surcis@inaf.it}
 \and
 Department of Space, Earth and Environment, Chalmers University of Technology, Onsala Space Observatory, SE-439 92 Onsala, Sweden
 \and
 Joint Institute for VLBI ERIC, Oude Hoogeveensedijk 4, 7991 PD Dwingeloo, The Netherlands
 \and
 Sterrewacht Leiden, Leiden University, Postbus 9513, 2300 RA Leiden, The Netherlands
 \and
 Max-Planck Institut f\"{u}r Radioastronomie, Auf dem H\"{u}gel 69, 53121 Bonn, Germany
 \and
 National Astronomical Research Institute of Thailand, Ministry of Science and Technology, Rama VI Rd., Bangkok 10400, Thailand
 \and
 Centre for Astronomy, Faculty of Physics, Astronomy and Informatics, Nicolaus Copernicus University, Grudziadzka 5, 87-100 Torun, Poland
   }

\date{Received ; accepted}
\abstract
{Magnetohydrodynamical simulations show that the magnetic field can drive molecular outflows during the 
formation of massive protostars. The best probe to observationally measure both the morphology and the strength of this
magnetic field at scales of 10-100~au is maser polarization.}
{We measure the direction of magnetic fields at milliarcsecond resolution around a sample of massive star-forming regions to determine whether there is a relation between the orientation of the magnetic field and of the
outflows. In addition, by estimating the magnetic field strength via the Zeeman splitting measurements, the
role of magnetic field in the dynamics of the massive star-forming region is investigated.} 
{We selected a flux-limited sample of 31 massive star-forming regions to perform a statistical analysis of 
the magnetic field properties with respect to the molecular outflows characteristics. We report the linearly and
circularly polarized emission of 6.7~GHz \meth ~masers towards seven massive star-forming regions of the total sample
with the European VLBI Network. The sources are: G23.44-0.18, G25.83-0.18, G25.71-0.04, G28.31-0.39, G28.83-0.25, 
G29.96-0.02, and G43.80-0.13.
}
{We identified a total of 219 \meth ~maser features, 47 and 2 of which showed linearly and circularly polarized 
emission, respectively. We measured well-ordered linear polarization vectors around all the massive young stellar 
objects and Zeeman splitting towards G25.71-0.04 and G28.83-0.25. Thanks to recent theoretical results, we were able to provide
lower limits to the magnetic field strength from our Zeeman splitting measurements. }
{We further confirm (based on $\sim80\%$ of the total flux-limited sample) that the magnetic field on scales of 10-100~au is
preferentially oriented along the outflow axes. The estimated magnetic field strength of $|B_{||}|>61$~mG
and $>21$~mG towards G25.71-0.04 and G28.83-0.2, respectively, indicates that it dominates the dynamics of the gas in both
regions.}
\titlerunning{EVN observations of 6.7~GHz methanol maser polarization in MSFRs IV.}
\authorrunning{Surcis et al. 2019}

\maketitle
\section{Introduction}
\label{intro}
%
Several theoretical and observational efforts are advancing our understanding of the formation of 
high-mass stars ($M>8$~\solmass). In the last twenty years several models were developed, among which we mention the Core
Accretion model (e.g., McKee \& Tan \cite{mck03}), the Competitive Accretion model (e.g., Bonnell et al. \cite{bon01}),
and the hybrid model of these two models (e.g., Tan et al. \cite{tan14} and references therein). From a theoretical point of view,
intensive simulation campaigns have been carried out by several authors in the last decade (e.g., Krumholz et al.
\cite{kru09}; Peters et al. \cite{pet10, pet11}; Hennebelle et 
al. \cite{hen11}; Klessen et al. \cite{kle11}; Seifried et al. \cite{sei11, sei12, sei15}; Klassen et al. \cite{kla12, 
kla14, kla16}; Kuiper et al. \cite{kui11, kui15, kui16}). The simulations are mainly focused on understanding the role 
of feedback before and after the formation of cores (e.g., Myers et al. \cite{mye13, mye14}) and young stellar objects 
(YSOs; e.g., Peters et al. \cite{pet12}; Kuiper et al. \cite{kui16}), and  the effects of more than two feedback
mechanisms are rarely taken into consideration within the same simulation (for more details see Tan et al. \cite{tan14}).\\
\indent Magnetohydrodynamical (MHD) simulations show that jets and molecular outflows are common in massive YSOs
and that they have an important role in the formation process (e.g., Tan et al. \cite{tan14}; Matsushita et al. 
\cite{mat18}). For instance, Banerjee \& Pudritz (\cite{ban07}) showed that early outflows can reduce the radiation 
pressure allowing the further growth of the protostar.  The outflows are found to be initially poorly collimated 
and became more collimated only when a nearly Keplerian disk is formed and fast jets are generated (Seifried et al. 
\cite{sei12}). The mass ejection rate through a MHD outflow and the accretion rate through an accretion disk are 
comparable only when the magnetic energy ($E_{\rm{B}}$) of the initial core is comparable to the gravitational energy 
($E_{\rm{G}}$; Matsushita et al. \cite{mat17}). Furthermore, when the magnetic field is strong 
($\mu<5$\footnote{$\mu=(M/\Phi)/(M/\Phi)_{\rm{crit}}$, where $(\rm{M/\Phi})$ is the mass-to-magnetic flux ratio and 
$(M/\Phi)_{\rm{crit}}\approx0.12/\sqrt{G}$ is the critical value of this ratio, where $G$ is the gravitational constant
(Tomisaka et al. \cite{tom88}). The critical value indicates the maximum mass supported by the magnetic
field (Tomisaka et al. \cite{tom88}). The stronger the magnetic field, the lower $\mu$.}) the outflow is slow and
poorly collimated (Seifried et al. \cite{sei12}) and its structure is 
determined by the large-scale geometry of the magnetic field lines (Matsushita et al. \cite{mat17}). When 
the magnetic field is weak ($\mu>10$), in addition to leading to the fragmentation of the parental cloud core, the outflow is 
fast and well collimated (Hennebelle et al. \cite{hen11}), and it might even disappear during the formation of the YSO,
leaving the scene to a magnetically supported toroid-like structure (Matsushita et al. \cite{mat17}).\\
\indent It needs to be mentioned that other theoretical studies investigated the possibility that the outflows
are driven by other mechanisms than magnetic fields (e.g., Yorke \& Richling \cite{yor02}; Vaidya et al. 
\cite{vai11}; Peters et al.  \cite{pet11}; Kuiper et al. \cite{kui15}). In particular, if the outflows are driven by 
the ionization feedback they appear to be uncollimated (e.g., Peters et al. \cite{pet11}). In addition, Peters et al. 
(\cite{pet14}) also found that stars that form in a common accretion flow tend to have aligned outflows, which can 
combine to form a collective outflow similar to what is observed in W75N(B) (e.g., Surcis et al. \cite{sur14a}).\\
\indent From an observational point of view, the presence of molecular outflows in massive star-forming regions (SFRs)
is nowadays a fact (e.g., Tan et al. \cite{tan14} and references therein). Also, measuring the morphology of the
magnetic field around massive YSOs is now regularly done  by using polarized dust emission (on a scale of 
$10^3$~au; e.g., Zhang et al. \cite{zha14a}; Girart et al. \cite{gir16}) and polarized maser 
emission (on a scale of tens of au; e.g., Vlemmings et al. \cite{vle10}; Surcis et al. \cite{sur12, sur13, sur15}, 
hereafter Papers~I-III; Sanna et al. \cite{san17}; Dall'Olio et al. \cite{dal17}). However, the findings at the
two different scales conflict. On the large scale (arcsecond resolution) magnetic fields appear to be
randomly distributed with respect to the outflow axis (Zhang et al. \cite{zha14}), while on the small scale 
(milliarcsecond resolution) magnetic fields, estimated from the polarized emission of the 6.7~GHz \meth ~masers, are
preferentially oriented along the outflow (Paper~III).  We note that the conclusions reported in Paper~III are based on  
60\% of the flux-limited sample (19 out of 31 sources), whereas those in Zhang et al. (\cite{zha14}) are based on 
a total of 21 sources. There are no sources  in common between the two samples (Zhang et al. \cite{zha14}; Paper~III). 
Nonetheless, the comparison around the same YSO, when possible, of the magnetic field morphology at the two scales has 
shown consistency (e.g., Surcis et al. \cite{sur14}).\\
\indent Measurements of the magnetic field strength close to YSOs are less common. It has been possible,
so far, only through the analysis of the circularly polarized emission of \water ~masers (e.g., Surcis et al. 
\cite{sur14a}; Goddi et al. \cite{god17}). Because of the shock-nature of the \water ~masers, the strength of the 
magnetic field is estimated in the post-shock compressed gas, even if it can be possible to derive the pre-shock 
magnetic field strength (e.g., Imai et al. \cite{ima03}; Vlemmings et al. \cite{vle06}; Goddi et al. 
\cite{god17}). Although the circularly polarized emission of the 6.7~GHz \meth ~maser has been regularly detected
(e.g., Papers~I-III), no estimates of the magnetic field strength have been possible due to the unknown Land\'{e}
g-factors (Vlemmings et al. \cite{vle11}). Very recently, Lankhaar et al. (\cite{lan18}) theoretically calculated the 
g-factors for all the \meth ~maser transitions making it possible to estimate at least a lower limit of the magnetic field
strength from the Zeeman splittings measurements.\\
\indent Here, in the fourth paper of the series after Papers~I-III, we present the results of the next seven observed 
sources of the flux-limited sample, briefly described in Sect.~\ref{SEVNG}. In Sect.~\ref{obssect}, in addition to  reporting the
observations, we describe the changes made to the adapted full radiative transfer method (FRTM) code for the 6.7~GHz
\meth ~maser emission. The results are presented in Sect.~\ref{res} and discussed in Sect.~\ref{discussion}, where we
also briefly update the previous statistics (see Paper~III).
\section{Massive star-forming regions}
\label{SEVNG}
We selected a flux-limited sample of 31 massive SFRs with declination $>-9$\d ~and a total \meth ~maser 
single-dish flux density greater than 50~Jy from the 6.7~GHz \meth ~maser catalog of Pestalozzi et al. (\cite{pes05}), and that
in more recent single-dish observations showed a total flux density $>$20~Jy (Vlemmings et al. \cite{vle11}). We already
observed and analyzed 19 of these sources (Vlemmings et al. \cite{vle10}; Surcis et al. \cite{sur09, sur11a, sur14}; 
Papers~I-III). Seven more sources, which are described below in Sects.~\ref{G23_intro}-\ref{G43_intro}, have been
observed at 6.7~GHz with the European VLBI Network\footnote{The European VLBI Network is a joint facility of European,
Chinese, South African, and other radio astronomy institutes funded by their national research councils.} (EVN). The last
five sources in the sample will be presented in the next paper of the series.
\subsection{\object{G23.44-0.18}}
\label{G23_intro}
G23.44-0.18 is a high-mass SFR at a heliocentric distance of $5.88^{+1.37}_{-0.93}$~kpc 
(Brunthaler et al. \cite{bru09}) in the Norma arm of our Galaxy (Sanna et al. \cite{san14}) with a systemic
velocity of $V_{\rm{lsr}}^{\rm{^{CS(2-1)}}}=+104.2$~\kms ~(Bronfman et al. \cite{bro96}).
The region contains two millimeter dust continuum cores, named MM1 and MM2, separated by $\sim14''$ ($8\cdot10^4$~AU), 
suggesting the presence of two YSOs prior to forming an ultra-compact \hii ~(UC~\hii) region (Ren et al. \cite{ren11}). 
Furthermore, a strong bipolar CO outflow 
($\rm{PA_{outflow}^{^{12}CO}}=-40$\d) originates from MM2 (Ren et al. \cite{ren11}). The CO outflow consists 
of a low-velocity component (LVC) and a high-velocity component (HVC) whose blue-shifted
(+83~\kms$<\rm{V_{blue}^{LVC}}<$+93~\kms ~and +65~\kms$<\rm{V_{blue}^{HVC}}<$+75~\kms) and red-shifted
(+113~\kms$<\rm{V_{red}^{LVC}}<$+123~\kms ~and +140~\kms$<\rm{V_{red}^{HVC}}<$+170~\kms) lobes are oriented 
northwest and southeast, respectively (Ren et al. \cite{ren11}). Two groups of 6.7 GHz \meth ~masers have
been detected around  MM1 and MM2 (Walsh et al. \cite{wal98}; Fujisawa et al. \cite{fuj14}; Breen et al.
\cite{bre15}). A 12 GHz \meth ~maser emission has been detected with a velocity coverage consistent
with the two groups of 6.7~GHz \meth ~masers, while the OH maser emission is likely associated with MM2 (e.g., 
Breen et al. \cite{bre16}; Caswell et al. \cite{cas13}). Vlemmings et al. (\cite{vle11}) measured a Zeeman splitting of the
6.7~GHz \meth ~maser of $0.43 \pm 0.06$~\ms ~with the Effelsberg telescope.

\subsection{\object{G25.83-0.18}}
\label{G258_intro}
G25.83-0.18 is a very young SFR in an evolutionary stage prior to the UC~\hii ~region phase at a kinematic 
distance of $5.0\pm0.3$~kpc (Araya et al. \cite{ara08}; Andreev et al. \cite{and17}). Both 6.7~GHz and 12~GHz
\meth ~maser emissions were detected (e.g., Walsh et al. \cite{wal98}; B\l{}aszkiewicz \& Kus \cite{bla04}; 
Breen et al. \cite{bre15, bre16}),
which are $\sim2''$ north of the 4.8 GHz H$_2$CO maser detected at the center of an IR dark cloud (Araya et 
al. \cite{ara08}). The velocities of all the detected maser species, including the \water ~masers (Breen 
\& Ellingsen \cite{bre11}), are close to the systemic velocity of the 
region, i.e., $V_{\rm{lsr}}^{\rm{{^{18}CO}}}=+93.2$~\kms ~(de Villiers et al. \cite{dev14}). A 
$^{13}$CO outflow has been measured with the James Clerk Maxwell Telescope (JCMT), and its red-shifted
(+91.8~\kms$<\rm{V_{red}^{^{13}\rm{CO}}}<$+101.8~\kms) and blue-shifted
(+83.8~\kms$<\rm{V_{blue}^{^{13}\rm{CO}}}<$+91.8~\kms) lobes are oriented north-south with a position angle of
$\rm{PA_{outflow}^{^{13}\rm{CO}}}\sim+10$\d ~(de Villiers et al. \cite{dev14}). No radio continuum emission 
has been detected towards the \meth ~maser clumps (Walsh et al. \cite{wal98}).\\
\indent A Zeeman splitting of the 6.7~GHz \meth ~maser emission of \dvz$=(0.99 \pm 0.26)$~\ms ~was measured 
with the Effelsberg 100~m telescope (Vlemmings et al. \cite{vle11}).
\subsection{\object{G25.71-0.04}}
\label{G257_intro}
The massive SFR G25.71-0.04, also known  as IRAS\,18353-0628, is located at a distance of $10.1\pm0.3$~kpc from 
the Sun (Green \& McClure-Griffiths \cite{gre11});  it is associated with 6.7~GHz and 12~GHz \meth
~masers and OH masers (Walsh et al. \cite{wal97};  Fujisawa et al. \cite{fuj14}; Breen et al.
\cite{bre15, bre16}; Szymczak \& G\'{e}rard \cite{szy04}). Neither 
radio continuum emission nor UC~\hii ~region are  observed at the position of the \meth ~maser clump (Walsh 
et al. \cite{wal98}), a warm dust sub-millimeter source has  instead been detected (Walsh et al. 
\cite{wal03}). Its bright sub-mm peak suggests that the maser site is likely to be in a stage of evolution before 
the UC~\hii ~region has been created (Walsh et al. \cite{wal03}). De Villiers et al. (\cite{dev14}) detected a 
$^{13}$CO outflow of which the blue-shifted lobe ($\rm{PA_{outflow, blue}^{\rm{^{13}CO}}}=-90$\d) coincides in 
position and velocity (+92.3~\kms$<\rm{V_{blue}^{^{13}\rm{CO}}} <$+101.3~\kms) with the \meth ~masers (Fujisawa 
et al. \cite{fuj14}). The red-shifted lobe is instead oriented on the plane of the sky with an angle of 
$\rm{PA_{outflow, red}^{\rm{^{13}CO}}}=-27$\d ~and its velocity range is 
+101.3~\kms$<\rm{V_{red}^{^{13}\rm{CO}}}<$+103.8~\kms.\\
\indent Vlemmings et al. (\cite{vle11}) detected circularly polarized emission of the 6.7~GHz \meth ~maser 
with the 100~m Effelsberg telescope, which provided a Zeeman-splitting of $0.81\pm0.10$~\ms, though 
few years earlier this emission was not detectable (Vlemmings \cite{vle08}).
\subsection{\object{G28.31-0.39}}
\label{G283_intro}G28.31-0.39 (also known as IRAS\,18416-0420a) is a massive YSO, at a parallax distance of
$4.52^{+0.5}_{-0.4}$~kpc (Li et al. 2018 \textit{in prep}), associated with the UC~\hii ~regions field G28.29-0.36
(Thompson et al. \cite{tho06}).
Walsh et al. (\cite{wal97,wal98}) detected 6.7~GHz \meth ~maser emission coinciding with the center of an 
east-west sub-millimeter dust-emission (Walsh et al. \cite{wal03}), but with no evident association with any of the \hii 
~regions (Thompson et al. \cite{tho06}). Some of the \meth ~maser features showed short-lived 
bursts suggesting a region of weak and diffuse maser emission, probably unsaturated, located far from the central 
core structure (Szymczak et al. \cite{szy18}). A $^{13}$CO outflow with a 
$\rm{PA_{outflow}^{\rm{^{13}CO}}}=-52$\d ~was detected at the position of the \meth ~masers (de Villiers et al.
\cite{dev14}). The velocity range of the blue-shifted lobe 
(+80.4~\kms$<\rm{V_{blue}^{^{13}\rm{CO}}} <$+85.9~\kms) is consistent with the velocities of most of the \meth 
~masers (e.g., Walsh et al. \cite{wal98}; Breen et al. \cite{bre15, bre16}; Szymczak et al. \cite{szy18}). 
The red-shifted lobe has a velocity 
range of +85.9~\kms$<\rm{V_{blue}^{^{13}\rm{CO}}} <$+88.9~\kms\ (de Villiers et al. \cite{dev14}). No \meth ~maser
polarization observations had been conducted until now.
\subsection{\object{G28.83-0.25}}
\label{G288_intro}
The extended green object (EGO) G28.83-0.25 is located at the edge of the mid-infrared bubble N49 at a 
kinematic distance of $4.6\pm0.3$ kpc (Churchwell et al. \cite{chu06}; Cyganowski et al. \cite{cyg08}; Green \&
McClure-Griffiths \cite{gre11}). Two faint continuum radio sources have been identified at 3.6~cm with G28.83-0.25,
named CM1 and CM2 (Cyganowski et al. \cite{cyg11}). CM2 is coincident with the linearly distributed 6.7~GHz \meth 
~masers ($\rm{PA_{CH_3OH}}\approx-45$\d, Cyganowski et al. \cite{cyg09}; Fujisawa et al. \cite{fuj14}) and  is
surrounded by 44~GHz \meth ~masers (Cyganowski et al. \cite{cyg09}). No 25~GHz \meth ~maser emission has 
been detected (Towner et al. \cite{tow17}). While the 44~GHz \meth ~masers are located at the edges of a 
$^{13}$CO-bipolar outflow whose axis is oriented close to the line of sight, the 6.7~GHz \meth ~masers coincide
with the peak emission of the blue-shifted lobe of the outflow that is about $5''$ southwest  from the peak of the 
red-shifted lobe (de Villiers et al. \cite{dev14}). Although the outflow is close to the line of sight, the small 
misalignment of the two lobes implies an orientation on the plane on the sky of 
$\rm{PA_{outflow}^{\rm{^{13}CO}}}=-40$\d ~(de Villiers et al. \cite{dev14}). The velocities of the lobes are
+77.4~\kms$<\rm{V_{blue}^{^{13}\rm{CO}}} <$+88.4~\kms ~and +88.4~\kms$<\rm{V_{red}^{^{13}\rm{CO}}} <$+96.4~\kms
~(de Villiers et al. \cite{dev14}). However, the velocity range of the 6.7~GHz and 12~GHz \meth ~maser emissions 
agrees with that of the red-shifted lobe (e.g., Fujisawa et al. \cite{fuj14}; Breen et al. \cite{bre15, bre16}).\\
\indent Bayandina et al. (\cite{bay15}) detected both 1665~MHz and 1667~MHz OH masers at the same location, within 
the uncertainties, of the 6.7~GHz \meth ~maser emission. By measuring the Zeeman splitting of the OH maser emissions 
they determined magnetic field strengths of 6.6~mG (1665~MHz OH) and of 5.1~mG (1667~MHz OH). No polarization 
observations of the \meth ~maser emissions had been made until now.
\subsection{\object{G29.96-0.02 (W43\,S)}}
\label{G29_intro}
G29.96-0.02 is a well-studied high-mass star-forming cloud (e.g., Cesaroni et al. \cite{ces94}; De Buizer et al. 
\cite{deb02}; Pillai et al. \cite{pil11}; Beltr\'{a}n et al. \cite{bel13}) located in the massive SFR W43-South 
(W43\,S) at a parallax distance of $5.26^{+0.62}_{-0.50}$~kpc (Zhang et al. \cite{zha14}). G29.96-0.02 contains a 
cometary UC~\hii ~region and a hot molecular core (HMC) located in front of the cometary arc (e.g., Wood \& Churchwell 
\cite{woo89}; Olmi et al. \cite{olm03}; Beuther et al. \cite{beu07}; Cesaroni et al. \cite{ces98,ces17}). \water,
OH, and $\rm{H_2CO}$ maser emissions (e.g., Hofner \& Churchwell \cite{hof96}; Hoffman et al. \cite{hof03};
Breen \& Ellingsen \cite{bre11}; Caswell et al. \cite{cas13}), and several \meth ~maser lines (Minier et al. \cite{min00,
min02}; Breen et al. \cite{bre15, bre16}) were detected towards the HMC. De Villiers et al. (\cite{dev14}) reported a
$^{13}$CO outflow from the HMC with a
PA$=+50$\d. The velocity range of the blue- (southeast)  and red-shifted (northwest)  lobes of the CO outflow are 
+92.1~\kms$<\rm{V_{blue}^{^{13}\rm{CO}}} <$+97.6~\kms ~and +97.6~\kms$<\rm{V_{red}^{^{13}\rm{CO}}} <$+106.6~\kms
~(De Villiers et al. \cite{dev14}). At an angular resolution of 0.2 arcseconds, Cesaroni et al. (\cite{ces17}) were
able to detect with the Atacama Large Millimeter/submillimeter Array (ALMA) a SiO bipolar jet (PA$=-38$\d) and a rotating disk perpendicular to it. Both are associated with the HMC. The velocity range of the SiO bipolar jet are
+83.5~\kms$<\rm{V_{blue}^{SiO}} <$+91.6~\kms ~and +102.4~\kms$<\rm{V_{red}^{SiO}} <$+110.5~\kms ~(Cesaroni et al. 
\cite{ces17}). The 6.7~GHz \meth ~masers are associated with this system, which likely harbors a massive YSO of 
$\sim$10~\solmass ~(Sugiyama et al. \cite{sug08}; Cesaroni et al. \cite{ces17}).\\
\indent A Zeeman-splitting of the 6.7~GHz \meth ~maser line of \dvz$=-0.33\pm0.11$~\ms ~was measured with the 100~m 
Effelsberg telescope (Vlemmings et al. \cite{vle11}).
\begin {table*}[th!]
\caption []{Observational details.} 
\begin{center}
\scriptsize
\begin{tabular}{ l c c c c c c c c c c c}
\hline
\hline
Source               & Observation      & Calibrator   & Polarization & Beam size        & Position & rms     &$\sigma_{\rm{s.-n.}}$\tablefootmark{d}  & \multicolumn{4}{c}{Estimated absolute position using FRMAP} \\ 
                     & date             &              &  angle       &                  & Angle    &         &         &    $\alpha_{2000}$           & $\delta_{2000}$            & $\Delta\alpha$\tablefootmark{e} & $\Delta\delta$\tablefootmark{e}     \\ 
                     &                  &              &  (\d)        &(mas~$\times$~mas)& (\d)     & ($\frac{\rm{mJy}}{\rm{beam}}$) &   ($\frac{\rm{mJy}}{\rm{beam}}$) &($\rm{^{h}:~^{m}:~^{s}}$) & ($\rm{^{\circ}:\,':\,''}$) &     (mas)             & (mas) \\ 
\hline
G23.44-0.18          & 01 March 2014    & J2202+4216\tablefootmark{a} & $-33\pm 4$   & $9.6\times4.2$   & -26.47   & 3       & 31      & +18:34:39.187             & -08:31:25.441              & 0.8 & 5.4 \\
G25.83-0.18          & 02 March 2014    & J2202+4216\tablefootmark{b} & $-33\pm 5$   & $8.7\times5.5$   & -8.76    & 4       &  43     & +18:39:03.630             & -06:24:11.163              & 0.8 & 6.0\\
G25.71-0.04          & 03 March 2014    & J2202+4216\tablefootmark{c} & $-33\pm 4$   & $16.5\times6.5$  & +4.98    & 3       &  9      & +18:38:03.140             & -06:24:15.453              & 1.7 & 12.3\\
G28.31-0.39          & 12 June  2014    & J2202+4216\tablefootmark{c} & $-33\pm 4$   & $12.9\times3.0$  & -44.20   & 3       & 13     & +18:44:22.030             & -04:17:38.304              & 0.6 & 6.3 \\
G28.83-0.25          & 13 June  2014    & J2202+4216\tablefootmark{c} & $-33\pm 4$   & $10.5\times3.5$  & -36.32   & 4       & 10     & +18:44:51.080             & -03:45:48.494              & 0.4 & 6.4 \\
G29.96-0.02          & 14 June  2014    & J2202+4216\tablefootmark{c} & $-33\pm 4$   & $10.1\times3.2$  & -39.10   & 4       & 55     & +18:46:03.740             & -02:39:22.299              & 0.3 & 7.8 \\
G43.80-0.13          & 15 June  2014    & J2202+4216\tablefootmark{c} & $-33\pm 4$   & $9.9\times3.4$   & -30.83   & 7       & 16     & +19:11:53.990             & +09:35:50.300              & 0.3 & 1.0\\
\hline
\end{tabular}
\end{center}
\tablefoot{
\tablefoottext{a}{Calibrated using 3C286 ($I=0.39$~\jyb, $P_{\rm{l}}=4.3\%$).}
\tablefoottext{b}{Calibrated using 3C286 ($I=0.49$~\jyb, $P_{\rm{l}}=4.8\%$).}
\tablefoottext{c}{Calibrated using results from G23.44-0.18.}
\tablefoottext{d}{Self-noise in the maser emission channels (e.g., Sault \cite{sau12}). When no circularly 
polarized emission is detected we consider the self-noise of the brightest maser feature.}
\tablefoottext{e}{Formal errors of the fringe rate mapping.}
}
\label{Obs}
\end{table*}
\subsection{\object{G43.80-0.13} (OH~43.8-0.1)}
\label{G43_intro}
G43.80-0.13, better known as OH\,43.8-0.1, is a massive star-forming region associated with an infrared source 
(IRAS\,19095+0930) and an UC\hii ~region at a parallax distance of $6.0^{+0.19}_{-0.18}$~kpc (Braz \& Epchtein 
\cite{bra83}; Kurtz et al. \cite{kur94}; Wu et al. \cite{wu14}). OH, \water, and \meth ~masers have been detected
at VLBI scale towards the UC\hii ~region within the same velocity range (e.g., Fish et al. \cite{fis05}; Sarma et al.
\cite{sar08}; Sugiyama et al. \cite{sug08}). L\'{o}pez-Sepulcre et al. (\cite{lop10}) detected an HCO$^+$-outflow
oriented NE-SW ($\rm{PA_{outflow}^{HCO^+}}=+38$\d) with the blue-shifted lobe 
(+32.5~\kms$<\rm{V_{blue}^{HCO^+}}<$+38.5~\kms) and the red-shifted lobe (+49.0~\kms$<\rm{V_{red}^{HCO^+}}<$+53.5~\kms) directed towards the  northeast and southwest, respectively.\\
\indent Magnetic field strengths were measured at VLBI scale via Zeeman-splitting of both OH and \water ~masers.
These are $|B^{\rm{OH}}|=3.6~\rm{mG}$ and $0.3~\rm{mG}<|B_{||}^{\rm{H_2O}}|<22~\rm{mG}$
(Fish et al. \cite{fis05}, Sarma et al. \cite{sar08}). No polarization observations of \meth ~maser emissions had 
been performed until now.
\section{Observations and analysis}
\label{obssect}
The second group of seven massive SFRs was observed in full polarization spectral mode at 6.7 GHz with 
eight of the EVN antennas (Ef, Jb, On, Mc, Nt, Tr, Wb, and Ys) between March and June 2014; the Medicina 
antenna (Mc) was not available in June 2014 (program code: ES072). The total observing time was 49~h. We 
covered a velocity range of $\sim$100~\kms ~by observing a bandwidth of 2~MHz. The correlation of the data was 
made with the EVN software correlator (SFXC, Keimpema et al. \cite{kei15}) at the Joint Institute for VLBI 
ERIC (JIVE, the Netherlands) by using 2048 channels and generating all four polarization combinations (RR, LL, RL, 
LR) with a spectral resolution of $\sim$1~kHz ($\sim$0.05~\kms). In Table~\ref{Obs} we present all the 
observational details. Here, the target sources and the date of the observations are listed in Cols.~1 and 2, 
respectively;
in Cols.~3 and 4 the polarization calibrators with their polarization angles are given. From Col.~5 to Col.~7 
some of the image parameters are listed; in particular, the restoring beam size and 
corresponding position angle are in Cols.~5 and 6 and the thermal noise in Col.~7. In Col.~9 we also show the 
self-noise  in the maser emission channels (see Paper~III for  details). Finally, the estimated absolute 
position of the reference maser and the FRMAP uncertainties are listed from Col.~9 to Col.~12 (see Paper III
for  details).\\
\indent The Astronomical Image Processing Software package (AIPS) was used for calibrating and imaging the 
data. Following the same calibration procedure reported in Papers~I-III, the bandpass, the delay, the phase, 
and the polarization calibration were performed on the calibrators listed in Col.~3 of Table~\ref{Obs}.  
Fringe-fitting and self-calibration were subsequently performed on the brightest maser feature of each SFR that
is identified as the reference maser feature in Tables~\ref{G23_tab}--\ref{G43_tab}. The cubes of the four 
Stokes parameters (\textit{I}, \textit{Q}, \textit{U}, and \textit{V}) were imaged using the AIPS task IMAGR. 
The polarized intensity ($POLI=\sqrt{Q^{2}+U^{2}}$) and polarization angle ($POLA=1/2\times~\rm{atan}(U/Q)$) cubes 
were produced by combining the \textit{Q} and \textit{U} cubes. During the observations we observed a 
primary polarization calibrator (J2202+4216;  Col. 3 of Table~\ref{Obs}) and a well-known polarized 
calibrator (3C286). For G23.44-0.18 and G25.83-0.18 the signal-to-noise ratio of the 3C286 maps were so good that 
we were able to calibrate the polarization angle of J2202+4216 by using the calibration of 3C286. As expected,
within the errors this was consistent with the constant polarization angle measured between 
2005\footnote{http://www.vla.nrao.edu/astro/calib/polar/} and
2012\footnote{http://www.aoc.nrao.edu/$\sim$smyers/evlapolcal/polcal\_master.html}, i.e., $-31$\d$\pm4$\d.  For the other
five sources we assumed that the polarization angle of J2202+4216 did not change from March to June 2014. 
Therefore, we calibrated the linear polarization angles of the maser features by comparing the linear polarization 
angle of J2202+4216 measured by us with its angle obtained during the calibration of G23.44-0.18. The formal error 
on $POLA$ due to the thermal noise is given by $\sigma_{POLA}=0.5 ~(\sigma_{P}/POLI) \times (180^{\circ}/\pi)$ 
(Wardle \& Kronberg \cite{war74}), where $\sigma_{P}$ is the rms error of POLI.\\
\indent As for Paper~III, the observations were not performed in phase-referencing mode and the absolute 
position of the brightest maser feature of each source was estimated through fringe rate mapping (AIPS task 
FRMAP). The results and the formal errors of FRMAP are listed from Col.~9 to Col.~12 of Table~\ref{Obs}. The
phase fluctuations dominated the absolute positional uncertainties and, from our experience 
with other experiments and varying the task parameters, we estimate that the absolute position uncertainties
are on the order of a few mas.\\
\begin{figure*}[th!]
\centering
\includegraphics[width = 9 cm]{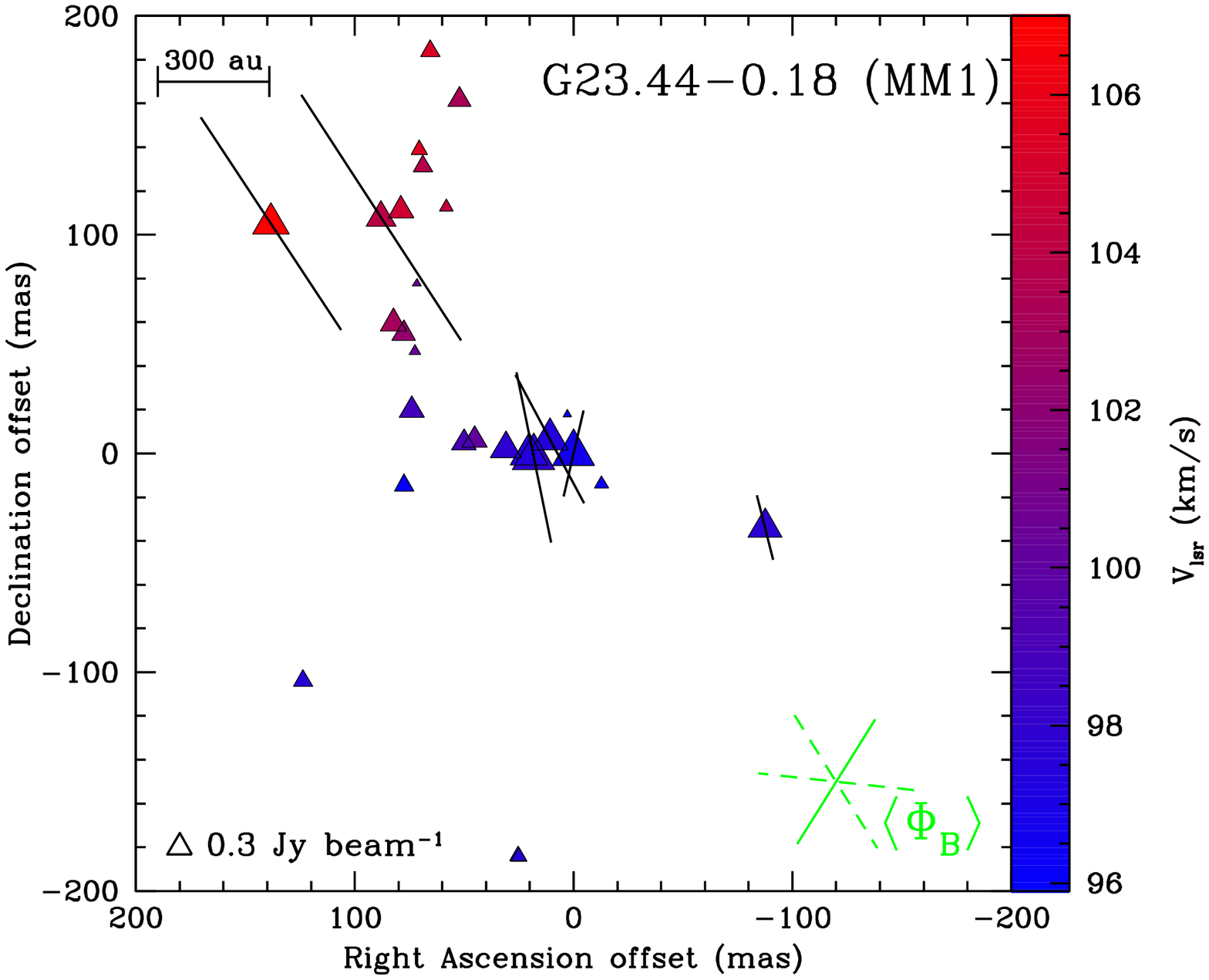}
\includegraphics[width = 9 cm]{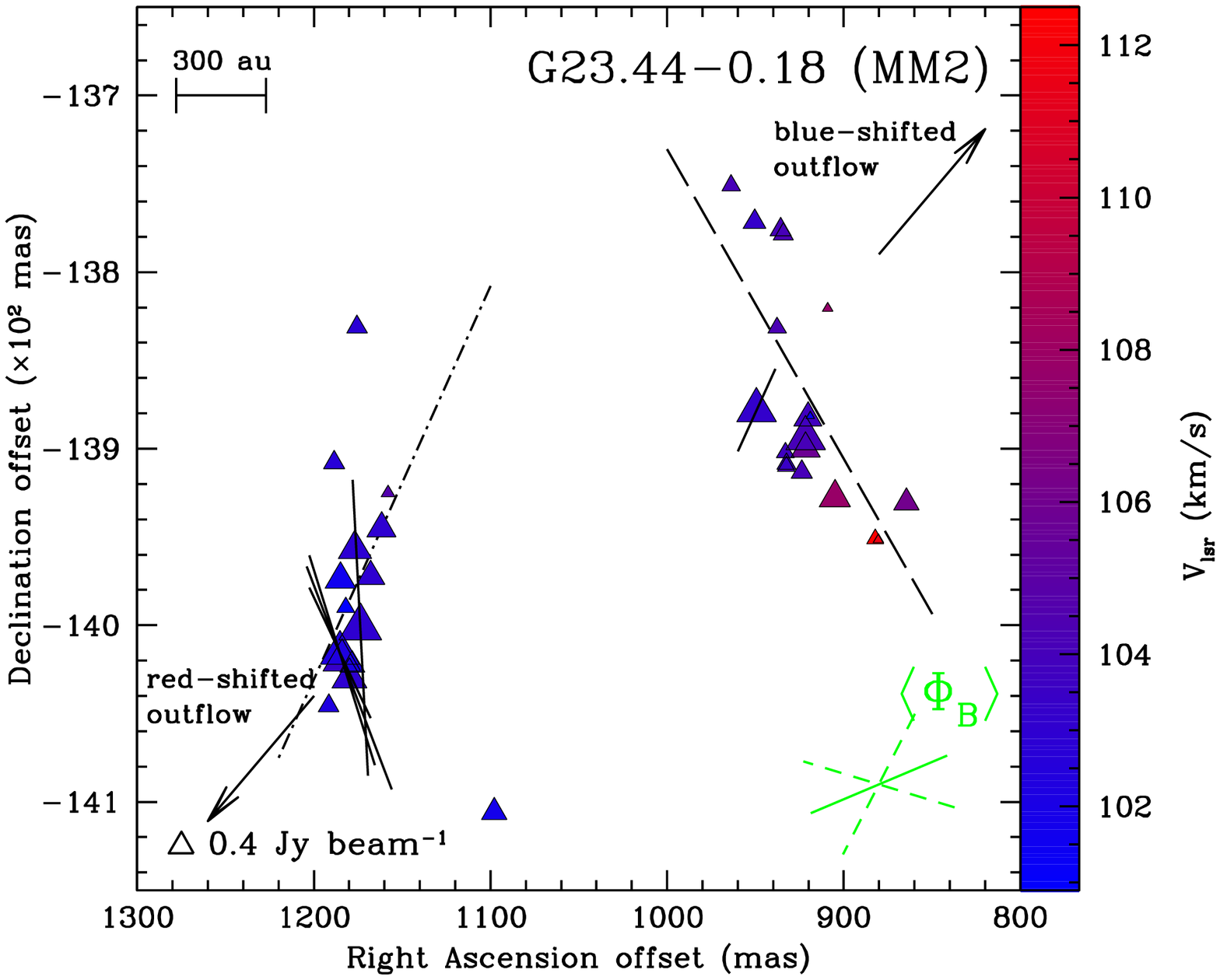}
\caption{View of the \meth ~maser features detected around G23.44-0.18 MM1 (left panel) and MM2 (right panel).
The reference position is the estimated absolute position from Table~\ref{Obs}.
Triangles  identify \meth ~maser features whose side length is scaled logarithmically according to their peak flux 
density (Table~\ref{G23_tab}). 
Maser local standard of rest radial velocities are indicated by color (the assumed velocity of the region is 
$V_{\rm{lsr}}^{\rm{^{CS(2-1)}}}=+104.2$~\kms,
 Bronfman et al. \cite{bro96}). 0.3~\jyb ~and 0.4~\jyb ~symbols are plotted for comparison.
The linear polarization vectors, scaled logarithmically according to the polarization 
fraction $P_{\rm{l}}$ ($P\rm{_l^{MM1}}=1.1-6.5\%$ and $P\rm{_l^{MM2}}=1.4-8.3\%$; see
Table~\ref{G23_tab}), are overplotted. In the bottom right corner the corresponding error-weighted 
orientation of the magnetic field ($\langle\Phi_{\rm{B}}\rangle$, see Sect.\ref{Borient}) is also shown,
the two dashed segments indicate the uncertainty. 
The two arrows in the right panel indicate the direction, but not the actual position, of the 
red- and blue-shifted lobes of the $\rm{^{12}CO(2-1)}$ outflow associated with G23.44-0.18~(MM2)  
($\rm{PA_{outflow}^{^{12}CO}}=-40$\d; Ren et al. \cite{ren11}). The dashed and dash-dotted lines are the best 
least-squares linear fit of the western ($\rm{PA_{CH_{3}OH}^{\rm{west}}}=+30^{\circ}\pm26$\d) and eastern 
($\rm{PA_{CH_{3}OH}^{\rm{east}}}=-24^{\circ}\pm90$\d) groups of \meth ~maser features in G23.44+0.18 (MM2).} 
\label{G23_cp}
\end{figure*}
\indent The analysis of the polarimetric data followed the procedure given in Paper~III and references 
therein. This consists in (1) identifying the individual \meth ~maser features by using the process described 
in Surcis et al. (\cite{sur11b}); (2) determining the mean linear polarization fraction ($P_{\rm{l}}$) and the 
mean linear polarization angle ($\chi$) across the spectrum of each \meth ~maser feature; (3) modeling the 
total intensity and the linearly polarized spectrum of every maser feature for which we were able to detect 
linearly polarized emission by using the adapted FRTM code for 6.7 GHz \meth 
~masers (Paper~III and references therein); and (4) measuring the Zeeman splitting by including the 
results obtained from point (3) for fitting the total intensity and circularly polarized spectra of the 
corresponding \meth ~maser feature. While  points (1) and (2) were applied as described and used in 
Paper~III (we refer the reader  there for more details), for  points (3) and (4) we had to modify 
the adapted FRTM code as follows:
\begin{itemize}
\item a new subroutine for calculating the Clebsch-Gordan coefficients was implemented, this provides more 
accurate values than previously;
\item  an error in the part of the code used for quantifying the Zeeman-splitting was corrected. We found that
the error led us to overestimate the Zeeman-splitting for the massive YSOs W51-e2, W48, IRAS\,06058+2138-IRS1,
S255-IR, IRAS\,20126+4104, G24.78+0.08, G29.86-0.04, and G213.70-12.6, and to underestimate for W3(OH) (in  three-quarters of the cases the 
corrections are less than a factor of 2 than those previously reported; 
Papers~I-III; Surcis et al. \cite{sur14}).
\end{itemize}
The outputs of the \code ~are the emerging brightness temperature (\tbo), the intrinsic thermal linewidth (\dvi), 
~and the angle between the magnetic field and the maser propagation direction ($\theta$). If $\theta>\theta_{\rm{crit}}=55$\d, 
where $\theta_{\rm{crit}}$ is the Van Vleck angle, the magnetic field appears to be perpendicular to the linear 
polarization vectors; otherwise, it is parallel (Goldreich et al. \cite{gol73}). The fits and models are performed 
for a sum of the decay and cross-relaxation rates of $\Gamma=1~\rm{s^{-1}}$. Since the emerging brightness 
temperature scales linearly with $\Gamma$ the fitted \tbo ~values can be adjusted by simply scaling according 
to $\Gamma$. Furthermore, when fitting the observed polarized \meth ~maser features we restricted our analysis 
to $10^6$~K~sr~<~\tbo~<~$10^{11}$~K~sr and to the most plausible range 0.5~\kms~<~\dvi~<~2.5~\kms, in steps of 0.05~\kms,
because it takes a prohibitively long time to fit for smaller \dvi ~values.  \\
\indent As in Paper~III, we consider  a detection of circularly polarized emission to be real only when the detected $V$ peak flux 
of a maser feature is both $V>5\cdot \rm{rms}$ and $V>3\cdot \sigma_{\rm{s.-n.}}$, where $\sigma_{\rm{s.-n.}}$ is 
the self-noise\footnote{The self-noise is high when the power contributed by the astronomical maser is a significant
portion of the total received power (Sault \cite{sau12}).} produced by the maser (Col.~8 of Table~\ref{Obs}; 
e.g., Sault \cite{sau12}). From the Zeeman effect theory we know that $\Delta V_{\rm{Z}}=\alpha_{\rm{Z}}\cdot B_{||}$, where
\dvz ~is the Zeeman-splitting, $B_{||}$ is the magnetic field strength along the line of sight, and $\alpha_{\rm{Z}}$ is
the Zeeman splitting coefficient that depends on the Land\'{e} g-factor(s). Following Lankhaar et al. (\cite{lan18}), who
identify for the 6.7 GHz methanol maser transition the hyperfine transition with the largest Einstein coefficient for
stimulated emission, i.e., $F=3\rightarrow4$, as
more favored among the eight hyperfine transitions that might contribute to the maser line, we estimated  
$B_{||}$ by assuming $\alpha_{\rm{Z}}=-0.051$~\kmsg~($\alpha_{\rm{Z}}=-1.135$~Hz~mG$^{-1}$, Lankhaar et al. \cite{lan18}). 
Considering that $\alpha_{\rm{Z}}=\mu_{\rm{N}}\cdot g_{\rm{l}}$, where $\mu_{\rm{N}}$ is the nuclear magneton and $g_{\rm{l}}$
is the Land\'{e} g-factor, and because $g_{\rm{l}}$ for $F=3\rightarrow4$ is the largest one among the eight hyperfine
transitions, our estimate of $B_{||}$ is therefore a lower limit. Even in the case of a combination of hyperfine
components the derived $B_{||}$ would be higher (e.g., Lankhaar et al. \cite{lan18}).

\section{Results}
\label{res}
In  Sects.~\ref{G23_sec}--\ref{G43_sec} the 6.7 GHz \meth ~maser distribution and the polarization results 
for each of the seven massive SFRs observed with the EVN are reported separately. The 
lists of all the maser features, with their properties, can be found in Tables~\ref{G23_tab}--\ref{G43_tab}.
\subsection{\object{G23.44-0.18}}
\label{G23_sec}
We detected a total of 61 \meth ~maser features, named G23.E01--G23.E61 in Table~\ref{G23_tab}, towards both
MM1 (27/61) and MM2 (34/61), see Fig.~\ref{G23_cp}, with the strongest maser features associated with MM1. We were
able to detect the weak maser features (e.g., G23.E29 and G23.E30) with velocities $>$+111~\kms ~previously detected with 
the Australia Telescope Compact Array (ATCA) by Breen et al. (\cite{bre15}).
The maser features distribution around the two cores are similar to what was
observed by Fujisawa et al. (\cite{fuj14}). The maser features associated with MM2 are distributed along two lines
separated by about 250~mas and with position angles of $\rm{PA_{MM2}^{east}}=+30$\d ~and $\rm{PA_{MM2}^{west}}=-24$\d 
~(see right panel of Fig.~\ref{G23_cp}). No velocity gradient is measured along them.\\
\indent Only linearly polarized emission has been detected, in particular from six maser features associated 
with MM1 ($P\rm{_l^{MM1}}=1.1-6.5\%$) and from five associated with MM2 ($P\rm{_l^{MM2}}=1.4-8.3\%$). The
FRTM code (see Sect.\ref{obssect}) was able to fit all of them and the outputs of the code are reported in
Cols.~9, 10, and 14 of Table~\ref{G23_tab}. We note that due to their high $P\rm{_l}$ the features G23.E25 and 
G23.E51 might be partially saturated. The estimated $\theta$ angles indicate that the 
magnetic field is perpendicular to the linear polarization vectors.\\
\indent Considering the rms and the $\sigma_{\rm{s.-n.}}$ (see Table~\ref{Obs}) for the strongest maser feature (G23.E03)
we would have been able to detect circularly polarized emission only if $P_{\rm{V}}>1\%$, which is twice the typical 
fraction ($P_{\rm{V}}\approx0.5\%$; e.g., Paper~III).
\begin{figure}[h!]
\centering
\includegraphics[width = 9 cm]{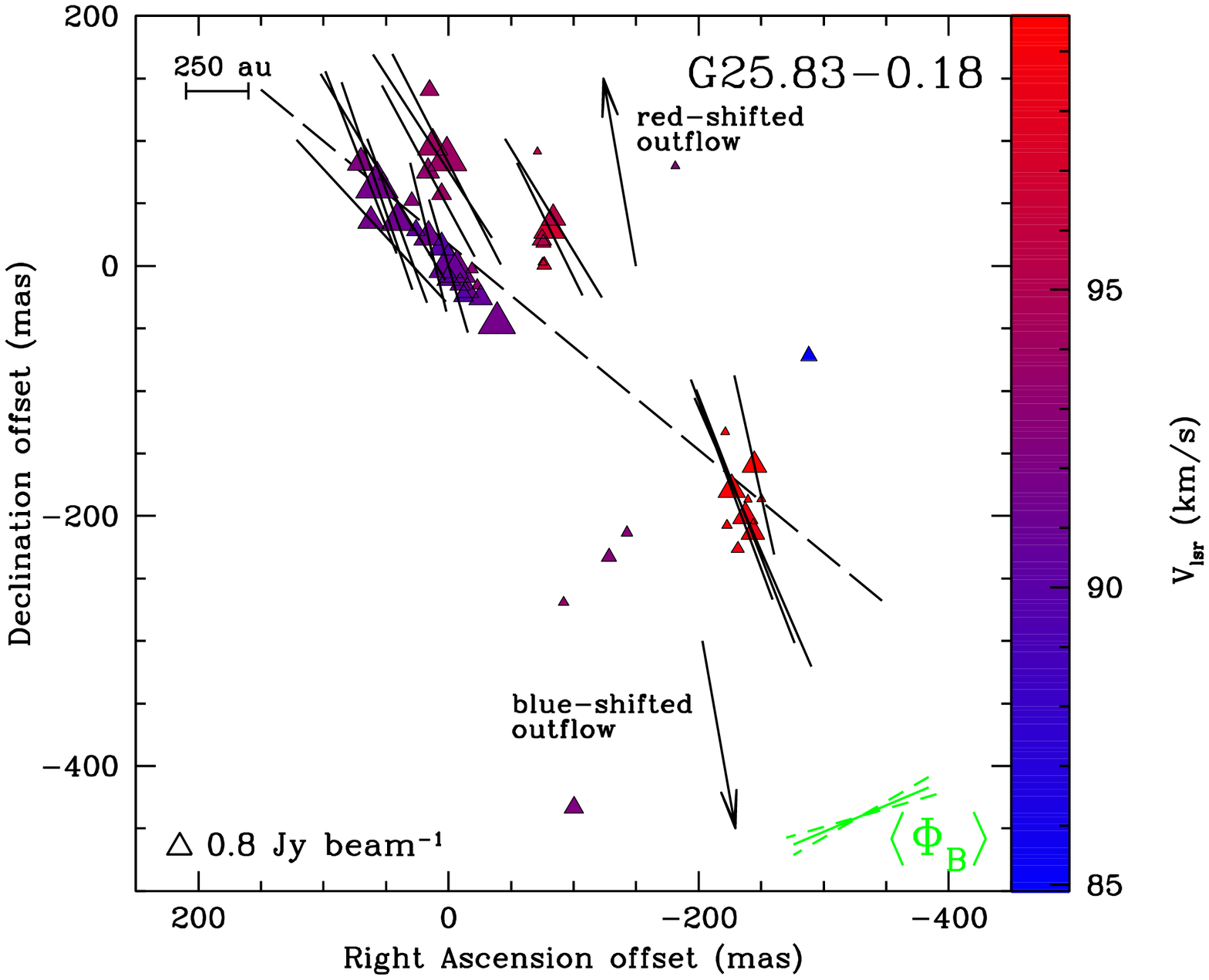}
\caption{View of the \meth ~maser features detected around G25.83-0.18 (Table~\ref{G258_tab}). Symbols are the same  
as in Fig.~\ref{G23_cp}. The polarization fraction is in the range $P\rm{_l}=2.3-9.7\%$
(Table~\ref{G258_tab}). The assumed velocity of the YSO is $V_{\rm{lsr}}^{\rm{{^{13}CO}}}=+93.2$~\kms 
~(de Villiers et al. \cite{dev14}). The two arrows indicate the direction, and not the actual position, of the 
red- and blue-shifted lobes of the bipolar outflow ($\rm{PA_{outflow}^{^{13}CO}}=+10$\d; de Villiers et 
al. \cite{dev14}). The dashed line is the best least-squares linear fit of the \meth ~maser features 
($\rm{PA_{CH_{3}OH}}=+51^{\circ}\pm7$\d).
} 
\label{G258_cp}
\end{figure}
\begin{figure}[h!]
\centering
\includegraphics[width = 9 cm]{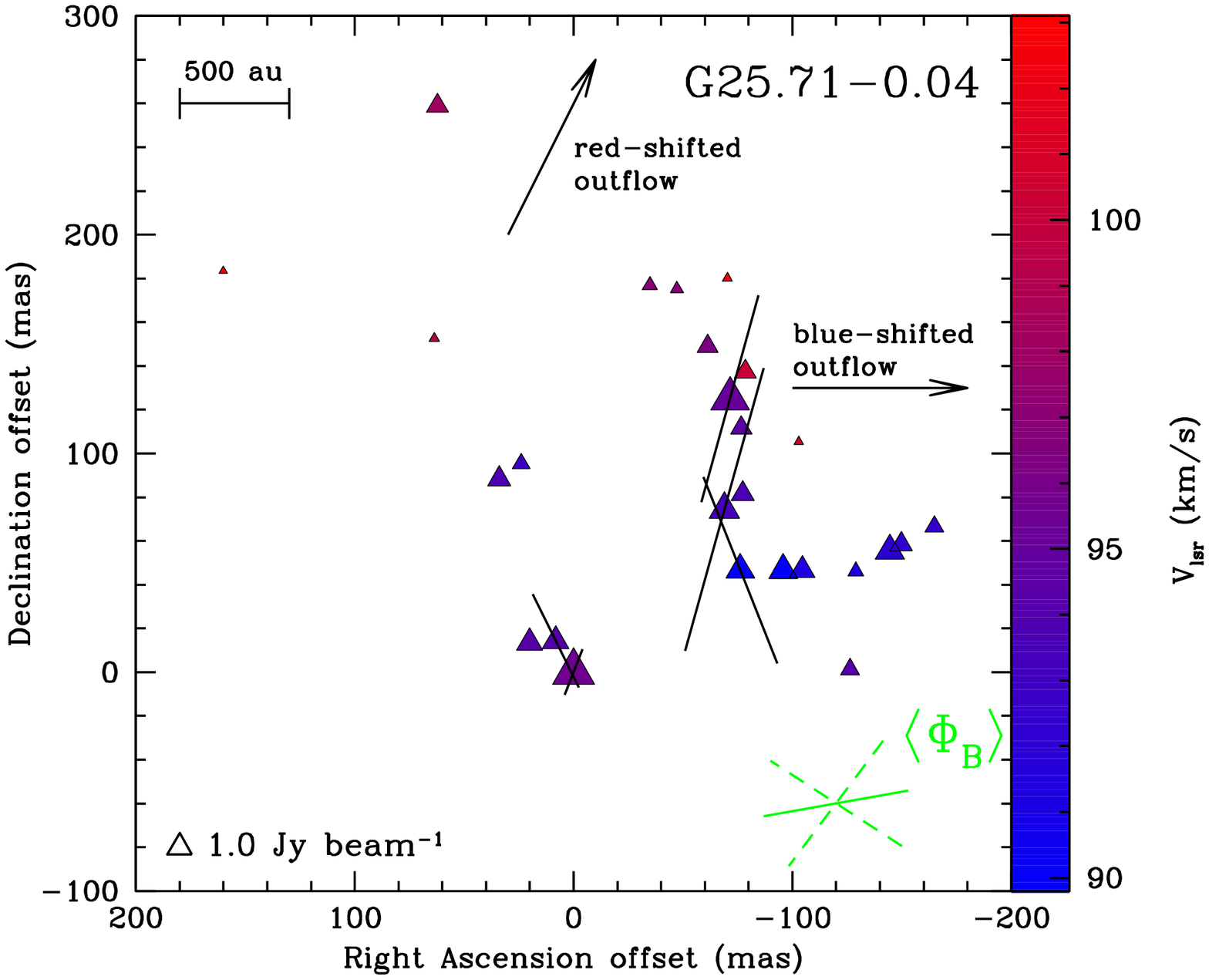}
\caption{View of the \meth ~maser features detected around G25.71-0.04 (Table~\ref{G257_tab}). Symbols are the same  
as in Fig.~\ref{G23_cp}. The polarization fraction is in the range $P\rm{_l}=0.4-1.6\%$ (Table~\ref{G257_tab}).
The assumed velocity of the YSO is $V_{\rm{lsr}}^{\rm{{^{13}CO}}}=101.3$~\kms ~(de Villiers 
et al. \cite{dev14}). The two arrows indicate the direction, and not the actual position, of the red- and blue-shifted
lobe of the bipolar outflow ($\rm{PA_{outflow, blue}^{\rm{^{13}CO}}}=-90$\d ~and $\rm{PA_{outflow, 
red}^{\rm{^{13}CO}}}=-27$\d, de Villiers et al. \cite{dev14}).
} 
\label{G257_cp}
\end{figure}
\subsection{\object{G25.83-0.18}}
\label{G258_sec}
In Table~\ref{G258_tab} we report the 46 \meth ~maser features detected towards G25.83-0.18 and named  
G258.E01--G258.E46. The very first VLBI distribution of these maser features is shown in Fig.~\ref{G258_cp}, where 
we also report the red- and blue-shifted orientation, but not at the actual position, of the $\rm{^{13}CO}$ outflow
(de Villiers et al. \cite{dev14}). The maser features, in particular the northwestern group, are aligned NW-SE 
($\rm{PA_{CH_{3}OH}}=+51^{\circ}\pm7$\d) without a clear velocity gradient. Their velocities, except for G258.E01 and 
G258.E28, are consistent with the red-shifted lobe of the outflow.\\
\indent Among the seven massive SFRs studied in the present work, G25.83-0.18 has the highest number of linearly
polarized features (16 out of 46).
The fractional linear polarization ranges from 2.3$\%$ to 9.7$\%$, which is one of the highest measured
so far. Although the high $P\rm{_l}$ indicates that five maser features might be partially saturated,
we were able to fit all of the polarized maser features with the \code. From the 
estimated $\theta$ angles (Col.~14 of Table~\ref{G258_tab}) the magnetic field is oriented perpendicular to the 
linear polarization vectors. No circular polarization was measured ($P_{\rm{V}}<0.9\%$).
\subsection{\object{G25.71-0.04}}
\label{G257_sec}
\begin{figure*}[th!]
\centering
\includegraphics[width = 8 cm]{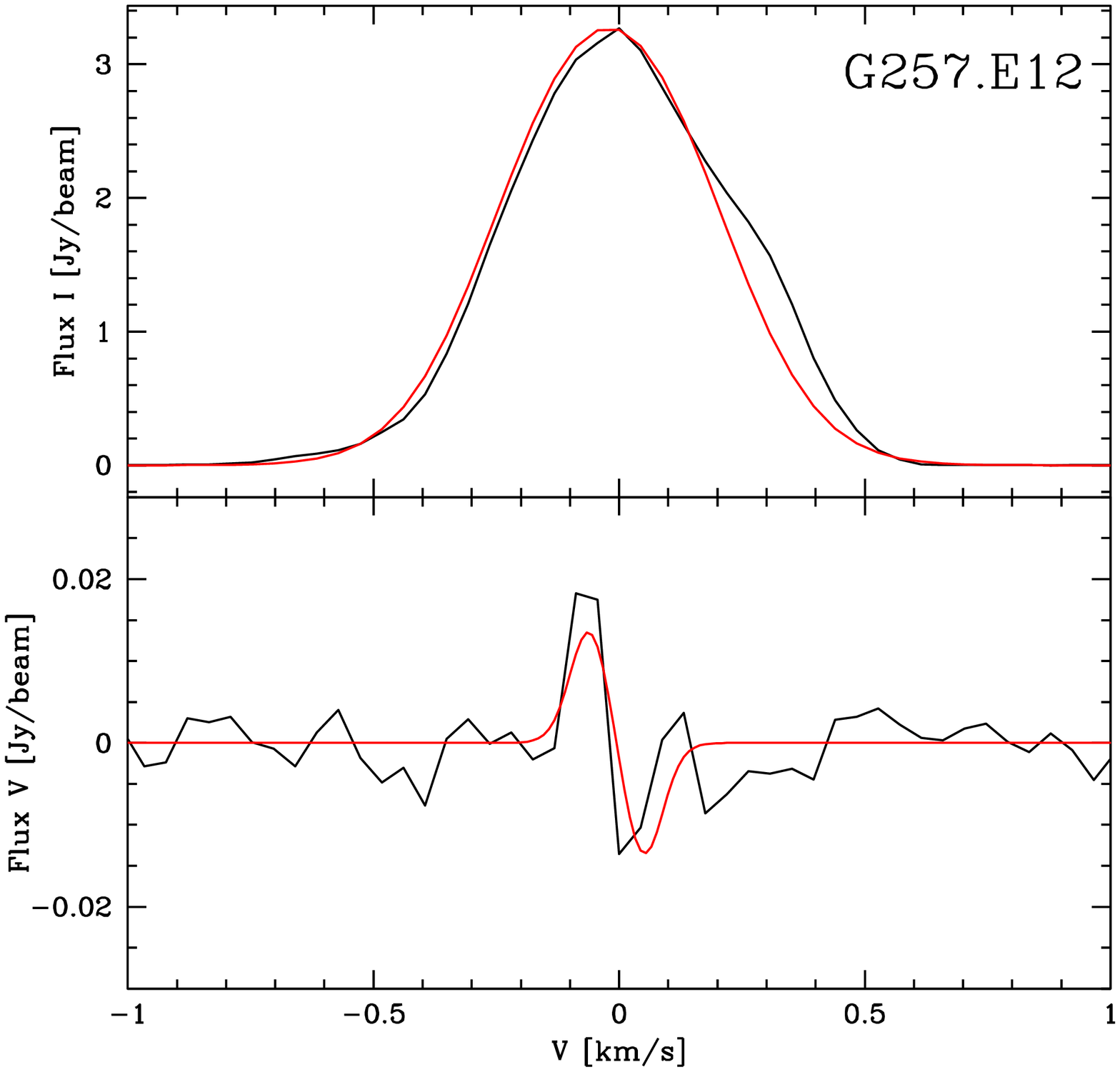}
\includegraphics[width = 8 cm]{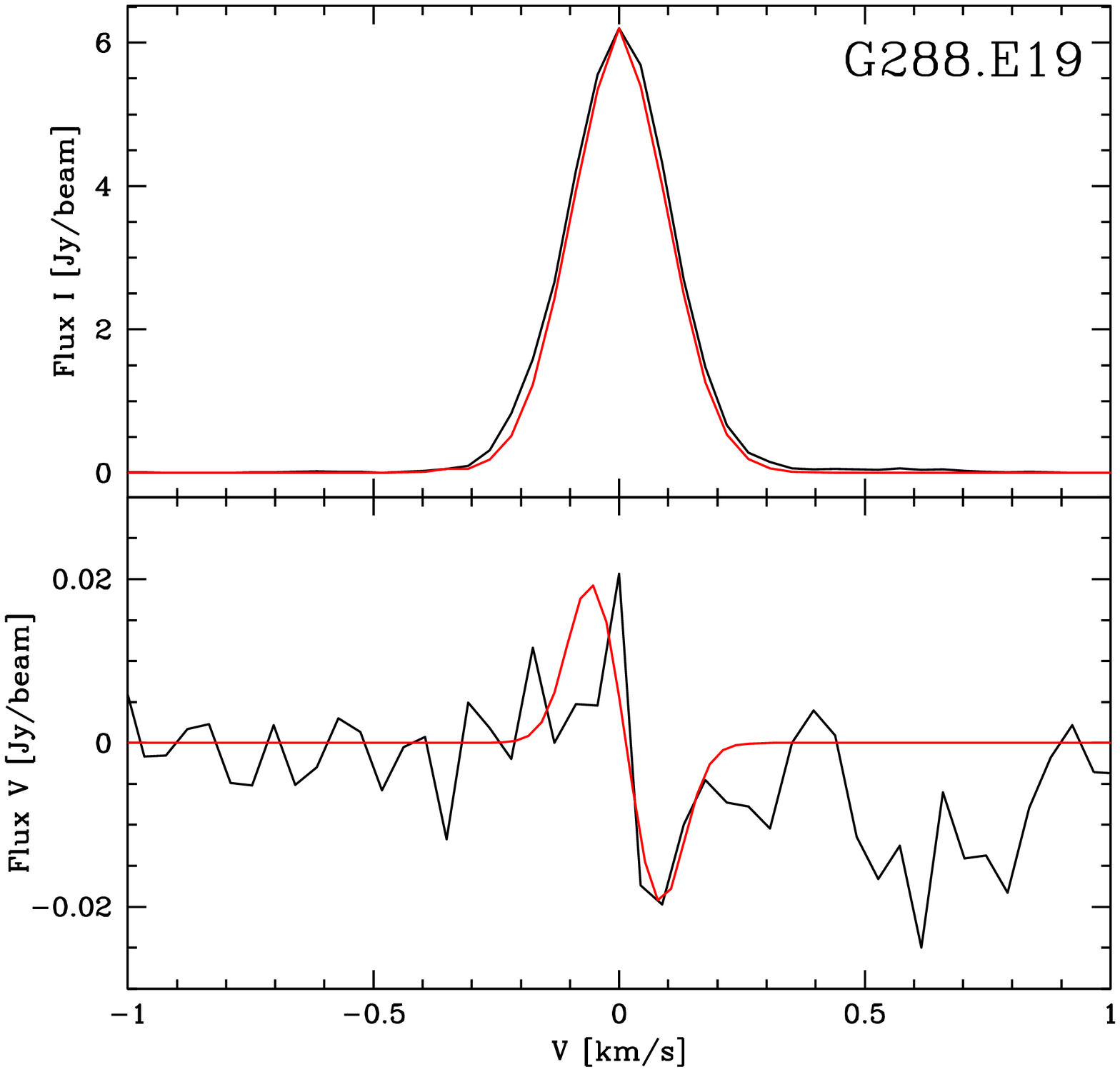}
\caption{Total intensity (\textit{I, upper panel}) and circularly polarized intensity (\textit{V, lower panel}) 
spectra for the \meth ~maser features named  G257.E12 and G288.E19 (see Tables~\ref{G257_tab}, \ref{G288_tab}).  
The thick red lines are the best-fit models of \textit{I} and \textit{V} emissions obtained using the adapted FRTM 
code (see Sect.~\ref{obssect}). The maser features were centered on zero velocity.}
\label{Vfit}
\end{figure*}

We identified 26 \meth ~maser features (named G257.E01--G257.E26 in Table~\ref{G257_tab}). Their complex 
distribution is shown in Fig.~\ref{G257_cp}. Here, the orientation of the red- and blue-shifted lobes 
of the $^{13}\rm{CO}$ outflow are also displayed (de Villiers et al. \cite{dev14}). The velocity range of 24
out of 26 \meth ~maser features (89~\kms$~<V_{\rm{lsr}}<$~100~\kms) is within the velocity range of the 
blue-shifted lobe (+92.3~\kms$<\rm{V_{blue}^{^{13}\rm{CO}}} <$+101.3~\kms; de Villiers et al. \cite{dev14}),
confirming their association with it.\\
\indent We detected linearly polarized emission towards five \meth ~maser features ($0.4\%<P_{\rm{l}}<1.6\%$), 
among which G257.E12 also showed circular polarization ($P_{\rm{V}}=0.8\%$). The 
error-weighted intrinsic thermal linewidth, $\langle$\dvi$\rangle=2.0^{+0.2}_{-0.4}$~\kms, is one of the largest 
ever provided by the \code, while the \tbo ~ values are within the typical estimated values for the 6.7~GHz \meth ~maser 
emission. Following Paper~III we found for the features G257.E13 (offset~$=-$71.456~mas; 125.005~mas) and 
G257.E19 (0~mas; 0~mas) that the magnetic field is parallel to
the linear polarization vectors since $|\theta^{\rm{+}}-55$\d$|<|\theta^{\rm{-}}-55$\d$|$, where 
$\theta^{\rm{\pm}}=\theta\pm\varepsilon^{\rm{\pm}}$, and with $\varepsilon^{\rm{\pm}}$ the errors associated with 
$\theta$. For all the other maser features the magnetic field is perpendicular. The circularly polarized emission
of G257.E12 was fitted with the \code ~by including the corresponding \dvi ~and \tbo ~fit values obtained from the
total and linearly polarized intensities. The fitted result is reported in Table~\ref{G257_tab} and shown in 
Fig.~\ref{Vfit}. The nondetection of circular polarization for the very bright maser features G257.E13 and G257.E19 is
likely due to the high $\sigma_{\rm{s.-n.}}$ (55 and 97 \mjyb, respectively). Considering the criterion 
$3\sigma_{\rm{s.-n.}}$, we have upper limits of $P_{\rm{V}}<0.8\%$ and $<0.6\%$ for G257.E13 and G257.E19, respectively.
\subsection{\object{G28.31-0.39}}
\label{G283_sec}
\begin{figure}[h!]
\centering
\includegraphics[width = 9 cm]{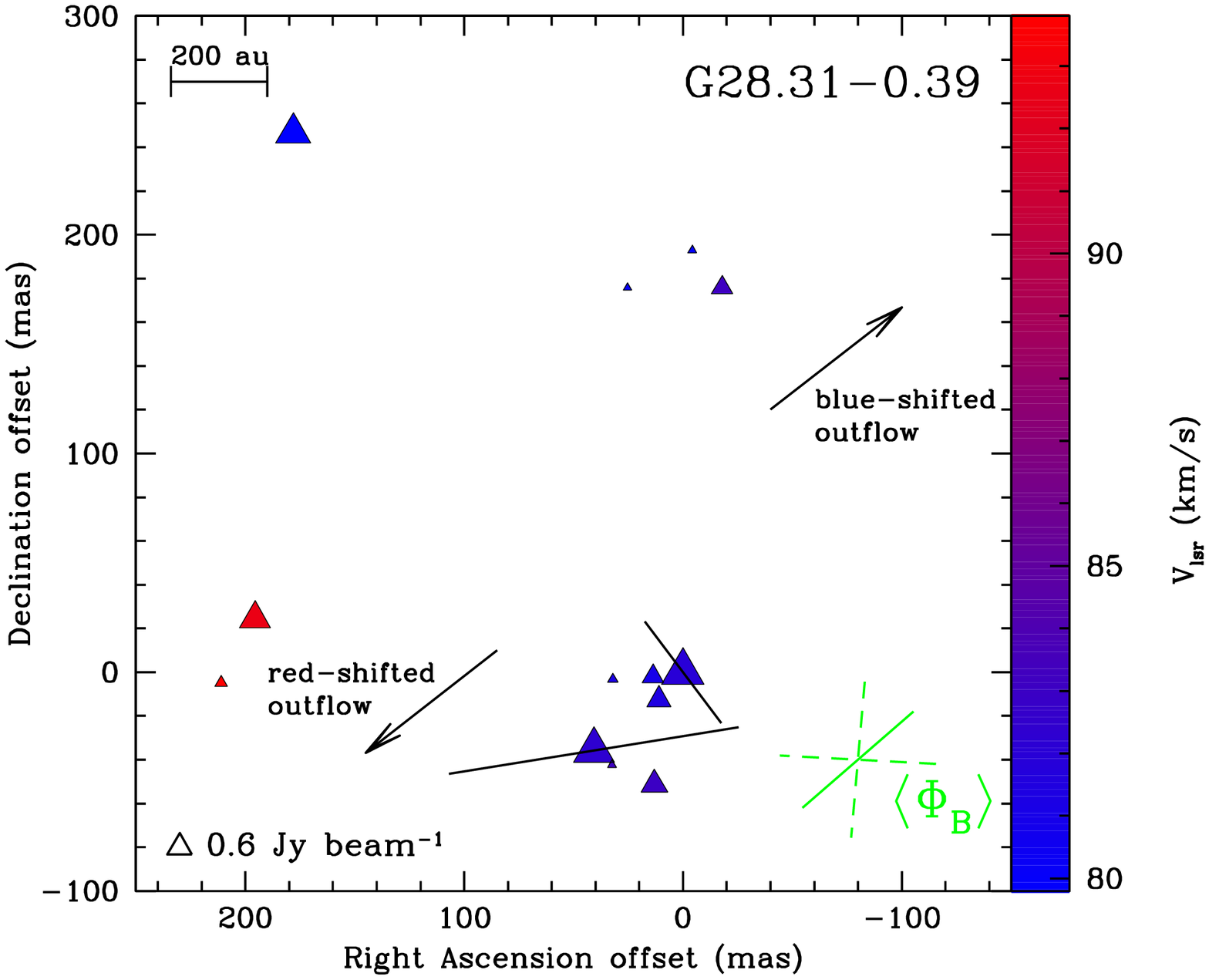}
\caption{View of the \meth ~maser features detected around G28.31-0.39 (Table~\ref{G283_tab}). Symbols are the same as in 
Fig.~\ref{G23_cp}. The polarization fraction is in the range $P\rm{_l}=1.6-4.3\%$ (Table~\ref{G283_tab}). 
The assumed velocity of the YSO is $V_{\rm{lsr}}^{\rm{{^{13}CO}}}=+85.9$~\kms ~(de Villiers 
et al. \cite{dev14}). The two arrows indicate the direction, and not the actual position, of the red- and blue-shifted lobe of
the bipolar outflow ($\rm{PA_{outflow}^{\rm{^{13}CO}}}=-52$\d,  de Villiers et al. \cite{dev14}).
} 
\label{G283_cp}
\end{figure}
Towards G28.31-0.39, we detected 13 \meth ~maser features (named G283.E01--G283.E13 in Table~\ref{G283_tab}).
The $V_{\rm{lsr}}$ of eleven features ranges from 79.91~\kms ~to 83.23~\kms, in accordance with 
+80.4~\kms$<V_{\rm{blue}}^{^{13}\rm{CO}} <$+85.9~\kms ~(de Villiers et al. \cite{dev14}). The maser features G283.E12 
and G283.E13 have higher velocities,  $V_{\rm{lsr}}=92.72$~\kms ~and $V_{\rm{lsr}}=93.81$~\kms, ~respectively, which 
are close to the highest velocity of the red-shifted lobe of the outflow ($\sim89$~\kms). In Fig.~\ref{G283_cp} 
the complex $3200~\rm{au}\times2400~\rm{au}$ distribution of the maser features, which resembles an X, is shown. \\
\indent The fit with the \code ~of the two linearly polarized maser features, G283.E03 and G283.E10, provided for both 
of them that $\theta>55$\d, i.e., that the magnetic field is perpendicular to the linear polarization vectors. Circular 
polarization has not been detected ($P_{\rm{V}}<0.5\%$).
\subsection{\object{G28.83-0.25}}
\label{G288_sec}
\begin{figure}[h!]
\centering
\includegraphics[width = 9 cm]{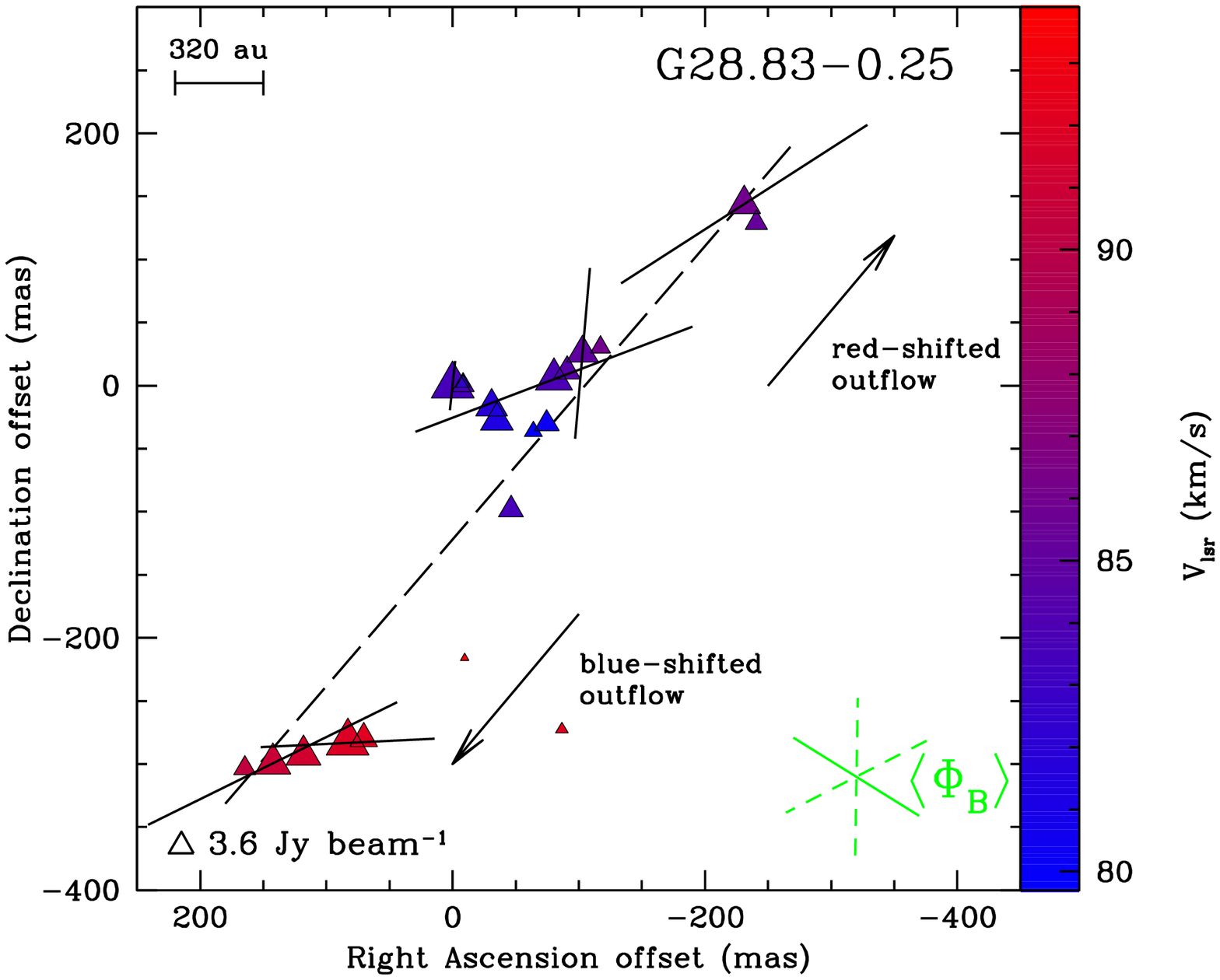}
\caption{View of the \meth ~maser features detected around G28.83-0.25 (Table~\ref{G288_tab}). Symbols are the same as in 
Fig.~\ref{G23_cp}. The polarization fraction is in the range $P\rm{_l}=0.5-3.3\%$ (Table~\ref{G288_tab}).
The assumed velocity of the YSO is  $V_{\rm{lsr}}^{\rm{{^{13}CO}}}=+88.4$~\kms ~(de Villiers 
et al. \cite{dev14}). The dashed line is the best least-squares  linear fit of the \meth ~maser features ($\rm{PA_{CH_{3}OH}}=-41^{\circ}\pm10$\d). The two arrows indicate the direction, and not the actual position, of the red- and blue-shifted lobe of the bipolar outflow ($\rm{PA_{outflow}^{\rm{^{13}CO}}}=-40$\d, de Villiers et al. \cite{dev14}). 
} 
\label{G288_cp}
\end{figure}

At VLBI scales we detected 21 6.7 GHz \meth ~maser features (named G288.E01--G288.E21 in Table~\ref{G288_tab}) linearly distributed
($\rm{PA_{CH_{3}OH}}=-41^{\circ}\pm10$\d; see Fig.~\ref{G288_cp}) from southwest (the most red-shifted) to northeast (around systemic 
velocity $V_{\rm{lsr}}^{\rm{{^{13}CO}}}=+88.4$~\kms; de Villiers et al. \cite{dev14}). The maser features at the 
center of the linear distribution show the most blue-shifted velocities, in accordance with Cyganowski et al. 
(\cite{cyg09}) and Fujisawa et al. (\cite{fuj14}). The velocity distribution of the maser features reflects an almost 
perfect overlap of the red- and blue-shifted lobe emissions of the $^{13}$CO outflow (de Villiers et al. 
\cite{dev14}). Therefore, it is difficult to associate the maser features with either the outflow or an accretion 
disk.\\
\indent We measured fractional linear polarization between 0.5\% and 3.3\% from six \meth ~maser features. 
For the brightest maser feature G288.E16, by fitting its polarized 
emission with the \code, we found that $\theta=61^{+3}_{-47}$. This implies that the magnetic field is more likely  
parallel to the linear polarization vector of G288.E16 (see Paper~III). For the other maser features this is instead 
perpendicular. Furthermore, we detected circularly polarized 
emission ($P_{\rm{V}}=0.6\%$) from G288.E19 that does not show linearly polarized emission. Hence, to 
model the circularly polarized emission we assumed that its emerging brightness temperature is equal to the 
error-weighted value $\langle$\tbo$\rangle=9.4\cdot10^8$~K~sr of the region;  for the intrinsic thermal linewidth we 
determined that \dvi$=1.1$~\kms ~is the value that best fits the total intensity emission (see right panel of 
Fig.~\ref{Vfit}). The $\sigma_{\rm{s.-n.}}$ for the bright maser features G288.E16 and G288.E18 are 81 and 64~\mjyb, which imply $P_{\rm{V}}<0.8\%$ and $<1.0\%$, respectively.
\subsection{\object{G29.96-0.02}}
\label{G29_sec}
\begin{figure}[h!]
\centering
\includegraphics[width = 9 cm]{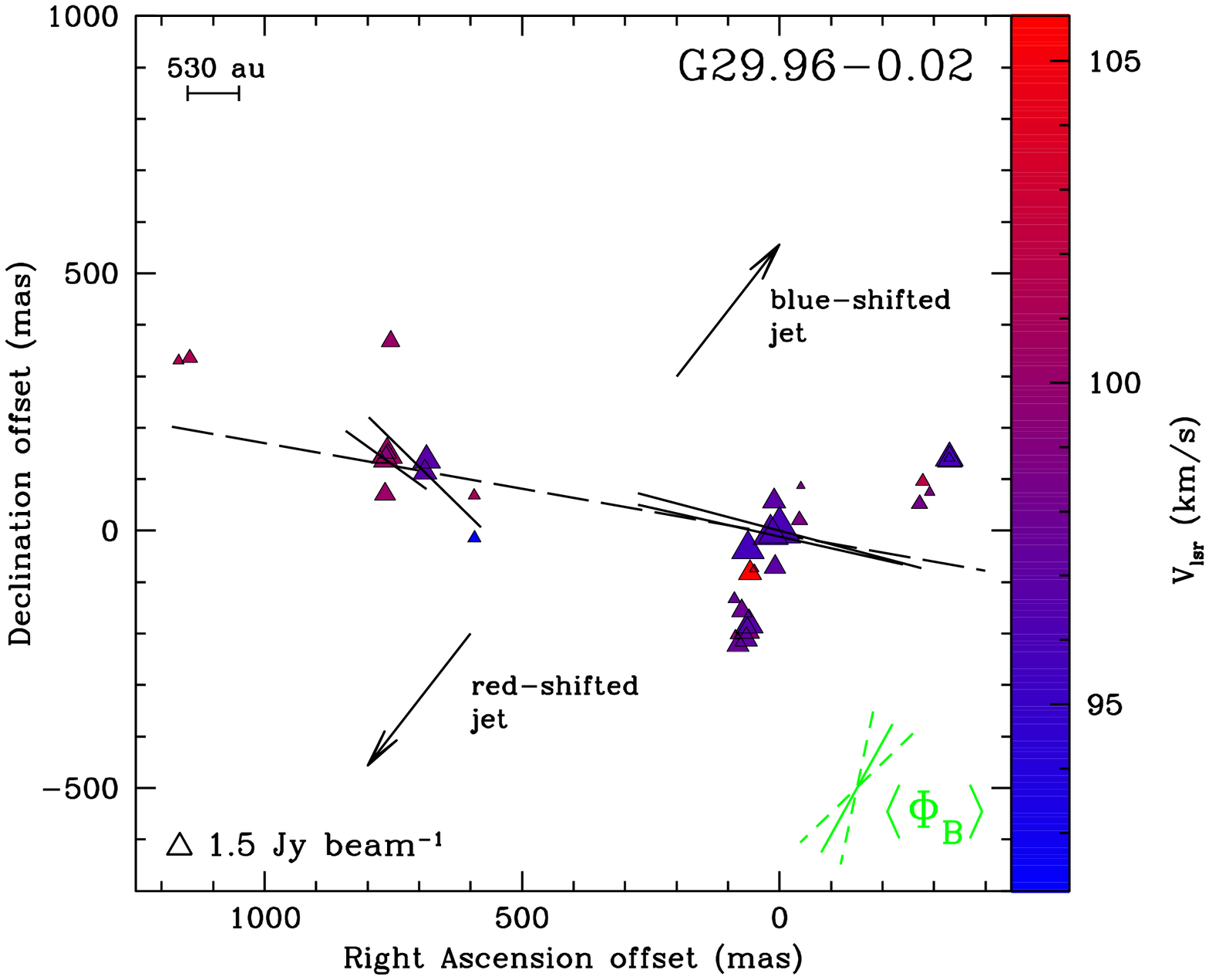}
\caption{View of the \meth ~maser features detected around G29.96-0.02 (Table~\ref{G29_tab}). Symbols are the same as in 
Fig.~\ref{G23_cp}. The polarization fraction is in the range $P\rm{_l}=1.4-5.7\%$ (Table~\ref{G29_tab}).
The assumed velocity of the YSO is $V_{\rm{lsr}}^{\rm{{^{13}CO}}}=+97.6$~\kms ~(de Villiers et al. 
\cite{dev14}). The dashed line is the best least-squares linear fit of the \meth ~maser features 
($\rm{PA_{CH_{3}OH}}=+80^{\circ}\pm3$\d). The two arrows indicate the direction, and not the actual position, of the 
red- and blue-shifted lobe of the bipolar jet ($\rm{PA_{jet}^{\rm{SiO}}}=-38$\d; Cesaroni et al. \cite{ces17}). 
} 
\label{G29_cp}
\end{figure}

We detected 34 6.7 GHz \meth ~maser features (named G29.E01--G29.E34 in Table~\ref{G29_tab}) towards the rotating disk 
of G29.96-0.02, distributed perpendicularly to the SiO jet from northeast to southwest 
($\rm{PA_{CH_{3}OH}}=+80^{\circ}\pm3$\d, Fig.~\ref{G29_cp}). Two of them (G29.E33 and G29.E34), the most northeastern, 
were previously undetected at VLBI scales (Sugiyama et al. \cite{sug08}). The velocities of the maser features 
ranges 
from 95.65~\kms ~to 105.75~\kms ~without an ordered distribution along the major axis of the rotating disk. Even though 
the association of most of the maser features with the rotating disk is plausible, some of them 
might be associated with the outflowing gas. This hypothesis can be verified only by measuring their proper motions.\\
\indent Linearly polarized emission ($P_{\rm{l}}=1.4\%-5.7\%$) was measured towards four maser features grouped two by
two: G29.E09 (the brightest) and G29.E12 towards the west and G29.E26 and G29.E30 towards the east. The west group shows 
 $P_{\rm{l}}>4\%$ and an error-weighted linear polarization angle of 
$\langle\chi\rangle_{\rm{west}}=+76^{\circ}\pm1$\d, while  the east group shows  $P_{\rm{l}}\lesssim2\%$ and 
$\langle\chi\rangle_{\rm{east}}=+49^{\circ}\pm6$\d. The total error-weighted linear polarization angle is 
$\langle\chi\rangle=+62^{\circ}\pm17$\d. The \code ~properly fits all of them and provided consistent outputs, 
though the \tbo ~is higher, as expected, for the west group, which  might be entering  the saturated state. 
The angle between the magnetic field and the maser propagation direction is greater than 55\d 
~for all  four maser features indicating that the magnetic field is perpendicular to the linear polarization 
vectors. No circularly polarized emission was detected ($P_{\rm{V}}<0.3\%$).  
\subsection{\object{G43.80-0.13}}
\label{G43_sec}
\begin{figure}[t!]
\centering
\includegraphics[width = 9 cm]{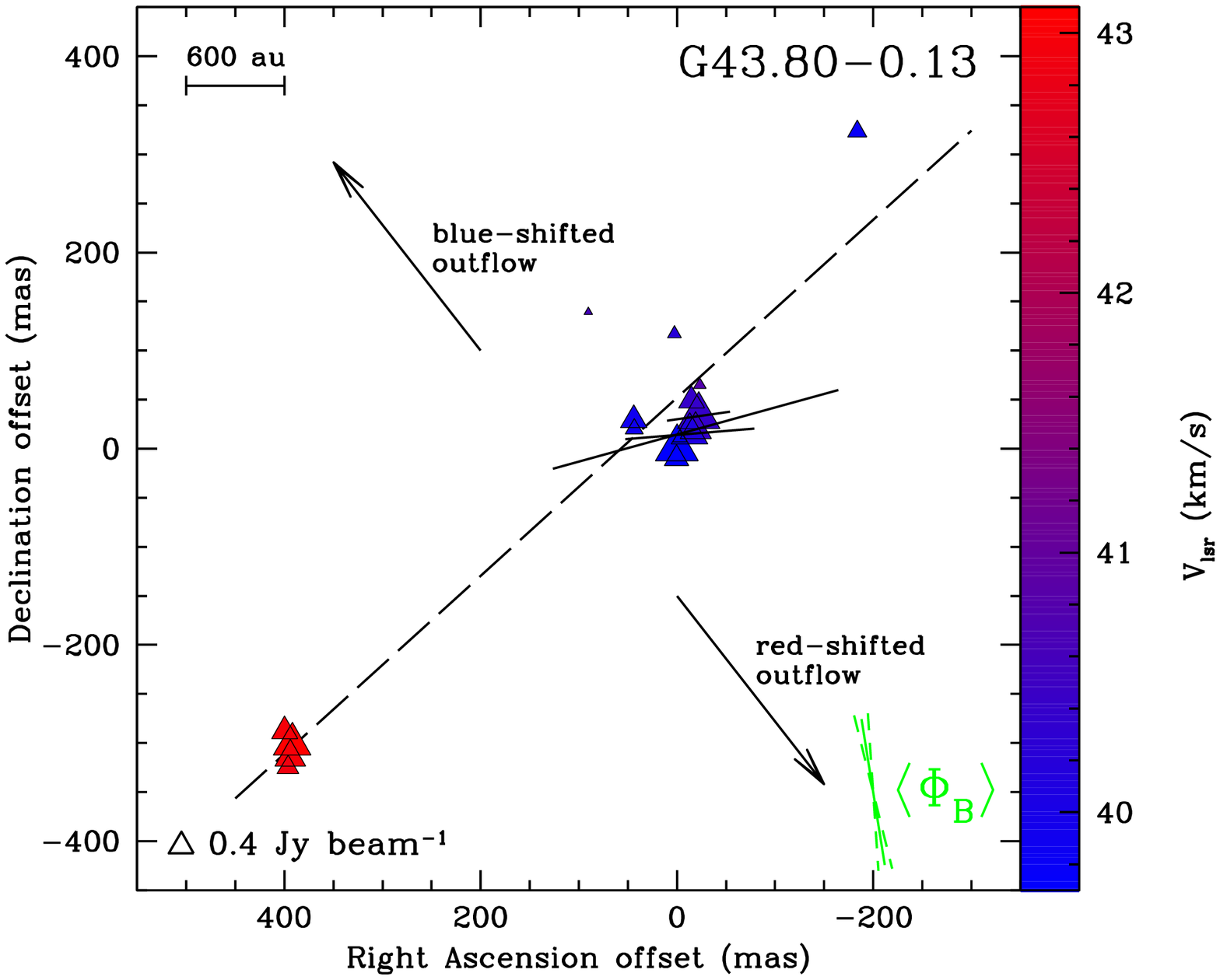}
\caption{View of the \meth ~maser features detected around G43.80-0.13 (Table~\ref{G43_tab}). Symbols are the same as in 
Fig.~\ref{G23_cp}. The polarization fraction is in the range $P\rm{_l}=1.1-4.4\%$ (Table~\ref{G43_tab}).
The assumed velocity of the YSO is $V_{\rm{lsr}}^{\rm{{C^{18}O}}}=+43.9$~\kms ~(L\'{o}pez-Sepulcre 
et al. \cite{lop10}). The dashed line is the best least-squares linear fit of the \meth ~maser features 
($\rm{PA_{CH_{3}OH}}=-48^{\circ}\pm5$\d). The two arrows indicate the direction, and not the actual position, of the 
red- and blue-shifted lobe of the bipolar outflow ($\rm{PA_{outflow}^{\rm{HCO^+}}}=+38$\d; L\'{o}pez-Sepulcre et al. 
\cite{lop10}). 
} 
\label{G43_cp}
\end{figure}
With the EVN we were able to detect twice the number of  6.7 GHz \meth ~maser features previously detected
by Sugiyama et al. (\cite{sug08}). These maser features are listed as G43.E01--G43.E18 in 
Table~\ref{G43_tab}, and are shown in Fig.~\ref{G43_cp} where the perpendicularity of their linear distribution 
($\rm{PA_{CH_{3}OH}}=-48^{\circ}\pm5$\d) to the HCO$^+$ outflow is clearly seen. The velocity range of the maser 
features (39.5~\kms$<V_{\rm{lsr}}<$43.2~\kms) is blue-shifted with respect to the systemic velocity of the region 
($V_{\rm{lsr}}^{\rm{{C^{18}O}}}=+43.9$~\kms, L\'{o}pez-Sepulcre et al. \cite{lop10}) and it is consistent with the 
range reported in Sugiyama et al. (\cite{sug08}). In particular, the velocities of the maser features located to the  
southeast (G43.E15--G43.E18) are closer to the systemic velocity, and the most blue-shifted are located
towards the  northeast.\\
\indent Three maser features out of 18 showed linearly polarized emission with $1\%<P_{\rm{l}}<4.5\%$, i.e., they are
likely unsaturated. Despite  this finding, the \code ~was able to provide only an upper limit of \dvi ~for 
G43.E04. The estimated $\theta$ angles indicate that the magnetic field is oriented perpendicular to the linear 
polarization vectors. No circularly polarized emission was detected, likely due to the weakness of the maser features
($\rm{I}<6$~\jyb; $P_{\rm{V}}<0.8\%$).
\section{Discussion}
\label{discussion}
\subsection{Magnetic field orientations}
\label{Borient}
If the linearly polarized emission passes through a medium where a magnetic field is present before reaching the 
observer, its linear polarization vector suffers a rotation known as Faraday rotation. In the case
of  polarized \meth ~maser emission, two main Faraday rotations can affect the linear polarization vectors that
we measure:  internal rotation ($\Phi_{\rm{i}}$) and  foreground Faraday rotation ($\Phi_{\rm{f}}$). An analysis 
of these two components of Faraday rotation was undertaken in the previous papers in the series
(Papers~I, II, and III); this analysis helps  understand why we do not consider them important here. However, in Col.~2 of Table~\ref{Comp_ang} 
we list $\Phi_{\rm{f}}$ for each source.\\
\indent The estimated orientation of the magnetic field\footnote{$\langle\Phi_{\rm{B}}\rangle$ is the mean error-weighted
orientation of the magnetic field measured considering all the magnetic field vectors measured in a
source. The weights are $1/e_{\rm{i}}$, where $e_{\rm{i}}$ is the error of the $i$th measured vector. The error on
$\langle\Phi_{\rm{B}}\rangle$ is the standard deviation. The position angle of the magnetic field vectors
are measured with respect to  the north, clockwise (negative) and counterclockwise (positive), as the PA of the outflows.} in the
seven massive SFRs under investigation here are separately discussed below.\\

\noindent\textit{\textbf{G23.44-0.18.}} The magnetic fields around MM1 and MM2 are oriented SE-NW with 
error-weighted orientation of $\langle\Phi_{\rm{B}}^{\rm{MM1}}\rangle=-32^{\circ}\pm64$\d ~and 
$\langle\Phi_{\rm{B}}^{\rm{MM2}}\rangle=-67^{\circ}\pm40$\d. It should be noted that the CO outflow on the plane of the 
sky is oriented with an angle of $\rm{PA_{outflow}^{^{12}CO}}=-40$\d ~and shows an opening 
angle of $\sim 30$\d ~(Ren et al. \cite{ren11}). However, the velocity range of the masers both in MM1 
and in MM2 falls between the velocity ranges of the LVC and HVC CO outflows  indicating 
that the masers are not associated with either of them. The eastern maser group of MM2 shows a linear 
distribution perpendicular to the CO outflow (see Table~\ref{Comp_ang}) suggesting a possible disk structure, 
though the velocity distribution of the masers does not.\\

\noindent\textit{\textbf{G25.83-0.18.}} The magnetic field is oriented with an angle on the plane of the sky of 
$\langle\Phi_{\rm{B}}\rangle=-67^{\circ}\pm7$\d, which is almost perpendicular both to the linear distribution of 
the 6.7~GHz \meth ~masers ($\rm{PA_{CH_{3}OH}}=+51^{\circ}\pm7$\d) and to the $\rm{^{13}CO}$ outflow
($\rm{PA_{outflow}^{^{13}\rm{CO}}}\sim+10$\d, de Villiers et al. \cite{dev14}). \\

\noindent\textit{\textbf{G25.71-0.04.}} Taking into account the different orientation of the magnetic field with 
respect to the linear polarization vectors of the maser features, we measured an error-weighted  magnetic field 
orientation of $\langle\Phi_{\rm{B}}\rangle=-80^{\circ}\pm43$\d, implying that it is aligned with the  
blue-shifted lobe of the $\rm{^{13}CO}$ outflow ($\rm{PA_{outflow}^{^{13}CO}}=-90$\d; de Villiers et al. 
\cite{dev14}). The location and velocities of the \meth ~masers also suggest that the masers probe the magnetic field along the blue-shifted lobe of the outflow.\\

\noindent\textit{\textbf{G28.31-0.39.}} We were able to determine the orientation of the magnetic field on the
plane of the sky from two blue-shifted \meth ~masers. Also in this case, the magnetic field 
($\langle\Phi_{\rm{B}}\rangle=-50^{\circ}\pm44$\d) is aligned with the $^{13}$CO outflow 
($\rm{PA_{outflow}^{^{13}CO}}=-52$\d, de Villiers et al. \cite{dev14}).\\

\noindent\textit{\textbf{G28.83-0.25.}}
Considering the different orientation of the magnetic field with respect to the linear polarization vector of 
G288.16, the error-weighted orientation of the magnetic field on the plane of the sky is 
$\langle\Phi_{\rm{B}}\rangle=+58^{\circ}\pm59$\d. This is perpendicular to the linear distribution of the masers
($\rm{PA_{CH_{3}OH}}=-41^{\circ}\pm10$\d) and to the $^{13}$CO outflow ($\rm{PA_{CH_{3}OH}}=-40$\d), even 
though the outflow is almost along the line of sight. Therefore, the interpretation of the morphology of the 
magnetic field is not straightforward.\\
\begin {table}[h!]
\caption []{$|B_{\rm{crit}}|$ values for the ranges of $n_{\rm{H_{2}}}$ and $T_{\rm{k}}$.} 
\begin{center}
\scriptsize
\begin{tabular}{ c c c c }
\hline
\hline
\multicolumn{4}{c}{$\beta=1$}\\
\hline
\multicolumn{2}{c|}{$T_{\rm{k}}=100$~K} & \multicolumn{2}{c}{$T_{\rm{k}}=200$~K}\\
\hline
\multicolumn{1}{c|}{$n_{\rm{H_{2}}}=10^5~\rm{cm^{-3}}$} & \multicolumn{1}{c|}{$n_{\rm{H_{2}}}=10^9~\rm{cm^{-3}}$} & \multicolumn{1}{c|}{$n_{\rm{H_{2}}}=10^5~\rm{cm^{-3}}$} & $n_{\rm{H_{2}}}=10^9~\rm{cm^{-3}}$\\
\hline
 0.2~mG   & \multicolumn{1}{c|}{18~mG} & 0.3~mG & 25~mG\\
\hline
\hline
\end{tabular}
\end{center}
\label{Bcrit}
\end{table}
\begin {table*}[th!]
\caption []{Magnetic fields measurements around massive YSOs determined by observing, with 
the EVN, the circularly polarized emission of 6.7~GHz \meth ~masers.} 
\begin{center}
\scriptsize
\begin{tabular}{ l c c c c c c c c c c c}
\hline
\hline
\,\,\,\,\,(1)&    (2)   &   (3)     &     (4)       &                (5)                   &                   (6)                     & (7)               & (8)          & (9)                 & (10)                           &    (11)              &\\
Source name  & Maser ID & Peak flux & $V_{\rm{lsr}}$& $\Delta V_{\rm{i}}\tablefootmark{a}$ & $T_{\rm{b}}\Delta\Omega\tablefootmark{a}$ & $\theta$          & $P_{\rm{V}}$ & $\Delta V_{\rm{Z}}$ & $|B_{||}|\tablefootmark{b}$    & $|B|\tablefootmark{c}$ & ref.$\tablefootmark{d}$ \\
             &          & Density(I)&               &                                      &                                            &                   &              &                     &                                &                      &                      \\ 
             &          & (Jy/beam) &  (km/s)       &               (km/s)                 &              (log K sr)                   & (\d)              &   ($\%$)     &  (m/s)              &  (mG)                          &      (mG)            &  \\ 
\hline
W75N(B)      & A4       & 47.58     &  +5.82        &             $-$                      &                 $-$                       &  $-$              &    $0.5$     & $+0.80\pm0.03\tablefootmark{e}$      & $>16\tablefootmark{e}$ &      $-$             & (1) \\  
             & A5       & 39.39     &  +5.12        &             $-$                      &                 $-$                       &  $-$              &    $0.5$     & $+0.75\pm0.13\tablefootmark{e}$      & $>15\tablefootmark{e}$ &      $-$             & (1) \\ 
             & B1       & 95.38     &  +7.23        &             $-$                      &                 $-$                       &  $-$              &    $0.4$     & $+0.53\pm0.04\tablefootmark{e}$      & $>10\tablefootmark{e}$ &      $-$             & (1) \\     
             &          &           &               &                                      &                                           &                   &              &                     &                                &                      &   \\      
NGC\,7538    & E02      & 16.82     &  -50.49       &             $1.0$                    &                 $9.41$                    &  $77^{+13}_{-11}$ &    $1.7$     &  $+2.7\pm0.3\tablefootmark{e}$       &  $>53\tablefootmark{e}$      &      $>104$ & (2) \\      
             & E26      & 95.15     &  -55.92       &             $0.5$                    &                 $11.39$                   &  $86^{+5}_{-19}$  &    $1.0$     &  $+1.6\pm0.3\tablefootmark{e}$       &  $>31\tablefootmark{e}$      &      $>62$ & (2) \\      
             & E48      & 23.68     &  -58.03       &             $1.2$                    &                 $9.26$                    &  $75^{+15}_{-35}$ &    $1.7$     &  $-2.7\pm0.3\tablefootmark{e}$       &  $>53\tablefootmark{e}$      &      $>65$ & (2) \\    
             &          &           &               &                                      &                                           &                   &              &                     &                                &                      &   \\           
W51-e2       & W51E.13  & 65.78     &  +57.86       &             $1.7$                    &                 $8.9$                     &  $77^{+13}_{-37}$ &    $0.4$     &   $-0.9\pm0.2$      &       $>17$                    &          $>21$       & (3) \\      
             & W51E.14  & 217.32    &  +59.26       &             $1.7$                    &                 $8.8$                     &  $67^{+11}_{-42}$ &    $0.1$     &  $+0.26\pm0.06$     &       $>5$                     &          $>5$       & (3) \\      
             & W51E.18  & 27.56     &  +59.35       &             $-$                      &                 $-$                       &  $-$              &    $1.1$     & $-2.1\pm0.5\tablefootmark{f}$ & $>42\tablefootmark{f}$ & $>51\tablefootmark{f}$    & (3) \\      
             & W51E.32  & 8.11      &  +57.86       &             $0.6$                    &                 $8.8$                     &  $75^{+14}_{-35}$ & $-\tablefootmark{g}$ & $-\tablefootmark{g}$ & $-\tablefootmark{g}$  & $-\tablefootmark{g}$ & (3) \\    
             &          &           &               &                                      &                                           &                   &              &                     &                                &                      &   \\           
W48          & W48.14   & 294.68    &  +44.49       &             $1.0$                    &                 $9.5$                     &  $73^{+16}_{-32}$ &    $0.7$     &  $-1.1\pm0.2$       &          $>22$                 &      $>27$           & (3) \\     
             &          &           &               &                                      &                                           &                   &              &                     &                                &                      &   \\          
W3(OH)       & W3OH.11  & 212.03    &  -42.60       &             $-$                      &                 $-$                       &  $-$              &    $2.9$     &   $-4.5\pm0.7$      & $>89\tablefootmark{h}$         &      $>111\tablefootmark{h}$   & (3) \\      
             & W3OH.22  & 2051.30   &  -45.41       &             $1.0$                    &                 $10.0$                    &  $73^{+10}_{-5}$  &    $2.1$     &   $+4.5\pm0.7$      & $>88$                          &      $>182$          & (3) \\      
             & W3OH.25  & 156.18    &  -43.74       &             $1.1$                    &                 $8.8$                     &  $68^{+7}_{-45}$  &    $2.3$     &   $+3.4\pm0.5$      & $>66$                          &      $>70$          & (3) \\      
             & W3OH.35  & 347.43    &  -42.86       &             $0.9$                    &                 $9.2$                     &  $71^{+18}_{-33}$ &    $3.8$     &   $+5.8\pm0.9$      & $>113$                         &      $>136$          & (3) \\      
             & W3OH.37  & 110.11    &  -42.51       &             $1.4$                    &                 $9.2$                     &  $76^{+13}_{-37}$ &    $5.0$     &  $-10.9\pm1.6$      & $>213$                         &      $>260$          & (3) \\      
             & W3OH.40  & 178.62    &  -45.14       &             $-$                      &                 $-$                       &  $-$              &    $1.3$     &   $-2.0\pm0.3$      & $>39\tablefootmark{i}$         &      $-$             & (3) \\      
             & W3OH.48  & 19.97     &  -41.81       &             $1.0$                    &                 $9.0$                     &  $76^{+11}_{-40}$ &    $1.5$     &   $+8.4\pm1.8$      & $>165$                         &      $>195$          & (3) \\  
             &          &           &               &                                      &                                           &                   &              &                     &                                &                      &   \\     
IRAS\,06058+2138-IRS\,1 & IRAS06.22 & 93.29 & +10.86 &            $1.2$                    &                 $9.6$                     &  $73^{+17}_{-7}$  &    $0.2$     &   $+0.4\pm0.2$      & $>9$                           &      $>17$           & (4) \\   
             &          &           &               &                                      &                                           &                   &              &                     &                                &                      &   \\    
S255-IR      & S255.30  & 10.64     &  +4.61        &             $1.1$                    &                 $9.5$                     &  $82^{+8}_{-18}$  &    $0.3$     &   $+0.4\pm0.1$      & $>9$                           &      $>16$           & (4) \\   
             &          &           &               &                                      &                                           &                   &              &                     &                                &                      &   \\    
IRAS\,20126+4104 & M05  & 27.84     &  -6.10        &             $2.0$                    &                 $8.8$                     &  $75^{+10}_{-43}$ &    $0.7$     &   $-1.8\pm0.3$      & $>36$                          &      $>41$           & (5) \\   
             &          &           &               &                                      &                                           &                   &              &                     &                                &                      &   \\    
G24.78+0.08  & G24.16   & 13.70     &  +110.41      &             $1.7$                    &                 $8.6$                     &  $82^{+8}_{-44}$  &    $0.3$     &   $-0.6\pm0.2$      & $>11$                          &      $>13$           & (6) \\      
             & G24.23   & 9.79      &  +107.69      &             $1.9$                    &                 $8.8$                     &  $67^{+9}_{-44}$  &    $0.7$     &   $+1.7\pm0.3$      & $>34$                          &      $>36$           & (6) \\      
             & G24.43   & 11.48     &  +114.23      &             $1.6$                    &                 $8.9$                     &  $88^{+1}_{-21}$  &    $0.7$     &   $-1.5\pm0.3$      & $>30$                          &      $>60$           & (6) \\      
             & G24.52   & 32.97     &  +113.40      &             $1.8$                    &                 $8.8$                     &  $73^{+2}_{-40}$  &    $0.3$     &   $-3.7\pm0.6$      & $>73$                          &      $>84$           & (6) \\    
             &          &           &               &                                      &                                           &                   &              &                     &                                &                      &   \\   
G29.86-0.04  & G29.09   & 19.26     &  +100.39      &             $1.8$                    &                 $8.6$                     &  $74^{+15}_{-37}$ &    $0.6$     &   $-1.2\pm0.2$      & $>24$                          &      $>28$           & (6) \\   
             &          &           &               &                                      &                                           &                   &              &                     &                                &                      &   \\    
G213.70-12.6 & G213.15  & 91.58     &  +12.57       &             $1.1$                    &                 $9.5$                     &  $82^{+8}_{-17}$  &    $0.7$     &   $-1.1\pm0.2$      & $>22$                          &      $>42$           & (6) \\  
\hline       
\end{tabular} \end{center}
\tablefoot{
\tablefoottext{a}{Values used to model the circularly polarized emission as evaluated from the linearly 
polarized emission.}
\tablefoottext{b}{The lower limit of the magnetic field strength along the line of sight was determined by modeling the 
circularly polarized emission assuming that the favored hyperfine transition is $F=3\rightarrow4$ (Lankhaar et 
al. \cite{lan18}). The Land\'{e} g-factor for this transition is $g_{\rm{l}}=-1.135~\rm{Hz~mG^{-1}}$ (or 
$\alpha_{\rm{Z}}=-0.051$~\kmsg).}
\tablefoottext{c}{$B=B_{||}/cos~(\theta+\varepsilon^{-}_{\rm{\theta}})$; considering that $\theta^{\varepsilon^{+}_{\rm{\theta}}}_{\varepsilon^{-}_{\rm{\theta}}}$.}
\tablefoottext{d}{References: (1) Surcis et al. (\cite{sur09}); (2) Surcis et al. (\cite{sur11a}); 
(3) Paper~I; (4) Paper~II; (5) Surcis et al. (\cite{sur14}); (6) Paper~III.}
\tablefoottext{e}{Due to the scarce spectral resolution, $B_{||}$ was determined from the cross-correlation between the 
RR and LL spectra, i.e., $\frac{B_{||}}{[\rm{mG}]}=(\frac{\Delta V_{\rm{Z}}}{[\rm{m~s^{-1}}]})\cdot(\frac{\alpha_{\rm{Z}}}{[\rm{km~s^{-1}~G^{-1}}]})^{-1}$.}
\tablefoottext{f}{To model the circularly polarized emission we considered the values of \tbo$=1.1\cdot10^9$~K~sr, \dvi$=0.7$~\kms, and $\theta={79^{\circ}}^{+11^{\circ}}_{-40^{\circ}}$
measured for W51-e2.}
\tablefoottext{g}{No 4$\sigma$ ~detection of the Zeeman splitting has been measured.}
\tablefoottext{h}{To model the circularly polarized emission we considered the error-weighted values of $\langle$\tbo$\rangle=1.8\cdot10^9$~K~sr, $\langle$\dvi$\rangle=0.6$~\kms, and 
$\langle\theta\rangle={78^{\circ}}^{+12^{\circ}}_{-37^{\circ}}$ measured for Group VI in W3(OH).}
\tablefoottext{i}{To model the circularly polarized emission we considered the values of \tbo$=2.2\cdot10^9$~K~sr and \dvi$=0.5$~\kms ~that best fit the total intensity spectrum.}
}
\label{B_tab}
\end{table*}

\noindent\textit{\textbf{G29.96-0.02.}}
The magnetic field is oriented on the plane of the sky along the  SiO jet ($\rm{PA_{jet}^{SiO}}=-38$\d, Cesaroni et al. 
\cite{ces17}) with an angle of $\langle\Phi_{\rm{B}}\rangle=-29^{\circ}\pm$17\d ~and perpendicular to the 
rotating disk (Cesaroni et al. \cite{ces17}). Considering separately the east and west group (
$\langle\Phi_{\rm{B}}\rangle_{\rm{east}}=-41^{\circ}\pm6$\d ~and 
$\langle\Phi_{\rm{B}}\rangle_{\rm{west}}=-14^{\circ}\pm1$\d), we note that the morphology of the magnetic field
coincides with the morphology of the red-shifted lobe of the SiO jet near to the massive YSO (see Fig.13 of 
Cesaroni et al. \cite{ces17}).\\

\noindent\textit{\textbf{G43.80-0.13.}}
The orientation of the magnetic field ($\langle\Phi_{\rm{B}}\rangle=+9^{\circ}\pm$5\d) derived from the masers, all 
of which belong to the most blue-shifted group, is misaligned by about 30\d ~compared to the HCO$^+$ outflow
($\rm{PA_{outflow}^{\rm{HCO^+}}}=+38$\d; L\'{o}pez-Sepulcre et al. \cite{lop10}).
\subsection{Magnetic field strength}
\label{Bstrength}
Thanks to the work of Lankhaar et al. (\cite{lan18}), we are able to estimate a lower limit of $B_{||}$ from the
measurements of the Zeeman splitting of the 6.7~GHz \meth ~maser by assuming $\alpha_{\rm{Z}}=-0.051$~\kmsg ~(Lankhaar et al. 
\cite{lan18}). A direct measurement of $B_{||}$ can indeed be only possible when the contribution of all eight hyperfine
transitions to the 6.7~GHz \meth ~maser emission will be determined properly by modeling the pumping mechanism of the maser.
However, by knowing the inclination of the magnetic field with respect to the line of sight (i.e., the $\theta$ angle), 
we can also estimate a lower limit for the magnetic field strength 
($B=\frac{B_{||}} {cos~(\theta+\varepsilon^{-}_{\rm{\theta}})}$; considering 
$\theta^{\varepsilon^{+}_{\rm{\theta}}}_{\varepsilon^{-}_{\rm{\theta}}}$). We were able to measure 
both $B_{||}$ and $B$ towards two of the seven massive YSOs reported in this work:  G25.71-0.04 and G28.83-0.25.\\

\noindent\textit{\textbf{G25.71-0.04.}} From the circularly polarized emission of G257.E12 we measured a Zeeman splitting of
$-3.1\pm0.7$~\ms, which implies a magnetic field along the line of sight of $B_{||}>61$~mG. By assuming $\theta_{\rm{min}}>43$\d
~(see Col.14 of Table~\ref{G257_tab}) the 3D magnetic field strength is $B>78$~mG.\\
\begin{table*}[th!]
\caption []{Comparison between position angle of magnetic field, \meth ~maser distribution, outflows, and linear polarization angles.} 
\begin{center}
\scriptsize
\begin{tabular}{ l c c c c c c c c c c}
\hline
\hline
\,\,\,\,\,(1) &(2)           & (3)                  & (4)                           & (5)                          & (6)                       & (7)        & (8)                                                   & (9)                                           &(10)                                               &(11)\\
Source & $\Phi_{\rm{f}}$\tablefootmark{a}& $\langle\chi\rangle$\tablefootmark{b} & $\langle\Phi_{\rm{B}}\rangle$\tablefootmark{b} & $\rm{PA}_{\rm{outflow}}$     & $\rm{PA}_{\rm{CH_{3}OH}}$ & $\rho$\tablefootmark{c}& $|\rm{PA}_{\rm{outflow}}-\langle\Phi_{\rm{B}}\rangle|$& $|\rm{PA}_{\rm{CH_{3}OH}}-\langle\chi\rangle|$& $|\rm{PA}_{\rm{CH_{3}OH}}-\rm{PA}_{\rm{outflow}}|$&ref.\tablefootmark{d} \\ 
       & (\d)                &  (\d)                & (\d)                          & (\d)                         & (\d)                      &            & (\d)                                                  &(\d)                                           & (\d)                                              & \\ 
\hline
G23.44-0.18 (MM2) & 13& $+23\pm40$& $-67\pm40$                    & $-40\pm15\tablefootmark{e}$            & $+30\pm26$\tablefootmark{f}                 &  +0.69     & $27\pm43$                                         & $13\pm48$                              & $70\pm30$                         & (1)\\
                 & &           &                               &                                        & $-24\pm91$\tablefootmark{g}                 &  -0.37     &                                               & $47\pm81$                              & $16\pm91$                         & \\
G25.83-0.18       & 11& $+23\pm7$& $-67\pm7$                    & $+10\pm15\tablefootmark{h}$            & $+51\pm7$                 &  +0.69     & $77\pm17$                                         & $28\pm10$                              & $41\pm17$                         & (2)\\
G25.71-0.04       & 23& $-51\pm77$& $-80\pm43\tablefootmark{i}$ & $-90\pm15\tablefootmark{h,j}$            & $-71\pm9$                 &  -0.41     & $10\pm46$                                         & $20\pm78$                              & $18\pm18$                         & (2)\\
G28.31-0.39       & 24& $+40\pm44$& $-50\pm44$     & $-52\pm15\tablefootmark{h}$            & $+85\pm22$                 &  +0.07     & $2\pm47$                                         & $45\pm49$                              & $84\pm27\tablefootmark{k}$                         & (2)\\
G28.83-0.25       & 10& $-45\pm34$& $+58\pm59\tablefootmark{i}$ & $-40\pm15\tablefootmark{h}$            & $-41\pm10$                 &  -0.83     & $82\pm61\tablefootmark{k}$                                         & $4\pm35$                           & $1\pm18$                         & (2)\\
G29.96-0.02       & 12& $+62\pm17$& $-29\pm17$ & $-38\pm15\tablefootmark{h}$            & $+80\pm3$                 &  0.51     & $9\pm23$                                         & $18\pm17$                           & $62\pm16\tablefootmark{k}$                         & (3)\\
G43.80-0.13       & 14& $-81\pm5$& $+9\pm5$ & $+38\pm15\tablefootmark{h}$            & $-48\pm5$                 &     -0.94    & $29\pm16$  & $33\pm7$                           & $86\pm16$                         & (4)\\
\hline
\multicolumn{11}{c}{From Paper~III\tablefootmark{l}}\\
\hline
IRAS\,20126+4104 & $4$       & $-70\pm16$           & $+20\pm16$                    & $-65\pm5$   & $+87\pm4$                 &  $+0.12$   & $85\pm17$                                             & $23\pm17$                    & $28\pm6$                         & (5)\\
G24.78+0.08-A2 &   $17$      & $-53\pm2$            & $+37\pm2$    & $-40\pm15$  & $-26\pm19$                &  $-0.77$   & $77\pm15$                                             & $79\pm19$                                     & $66\pm24$                                         & (5)\\
G25.65+1.05    &   $7$       & $-80\pm8$            & $-23\pm51$   & $-15\pm15$  & $-49\pm7$ &$-0.87$     & $8\pm53$                                              & $31\pm11$                                     & $64\pm17$                                         & (5)\\
G29.86-0.04    &   $17$      & $+46\pm41$           & $+82\pm56$   & $+6\pm15$ & $+8\pm7$ &$+0.73$     & $76\pm58$                                             & $38\pm42$                                     & $14\pm17$                                         & (5)\\
G35.03+0.35    &   $8$       & $-64\pm5$            & $+26\pm5$    & $+27\pm15$&$-26\pm19$                 &$-0.77$     & $1\pm16$                                              & $38\pm20$                                     & $53\pm24$                                         & (5)\\
G37.43+1.51    &   $4$       & $+90\pm3$            & $+90\pm3$   & $-4\pm15$   &$-64\pm5$ &$-0.87$     & $86\pm15$                           & $26\pm6$                  & $60\pm16$                                         & (5)\\
G174.20-0.08   &   $4$       & $-$                  & $-$                           & $-40\pm15$   &$-63\pm16$                 &$-0.45$     & $-$                                                   & $-$                                           & $23\pm22$                                         & (5)\\
G213.70-12.6-IRS3&$2$        & $+20\pm5$            & $-70\pm5$    & $+53\pm15$  &$+63\pm2$                  &$+0.95$     & $57\pm16$                            & $43\pm5$                                      & $10\pm15$                                         & (5)\\
Cepheus~A      &   $2$       & $-57\pm28$           & $+30\pm19$                    & $+40\pm4$                    & $-79\pm9$                 &  $-0.34$   & $10\pm19$                                             & $22\pm29$                                     & $61\pm10$                                         & (5)\\
W75N-group~A & $3$           & $-13\pm9$            & $+77\pm9$                     & $+66\pm15$                   & $+43\pm10$                &  $+0.96$   & $11\pm18$                                             & $56\pm14$                                     & $23\pm18$                                         & (5)\\
NGC7538-IRS1 &  $6$          &$-30\pm69$            & $+67\pm70$                    & $-40\pm10$                   & $+84\pm7$                 &  $+0.15$   & $73\pm71$                                               & $66\pm69$                                     & $56\pm12$                                         & (5)\\
W3(OH)-group II &  $4$       &$+21\pm45$            & $-47\pm44$                    & $-$                          & $-59\pm6$                 &  $-0.84$   & $-$                                                   & $80\pm45$                                     & $-$                                               & (5)\\
W51-e2    &  $12$            &$+33\pm16$            & $-60\pm21$                    & $-50\pm20$                   & $+57\pm8$                 &  $+0.70$   & $10\pm29$                                             & $24\pm18$                                     & $73\pm22$                                         & (5)\\
IRAS18556+0138 &  $5$        &$-2\pm11$             & $+88\pm11$                    & $+58\pm23$                   & $-40\pm2$                 &  $-0.99$   & $30\pm26$                                             & $42\pm11$                                     & $82\pm23$                                         & (5)\\
W48       & $7$              & $+23\pm7$            & $-67\pm7$                     & $-$                          & $+55\pm10$                &  $+0.70$   & $-$                                                   & $78\pm12$                                     & $-$                                               & (5) \\
IRAS06058+2138-NIRS1 &  $4$  &$+49\pm47$            & $-49\pm52$                    & $-50\pm15$                   & $+78\pm7$                 &  $+0.64$   & $1\pm54$                                              & $29\pm48$                                     & $52\pm17$                                         & (5)\\
IRAS22272+6358A &  $2$       &$-80\pm15$            & $+9\pm15$                     & $-40\pm15$                   & $-35\pm11$                &  $-0.87$   & $49\pm21$                                             & $45\pm19$                                     & $5\pm19$                                          & (5) \\
S255-IR   &  $4$             &$+36\pm12$            & $-54\pm12$                    & $+75\pm15$                   & $-63\pm49$                &  $-0.11$   &$51\pm19$                                              & $81\pm51$                                     & $42\pm51$                                         & (5) \\
S231      &  $4$             &$+28\pm49$            & $-62\pm49$                    & $-47\pm5$                    & $+28\pm8$                 &  $+0.97$   & $15\pm49$                                             & $0\pm50$                                      & $75\pm9$                                          & (5)\\
G291.27-0.70 &  $7$          &$-32\pm5$             & $+52\pm5$                     & $-$                          & $-77\pm14$                 &  $-$       & $-$                                                   & $45\pm15$                                     & $-$                                               & (5)\\
G305.21+0.21 &  $9$          &$-51\pm14$            & $28\pm14$                     & $-$                          & $+48\pm23$                &  $-$       &$-$                                                    & $81\pm27$                                     & $-$                                               & (5)\\
G309.92+0.47 &  $12$         &$+2\pm56$             & $-75\pm56$                    & $-$                          & $+35\pm5$                 &  $-$       &$-$                                                    & $33\pm56$                                     & $-$                                               & (5)\\
G316.64-0.08 &  $3$          &$-67\pm36$            & $+21\pm36$                    & $-$                          & $+34\pm29$                &  $-$       & $-$                                                   & $79\pm46$                                     & $-$                                               & (5)\\
G335.79+0.17 &  $8$          &$+44\pm28$            & $-41\pm28$                    & $-$                          & $-69\pm25$                &  $-$       & $-$                                                   & $67\pm38$                                     & $-$                                               & (5) \\
G339.88-1.26 &  $7$          &$+77\pm24$            & $-12\pm24$                    & $-$                          & $-60\pm17$                &  $-$       & $-$                                                   & $43\pm29$                                     & $-$                                               & (5) \\
G345.01+1.79 &  $5$          &$+5\pm39$             & $-86\pm39$                    & $-$                          & $+74\pm4$                 &  $-$       &$-$                                                    & $69\pm39$                                     & $-$                                               & (5) \\
NGC6334F (central) &  $5$    &$+77\pm20$            & $-13\pm20$                    & $+30\pm15$\tablefootmark{h}  & $-41\pm16$                &  $-$       &$43\pm25$                                              & $62\pm26$                                       & $71\pm41$                                         & (5)\\
NGC6334F (NW)&  $5$          &$-71\pm20$            & $+19\pm20$                    & $+30\pm15$\tablefootmark{h}  & $-80\pm38$                &  $-$       &$11\pm25$                                              & $9\pm43$                                      & $70\pm41$\tablefootmark{f}                        & (5)\\
\hline
\hline
\end{tabular}
\end{center}
\tablefoot{
\tablefoottext{a}{Foreground Faraday rotation estimated by using Eq.~3 in Paper~I.
\tablefoottext{b}{Because of the large uncertainties of the estimated $\Phi_{\rm{f}}$, the angles are not corrected for $\Phi_{\rm{f}}$.}
\tablefoottext{c}{Pearson product-moment correlation coefficient $-1\leq\rho\leq+1$; $\rho=+1$ ( $\rho=-1$) is total positive (negative) correlation, $\rho=0$ is no correlation.}}
\tablefoottext{d}{References: (1) Ren et al. (\cite{ren11}); (2) de Villiers et al. (\cite{dev14}); (3) Cesaroni et al. (\cite{ces17}); (4) L\'{o}pez-Sepulcre et al. \cite{lop10}; (5) Paper~III and references therein;}.
\tablefoottext{e}{We overestimate the errors by considering half of the opening angle of the outflow.}
\tablefoottext{f}{Calculated for the western group of masers.}
\tablefoottext{g}{Calculated for the eastern group of masers.}
\tablefoottext{h}{We consider an arbitrary conservative error of 15\d.}
\tablefoottext{i}{Before averaging we use the criterion described in Sect.~\ref{obssect} to estimate the orientation of the magnetic field with respect to the 
linear polarization vectors.}
\tablefoottext{j}{Since the 6.7~GHz \meth ~maser is associated with the blue-shifted lobe of the $^{13}\rm{CO}$ outflow we consider only its PA.} 
\tablefoottext{k}{The differences between the angles are evaluated taking into account that $\rm{PA}\equiv\rm{PA}\pm180$\d, $\langle\chi\rangle\equiv\langle\chi\rangle\pm180$\d, and $\langle\Phi_{\rm{B}}\rangle\equiv\langle\Phi_{\rm{B}}\rangle\pm180$\d.}
\tablefoottext{l}{Here we omit all the notes that are already indicated in Table~2 of Paper~III.}
}
\label{Comp_ang}
\end{table*}

\noindent\textit{\textbf{G28.83-0.25.}} The Zeeman splitting measured by modeling the circularly
polarized emission of G288.E19 is \dvz$=-1.1\pm0.3$~\ms, for which we have $B_{||}>21$~mG. Considering an error-weighted angle of $\langle\theta\rangle={81^{\circ}}^{+10^{\circ}}_{-34^{\circ}}$ the 3D magnetic field is $B>28$~mG. The
3D magnetic field measured from the \meth ~maser is four times larger than that measured from the OH maser
(Bayandina et al. \cite{bay15}). From the relation $|B|\propto
n_{\rm{H_2}}^{0.47}$ (Crutcher \cite{cru99}), we determine that the \meth ~maser are arising from a gas with a density
at least an order of magnitude higher than that of the OH maser. This agrees with the ranges of $n_{\rm{H_2}}$ for
the two maser species (e.g., Cragg et al. \cite{cra02}).\\

\indent From the estimated $B$ values we can investigate the importance of the magnetic field in the high-mass star-forming 
process. If the ratio $\beta$ between the thermal ($E_{\rm{T}}$) and the magnetic energies ($E_{\rm{B}}$) is 
lower than one ($\beta<1$), the magnetic field dominates in the high-density \meth ~maser environment. Following Eq.~11 
of Surcis et al. (\cite{sur11a}), we see that the ratio $\beta$ also depends, in addition to the magnetic field,  on the
characteristics of the gas where the masers arise, namely on the number density of the gas ($n_{\rm{H_{2}}}$) and on the
kinetic temperature of the gas ($T_{\rm{k}}$).  
Cragg et al. (\cite{cra02,cra05}) modeled the Class~II \meth ~maser emissions and  found that the masers arise when 
$10^5~\rm{cm^{-3}}<n_{\rm{H_{2}}}<10^9~\rm{cm^{-3}}$ and $100~\rm{K}<T_{\rm{k}}<200~\rm{K}$. Considering $\beta=1$ we 
determined $|B_{\rm{crit}}|$ values for all the possible combinations of the $n_{\rm{H_{2}}}$ and $T_{\rm{k}}$ 
extremes (see Table~\ref{Bcrit}). If $|B|>|B_{\rm{crit}}|$ the magnetic field dominates over the thermal motions. 
In the case of G25.71-0.04 and G28.83-0.25 the magnetic field dominates the dynamics 
independently of the characteristics of the gas and on the specific dominating hyperfine transition.\\
\indent In the past, we have reported erroneous values of Zeeman-splitting due to an error in the FRTM code (see 
Sect.~\ref{obssect}). We report the corrected values of Zeeman-splitting with the corresponding lower limits of $B_{||}$
and $B$ in Table~\ref{B_tab}. The varying values of $|B|$ measured from different maser features within the same source might
indicate different gas properties in the massive SFR due either to the association of the \meth ~masers with different YSOs 
(e.g., W3(OH) and G24.78+0.08; Papers~I and III) or  to the different locations of the masers in the associated protostar
(e.g., W51-e2 and G24.78+0.08-A1; Papers~I and III). Nevertheless, if at least one of the masers detected towards a massive YSO 
provides $|B|>25$~mG (see Table~\ref{Bcrit}) we can assume that  the magnetic field dominates there. The only sources for which
we cannot determine whether the magnetic field dominates  independently of the characteristics of the gas and on the specific
dominating hyperfine transition are IRAS\,06058+2138-IRS\,1 and S255-IR (Paper~II). 
\begin {table}[h!]
\caption []{Results of Kolmogorov-Smirnov test.} 
\begin{center}
\scriptsize
\begin{tabular}{ l c c c c }
\hline
\hline
\,\,\,\,\,\,\,\,\,\,\,\,\,\,\,(1)                      &(2)   & (3)  & (4)       & (5)          \\ 
\,\,\,\,\,\,\,\,\,\,Angle                              & $N$\tablefootmark{a}  & $D$\tablefootmark{b}  & $\lambda$\tablefootmark{c} & $Q_{\rm{K-S}}(\lambda)$\tablefootmark{d}\\
\hline
\\
$|\rm{PA}_{\rm{CH_{3}OH}}-\langle\chi\rangle|$         & 35   & 0.16 & 0.99      & 0.28 \\
$|\rm{PA}_{\rm{CH_{3}OH}}-\rm{PA_{\rm{outflow}}}|$     & 27   & 0.17 & 0.91      & 0.38 \\
$|\rm{PA}_{\rm{outflow}}-\langle\Phi_{\rm{B}}\rangle|$ & 25   & 0.27 & 1.42      & 0.03 \\
\\
\hline
\hline
\end{tabular}
\end{center}
\tablefoot{
\tablefoottext{a}{$N$ is the number of elements considered in the K-S test. }
\tablefoottext{b}{$D$ is the maximum value of the absolute difference between the data set composed of $N$ elements and the random
distribution.}
\tablefoottext{c}{$\lambda$ is a parameter given by $\lambda=(\sqrt{N}+0.12+0.11/\sqrt{N})\times D$.}
\tablefoottext{d}{$Q_{\rm{K-S}}(\lambda)=2\sum_{j=1}^{N} (-1)^{j-1}~ e^{-2j^2\lambda^2}$ is the significance level of the K-S test.}
}
\label{KS}
\end{table}
\subsection{Updated statistical results}
In Paper~III, since we were at the midpoint of our project\footnote{To determine whether there is any relation between
the morphology of the magnetic field and the ejecting direction of molecular outflow from massive YSOs on a scale of
tens of~au.}, we updated our first statistical results reported in Paper~II. Here, we would like to provide the updated
statistical results based on seven more sources with respect to Paper~III, i.e., 35 YSOs in total, also  including  the southern
hemisphere sources reported in the literature (Paper~II and references therein). Similarly to Papers~II and III, we
list in Table~\ref{Comp_ang} the sources of the flux-limited sample analyzed so far for which we were able to measure the
projection on the plane of the sky of the angles $|\rm{PA}_{\rm{outflow}}-\langle\Phi_{\rm{B}}\rangle|$, 
$|\rm{PA}_{\rm{CH_{3}OH}}-\langle\chi\rangle|$, and $|\rm{PA}_{\rm{CH_{3}OH}}-\rm{PA}_{\rm{outflow}}|$, where 
$\rm{PA}_{\rm{outflow}}$ is the orientation of the large-scale molecular outflow, $\langle\Phi_{\rm{B}}\rangle$ is the 
error-weighted orientation of the magnetic field, $\rm{PA}_{\rm{CH_{3}OH}}$ is the orientation of the \meth ~maser 
distribution, and $\langle\chi\rangle$ is the error-weighted value of the linear polarization angles (for more details
regarding Table~\ref{Comp_ang}, see the table notes and  Paper~III). In Table~\ref{KS} we report
the results of the Kolmogorov-Smirnov (K-S) test, which is a nonparametric test;  here it is used to compare  
samples of angles ($|\rm{PA}_{\rm{CH_{3}OH}}-\langle\chi\rangle|$, $|\rm{PA}_{\rm{CH_{3}OH}}-\rm{PA_{\rm{outflow}}}|$, and
$|\rm{PA}_{\rm{outflow}}-\langle\Phi_{\rm{B}}\rangle|$) with the random probability distribution (see Paper~II for more
details). 
Performing the statistical analysis we note that the probability that the angles
$|\rm{PA}_{\rm{CH_{3}OH}}-\langle\chi\rangle|$ are drawn from a random distribution is 28\%\footnote{In Paper~III
we erroneously reported a probability of $\sim80\%$.}.\\
\indent For $|\rm{PA}_{\rm{CH_{3}OH}}-\rm{PA}_{\rm{outflow}}|$ and 
$|\rm{PA}_{\rm{outflow}}-\langle\Phi_{\rm{B}}\rangle|$ we measured probabilities similar to those reported in 
Paper~III, these are 38\% and 3\% instead of 34\% and 10\% (Paper~III), respectively. The updated results reinforce our 
previous finding: the magnetic field close to the YSO is preferentially oriented along the outflow axis.
\section{Summary}
We observed seven massive SFRs at 6.7~GHz in full polarization spectral mode with the EVN;  our aim was to detect the
linearly and circularly polarized emission of \meth ~masers. We detected linearly polarized emission towards all the 
regions and circularly polarized emission towards G25.71-0.04 and G28.83-0.25.  We used the adapted \code ~to model both
the linear and the circular polarization of the masers. In particular, to estimate a lower limit of the magnetic field
along the line of sight we assumed that the dominant hyperfine component, the one with the largest
Einstein coefficient for stimulated emission, is $F=3\rightarrow4$ (Lankhaar et al. \cite{lan18}). 
By analyzing the linearly polarized emission
of the masers we were able to estimate the orientation of the magnetic field around eight massive YSOs (two are located 
within G23.44-0.18:  MM1 and MM2). We found that the magnetic fields are aligned with the outflows 
($|\rm{PA}_{\rm{outflow}}-\langle\Phi_{\rm{B}}\rangle|<30$\d) in five YSOs (G23.44-0.18-MM2, G25.71-0.04, G28.31-0.39,
G29.96-0.02, G43.80-0.13) and are perpendicular to the outflows in two YSOs (G25.83-0.18 and G28.83-0.25). The estimated 
magnetic field strengths along the line of sight for G25.71-0.04 and G28.83-0.25 are $B_{||}>61$~mG and 
$B_{||}>21$~mG, respectively. The magnetic field seems to dominate the dynamics of the gas in both YSOs.\\
\indent We further increased the number of sources  to 26, which is 80\% of the flux-limited sample; the 
projected angles of the magnetic field and of the outflows of these sources are known. Comparing these angles, we confirm the statistical
evidence that the magnetic fields around massive YSOs are preferentially oriented along the molecular outflows.
In particular, the probability that the distribution of angles $|\rm{PA}_{\rm{outflow}}-\langle\Phi_{\rm{B}}\rangle|$ is 
drawn from a random distribution is lower (3\%) than  was reported in Paper~III (10\%). \\

\noindent \small{\textit{Acknowledgements.} 
 We wish to thank the referee S. Ellingsen for the useful suggestions that have improved the paper. 
W.H.T.V. acknowledges support from the European Research Council through consolidator grant 614264. 
A.B. acknowledges support from the National Science Centre, Poland, through grant 2016/21/B/ST9/01455. 
G.S., W.H.T.V., and H.J.van L. thank Hans Engelkamp and his team for their help in attempting the laboratory 
measurements of the Land\'{e} g-factor of the \meth ~molecule by using the 30 Tesla Magnet at the High Field Magnet 
Laboratory of the Radboud University in Nijmegen (The Netherlands). 
The European VLBI Network is a joint facility of independent European, African, Asian, and North American radio 
astronomy institutes. Scientific results from data presented in this publication are derived from the following EVN 
project code(s): ES072.
The research leading to these results has received funding from the European Commission Seventh Framework Programme 
(FP/2007-2013) under grant agreement No. 283393 (RadioNet3).}

\bibliographystyle{aa}

\begin{appendix}
\normalsize
\section{Tables}
\label{appA}
In Tables~\ref{G23_tab}--\ref{G43_tab} we list the parameters of all the \meth ~maser features detected toward the seven 
massive SFRs observed with the EVN and reported in this work. The tables are organized as follows. In Col.~1 we give 
the name of the feature, and only in Table~\ref{G23_tab} the associated region is reported in Col.~1B.
The positions, Cols.~2 and 3, refer to the maser feature used for self-calibration. From Cols.~4 to 6 we give the peak 
flux density, the LSR velocity ($V_{\rm{lsr}}$), and the FWHM ($\Delta v\rm{_{L}}$) of the total intensity spectra of
the maser features that are obtained using a Gaussian fit. The mean linear polarization fraction ($P_{\rm{l}}$) and the 
mean linear polarization angles ($\chi$) are measured across the spectrum, and  are listed in Cols.~7 and 8. The
outcomes of the adapted FRTM code are listed in Cols.~9 (intrinsic thermal linewidth), 10 
(emerging brightness temperature), and 14 (angle between the magnetic field and the maser propagation 
direction). The errors were determined by analyzing the full probability distribution function. The value of $\theta$ in 
bold indicates that $|\theta^{\rm{+}}-55$\d$|<|\theta^{\rm{-}}-55$\d$|$, i.e., the magnetic field is assumed to be 
parallel to the linear polarization vector (see Papers~I-III). Finally, the circular polarization fraction ($P_{\rm{V}}$),
the Zeeman splitting (\dvz), and the lower limit of the magnetic field strength along the line of sight ($B_{\rm{||}}$)
determined by fitting the V Stokes spectra by using the best-fitting results ($\Delta V_{\rm{i}}$ and $T_{\rm{b}}\Delta\Omega$)
and the Land\'{e} g-factors calculated by Lankhaar et al. (\cite{lan18}) for the hyperfine transition $F=3\rightarrow4$ are  listed in Cols.~11, 12, and 13. 
\begin {table*}[t!]
\caption []{Parameters of the 6.7 GHz \meth ~maser features detected in G23.44-0.18.} 
\begin{center}
\scriptsize
\begin{tabular}{ l c c c c c c c c c c c c c c }
\hline
\hline
\,\,\,\,\,(1)&(1B) & (2)     & (3)      & (4)            & (5)       & (6)              & (7)         & (8)       & (9)                    & (10)                     & (11)         &(12)                  & (13)    & (14)     \\
Maser     & Associated & RA\tablefootmark{a}&Dec\tablefootmark{a}& Peak flux &$V_{\rm{lsr}}$& $\Delta v\rm{_{L}}$ &$P_{\rm{l}}\tablefootmark{b}$ &  $\chi\tablefootmark{b}$   & $\Delta V_{\rm{i}}\tablefootmark{c}$ & $T_{\rm{b}}\Delta\Omega\tablefootmark{c}$& $P_{\rm{V}}$ & $\Delta V_{\rm{Z}}$ & $B_{\rm{||}}$  &$\theta\tablefootmark{d}$\\
          & Region      & offset  &  offset  & Density(I)     &           &                  &             &              &                         &                         &              &                      &      \\ 
          &       & (mas)   &  (mas)   & (Jy/beam)      &  (km/s)   &      (km/s)      & (\%)        &   (\d)    & (km/s)                  & (log K sr)              &   ($\%$)     & (m/s) & (mG)               &(\d)       \\ 
\hline
G23.E01    & MM1   & -87.259 & -33.880  & $2.485\pm0.010$&  97.69    &      $0.26$      & $1.1\pm0.1$ & $14\pm8$  &  $1.0^{+0.2}_{-0.2}$    & $8.9^{+0.6}_{-0.6}$     & $-$         & $-$   & $-$                    &$80^{+9}_{-39}$ \\ 
G23.E02    & MM1   & -12.789 & -14.023  & $0.127\pm0.003$&  96.20    &      $0.26$      & $-$         & $-$       & $-$                     & $-$                     & $-$         & $-$   & $-$                    &$-$ \\ 
G23.E03    & MM1   & 0       & 0        & $9.206\pm0.007$&  96.68    &      $0.26$      & $1.4\pm0.3$ & $-13\pm1$ &  $1.1^{+0.2}_{-0.2}$    & $9.0^{+0.9}_{-0.7}$     & $-$          & $-$  & $-$                    &$74^{+14}_{-37}$ \\ 
G23.E04    & MM1   & 2.886   & 17.950   & $0.050\pm0.003$&  95.94    &      $0.28$      & $-$         & $-$       & $-$                     & $-$                     & $-$          & $-$  & $-$                    &$-$ \\ 
G23.E05    & MM1   & 10.752  & 6.477    & $3.832\pm0.011$&  97.52    &      $0.34$      & $2.1\pm0.1$ & $28\pm1$  &  $1.4^{+0.1}_{-0.1}$    & $9.2^{+0.6}_{-0.3}$     & $-$          & $-$  & $-$                    &$84^{+6}_{-39}$ \\ 
G23.E06    & MM1   & 18.222  & -1.747   & $9.038\pm0.010$&  98.09    &      $1.85$      & $2.6\pm0.1$ & $11\pm2$  &  $2.0^{+0.1}_{-0.7}$    & $9.3^{+0.6}_{-0.8}$     & $-$           & $-$ & $-$                    &$72^{+18}_{-31}$ \\ 
G23.E07    & MM1   & 20.202  & -0.244   & $4.614\pm0.010$&  97.43    &      $0.46$      & $-$         & $-$       & $-$                     & $-$                     & $-$           & $-$ & $-$                    &$-$ \\ 
G23.E08    & MM1   & 24.899  & -184.355 & $0.179\pm0.011$&  97.69    &      $0.26$      & $-$         & $-$       & $-$                     & $-$                     & $-$          & $-$  & $-$                    &$-$ \\ 
G23.E09    & MM1   & 25.238  & -184.031 & $0.185\pm0.011$&  97.61    &      $1.35$      & $-$         & $-$       & $-$                     & $-$                     & $-$          & $-$  & $-$                    &$-$ \\ 
G23.E10    & MM1   & 30.841  & 2.136    & $1.624\pm0.008$&  97.87    &      $0.38$      & $-$         & $-$       & $-$                     & $-$                     & $-$          & $-$  & $-$                    &$-$ \\ 
G23.E11    & MM1   & 45.158  & 6.119    & $0.622\pm0.003$&  100.11   &      $0.33$      & $-$         & $-$       & $-$                     & $-$                     & $-$          & $-$  & $-$                    &$-$ \\ 
G23.E12    & MM1   & 49.968  & 4.854    & $0.560\pm0.005$&  99.01    &      $0.42$      & $-$         & $-$       & $-$                     & $-$                     & $-$          & $-$  & $-$                    &$-$ \\ 
G23.E13    & MM1   & 52.005  & 161.655  & $0.479\pm0.051$&  103.18   &      $0.30$      & $-$         & $-$       & $-$                     & $-$                     & $-$           & $-$ & $-$                    &$-$ \\ 
G23.E14    & MM1   & 58.003  & 112.690  & $0.108\pm0.013$&  104.54   &      $1.24$      & $-$         & $-$       & $-$                     & $-$                     & $-$           & $-$ & $-$                    &$-$ \\ 
G23.E15    & MM1   & 65.303  & 183.970  & $0.271\pm0.003$&  105.42   &      $0.24$      & $-$         & $-$       & $-$                     & $-$                     & $-$           & $-$ & $-$                    &$-$ \\ 
G23.E16    & MM1   & 68.642  & 131.384  & $0.315\pm0.014$&  103.31   &      $0.81$      & $-$         & $-$       & $-$                     & $-$                     & $-$          & $-$  & $-$                    &$-$ \\ 
G23.E17    & MM1   & 70.339  & 138.926  & $0.166\pm0.007$&  105.68   &      $0.31$      & $-$         & $-$       & $-$                     & $-$                     & $-$          & $-$  & $-$                    &$-$ \\ 
G23.E18    & MM1   & 71.415  & 77.648   & $0.052\pm0.003$&  100.64   &      $0.42$      & $-$         & $-$       & $-$                     & $-$                     & $-$           & $-$ & $-$                    &$-$ \\ 
G23.E19    & MM1   & 72.263  & 46.728   & $0.078\pm0.003$&  100.28   &      $0.52$      & $-$         & $-$       & $-$                     & $-$                     & $-$          & $-$  & $-$                    &$-$ \\ 
G23.E20    & MM1   & 73.735  & 19.735   & $0.625\pm0.004$&  98.79    &      $0.34$      & $-$         & $-$       & $-$                     & $-$                     & $-$          & $-$  & $-$                    &$-$ \\ 
G23.E21    & MM1   & 77.243  & -14.578  & $0.275\pm0.003$&  96.03    &      $0.31$      & $-$         & $-$       & $-$                     & $-$                     & $-$           & $-$ & $-$                    &$-$ \\ 
G23.E22    & MM1   & 77.413  & 54.722   & $0.502\pm0.038$&  102.30   &      $0.28$      & $-$         & $-$       & $-$                     & $-$                     & $-$          & $-$  & $-$                    &$-$ \\ 
G23.E23    & MM1   & 78.658  & 111.210  & $0.720\pm0.005$&  104.94   &      $0.30$      & $-$         & $-$       & $-$                     & $-$                     & $-$           & $-$ & $-$                    &$-$ \\ 
G23.E24    & MM1   & 82.110  & 59.479   & $0.766\pm0.056$&  103.14   &      $0.25$      & $-$         & $-$       & $-$                     & $-$                     & $-$          & $-$  & $-$                    &$-$ \\ 
G23.E25    & MM1   & 87.769  & 107.803  & $1.300\pm0.017$&  103.93   &      $0.29$      & $6.5\pm0.4$ & $33\pm1$  & $0.7^{+0.4}_{-0.1}$     & $10.0^{+0.2}_{-0.6}$    & $-$            & $-$ & $-$                    &$85^{+5}_{-6}$ \\ 
G23.E26    & MM1   & 123.306 & -103.890 & $0.256\pm0.010$&  97.74    &      $0.22$      & $-$         & $-$       & $-$                     & $-$                     & $-$           & $-$ & $-$                    &$-$ \\ 
G23.E27    & MM1   & 138.019 & 105.028  & $3.663\pm0.006$&  106.96   &      $0.41$      & $4.9\pm0.6$ & $33\pm1$  & $1.4^{+0.2}_{-0.5}$     & $9.7^{+1.4}_{-0.9}$     & $-$          & $-$  & $-$                    &$87^{+3}_{-15}$ \\ 
\hline
G23.E28    & MM2   & 862.238 & -13930.504 & $0.396\pm0.005$& 106.12  &      $0.56$      & $-$         & $-$       & $-$                     & $-$                     & $-$           & $-$ & $-$                    &$-$ \\ 
G23.E29    & MM2   & 878.083 & -13951.554 & $0.049\pm0.003$& 111.80  &      $0.41$      & $-$         & $-$       & $-$                     & $-$                     & $-$           & $-$ & $-$                    &$-$ \\ 
G23.E30    & MM2   & 879.950 & -13951.012 & $0.113\pm0.003$& 112.49  &      $0.34$      & $-$         & $-$       & $-$                     & $-$                     & $-$           & $-$ & $-$                    &$-$ \\ 
G23.E31    & MM2   & 902.755 & -13927.719 & $1.025\pm0.007$& 107.83  &      $0.45$      & $-$         & $-$       & $-$                     & $-$                     & $-$           & $-$ & $-$                    &$-$ \\ 
G23.E32    & MM2   & 906.830 & -13820.572 & $0.045\pm0.006$& 107.75  &      $0.60$      & $-$         & $-$       & $-$                     & $-$                     & $-$           & $-$ & $-$                    &$-$ \\ 
G23.E33    & MM2   & 916.506 & -13881.450 & $0.034\pm0.003$& 101.16  &      $0.37$      & $-$         & $-$       & $-$                     & $-$                     & $-$           & $-$ & $-$                    &$-$ \\ 
G23.E34    & MM2   & 917.921 & -13882.568 & $0.602\pm0.013$& 103.75  &      $0.20$      & $-$         & $-$       & $-$                     & $-$                     & $-$           & $-$ & $-$                    &$-$ \\ 
G23.E35    & MM2   & 919.166 & -13894.001 & $3.270\pm0.024$& 104.15  &      $0.45$      & $-$         & $-$       & $-$                     & $-$                     & $-$           & $-$ & $-$                    &$-$ \\ 
G23.E36    & MM2   & 919.223 & -13899.982 & $0.675\pm0.006$& 105.77  &      $0.60$      & $-$         & $-$       & $-$                     & $-$                     & $-$           & $-$ & $-$                    &$-$ \\ 
G23.E37    & MM2   & 921.260 & -13913.116 & $0.217\pm0.023$& 104.15  &      $0.45$      & $-$         & $-$       & $-$                     & $-$                     & $-$           & $-$ & $-$                    &$-$ \\ 
G23.E38    & MM2   & 930.088 & -13910.046 & $0.140\pm0.018$& 104.24  &      $0.38$      & $-$         & $-$       & $-$                     & $-$                     & $-$           & $-$ & $-$                    &$-$ \\ 
G23.E39    & MM2   & 930.088 & -13908.653 & $0.152\pm0.020$& 103.49  &      $1.01$      & $-$         & $-$       & $-$                     & $-$                     & $-$           & $-$ & $-$                    &$-$ \\ 
G23.E40    & MM2   & 930.710 & -13902.115 & $0.132\pm0.013$& 103.71  &      $0.87$      & $-$         & $-$       & $-$                     & $-$                     & $-$           & $-$ & $-$                    &$-$ \\ 
G23.E41    & MM2   & 931.785 & -13778.488 & $0.202\pm0.018$& 104.19  &      $0.49$      & $-$         & $-$       & $-$                     & $-$                     & $-$           & $-$ & $-$                    &$-$ \\ 
G23.E42    & MM2   & 933.313 & -13775.871 & $0.229\pm0.021$& 104.06  &      $0.52$      & $-$         & $-$       & $-$                     & $-$                     & $-$          & $-$  & $-$                    &$-$ \\ 
G23.E43    & MM2   & 935.294 & -13831.359 & $0.143\pm0.020$& 104.28  &      $0.52$      & $-$         & $-$       & $-$                     & $-$                     & $-$          & $-$  & $-$                    &$-$ \\ 
G23.E44    & MM2   & 946.894 & -13878.128 & $3.299\pm0.032$& 103.23  &      $0.23$      & $1.4\pm0.6$ & $-25\pm5$ & $0.8^{+0.1}_{-0.2}$     & $9.0^{+0.8}_{-2.0}$     & $-$           & $-$ & $-$                    &$72^{+18}_{-35}$ \\ 
G23.E45    & MM2   & 948.083 & -13771.515 & $0.264\pm0.030$& 103.27  &      $0.33$      & $-$         & $-$       & $-$                     & $-$                     & $-$           & $-$ & $-$                    &$-$ \\ 
G23.E46    & MM2   & 961.268 & -13750.912 & $0.151\pm0.022$& 104.06  &      $0.47$      & $-$         & $-$       & $-$                     & $-$                     & $-$           & $-$ & $-$                    &$-$ \\ 
G23.E47    & MM2   & 1095.043 & -14105.755 & $0.424\pm0.010$& 101.82 &      $0.39$      & $-$         & $-$       & $-$                     & $-$                     & $-$          & $-$  & $-$                    &$-$ \\ 
G23.E48    & MM2   & 1155.083 & -13925.037 & $0.070\pm0.007$& 104.63 &      $0.64$      & $-$         & $-$       & $-$                     & $-$                     & $-$          & $-$  & $-$                    &$-$ \\ 
G23.E49    & MM2   & 1158.648 & -13945.259 & $0.718\pm0.046$& 103.01 &      $0.30$      & $-$         & $-$       & $-$                     & $-$                     & $-$           & $-$ & $-$                    &$-$ \\ 
G23.E50    & MM2   & 1164.816 & -13972.282 & $0.592\pm0.045$& 103.01 &      $0.32$      & $-$         & $-$       & $-$                     & $-$                     & $-$           & $-$ & $-$                    &$-$ \\ 
G23.E51    & MM2   & 1170.702 & -14001.327 & $5.269\pm0.045$& 103.01 &      $0.25$      & $8.3\pm0.3$ & $3.0\pm1$ & $<0.5$                  & $10.2^{+0.1}_{-0.3}$    & $-$            & $-$ & $-$                    &$90^{+7}_{-7}$ \\ 
G23.E52    & MM2   & 1172.513 & -13831.013 & $0.225\pm0.024$& 102.92 &      $0.38$      & $-$         & $-$       & $-$                     & $-$                     & $-$          & $-$  & $-$                    &$-$ \\ 
G23.E53    & MM2   & 1173.701 & -13956.898 & $1.274\pm0.041$& 102.92 &      $0.64$      & $-$         & $-$       & $-$                     & $-$                     & $-$          & $-$  & $-$                    &$-$ \\ 
G23.E54    & MM2   & 1175.172 & -14022.293 & $0.374\pm0.008$& 102.04 &      $0.34$      & $-$         & $-$       & $-$                     & $-$                     & $-$          & $-$  & $-$                    &$-$ \\ 
G23.E55    & MM2   & 1176.870 & -14029.843 & $1.632\pm0.044$& 102.96 &      $0.33$      & $5.1\pm0.3$ & $21\pm2$  & $1.0^{+0.2}_{-0.2}$     & $9.8^{+0.4}_{-0.4}$     & $-$          & $-$  & $-$                    &$82^{+6}_{-9}$ \\ 
G23.E56    & MM2   & 1178.794 & -13989.842 & $0.132\pm0.003$& 100.94 &      $0.25$      & $-$         & $-$       & $-$                     & $-$                     & $-$          & $-$  & $-$                    &$-$ \\ 
G23.E57    & MM2   & 1181.057 & -14019.821 & $2.422\pm0.026$& 103.36 &      $0.35$      & $4.3\pm0.5$ & $17\pm1$  & $1.0^{+0.3}_{-0.3}$     & $9.7^{+0.6}_{-0.9}$     & $-$          & $-$  & $-$                    &$82^{+8}_{-8}$ \\ 
G23.E58    & MM2   & 1181.850 & -13974.025 & $0.839\pm0.007$& 101.56 &      $0.24$      & $-$         & $-$       & $-$                     & $-$                     & $-$          & $-$  & $-$                    &$-$ \\ 
G23.E59    & MM2   & 1182.133 & -14015.575 & $3.089\pm0.021$& 102.30 &      $0.42$      & $2.3\pm0.2$ & $25\pm2$  & $1.3^{+0.4}_{-0.3}$     & $9.3^{+0.4}_{-0.4}$     & $-$          & $-$  & $-$                    &$84^{+6}_{-14}$ \\ 
G23.E60    & MM2   & 1185.358 & -13907.731 & $0.225\pm0.016$& 102.65 &      $0.59$      & $-$         & $-$       & $-$                     & $-$                     & $-$          & $-$  & $-$                    &$-$ \\ 
G23.E61    & MM2   & 1188.527 & -14045.452 & $0.201\pm0.020$& 102.30 &      $0.51$      & $-$         & $-$       & $-$                     & $-$                     & $-$          & $-$  & $-$                    &$-$ \\ 
\hline
\end{tabular} \end{center}
\tablefoot{
\tablefoottext{a}{The reference position is $\alpha_{2000}=18^{\rm{h}}34^{\rm{m}}39^{\rm{s}}\!.187$ and 
$\delta_{2000}=-08^{\circ}31'25''\!\!.441$ (see Sect.~\ref{res}).}
\tablefoottext{b}{$P_{\rm{l}}$ and $\chi$ are the mean values of the linear polarization fraction and the linear polarization angle measured across the spectrum, respectively.}
\tablefoottext{c}{The best-fitting results obtained by using a model based on the radiative transfer theory of methanol masers 
for $\Gamma+\Gamma_{\nu}=1~\rm{s^{-1}}$ (Vlemmings et al. \cite{vle10}, Surcis et al. \cite{sur11a}). The errors were determined 
by analyzing the full probability distribution function.}
\tablefoottext{d}{The angle between the magnetic field and the maser propagation direction is determined by using the observed $P_{\rm{l}}$ 
and the fitted emerging brightness temperature. The errors were determined by analyzing the full probability distribution function.}
}
\label{G23_tab}
\end{table*}

\begin {table*}[h!]
\caption []{Parameters of the 6.7 GHz \meth ~maser features detected in G25.83-0.18.} 
\begin{center}
\scriptsize
\begin{tabular}{ l c c c c c c c c c c c c c}
\hline
\hline
\,\,\,\,\,(1)&(2)   & (3)      & (4)            & (5)       & (6)              & (7)         & (8)       & (9)                     & (10)                    & (11)                        & (12)         &(13)      &(14)                     \\
Maser     & RA\tablefootmark{a}&Dec\tablefootmark{a}& Peak flux & $V_{\rm{lsr}}$& $\Delta v\rm{_{L}}$ &$P_{\rm{l}}\tablefootmark{b}$ &  $\chi\tablefootmark{b}$   & $\Delta V_{\rm{i}}\tablefootmark{c}$ & $T_{\rm{b}}\Delta\Omega\tablefootmark{c}$& $P_{\rm{V}}$ & $\Delta V_{\rm{Z}}$ & $B_{\rm{||}}$  &$\theta\tablefootmark{d}$\\
          &  offset &  offset  & Density(I)     &           &                  &             &            &                         &                         &              &                      &   &   \\ 
          &  (mas)  &  (mas)   & (Jy/beam)      &  (km/s)   &      (km/s)      & (\%)        &   (\d)    & (km/s)                  & (log K sr)              &   ($\%$)     &  (m/s) & (mG)               &(\d)       \\ 
\hline
G258.E01   & -287.740& 71.836   & $0.172\pm0.004$&  84.90    &      $0.70$      & $-$         & $-$       &  $-$                    & $-$                     & $-$     & $-$        & $-$             &$-$ \\ 
G258.E02   & -249.897& 186.352  & $0.055\pm0.004$&  98.73    &      $0.36$      & $-$         & $-$       &  $-$                    & $-$                     & $-$     & $-$        & $-$             &$-$ \\ 
G258.E03   & -244.321& 159.238  & $0.704\pm0.004$&  99.65    &      $0.23$      & $3.5\pm0.6$ & $13\pm3$  &  $<0.5$                 & $9.5^{+0.4}_{-2.0}$     & $-$           & $-$        & $-$             &$90^{+14}_{-14}$ \\ 
G258.E04   & -243.138& 213.008  & $0.611\pm0.004$&  98.68    &      $0.30$      & $9.7\pm0.9$ & $24\pm2$  &  $<0.5$                 & $10.2^{+0.7}_{-2.3}$    & $-$            & $-$        & $-$             &$90^{+7}_{-7}$ \\ 
G258.E05   & -239.225& 186.811  & $0.048\pm0.004$&  99.65    &      $0.27$      & $-$         & $-$       &  $-$                    & $-$                     & $-$     & $-$        & $-$             &$-$ \\ 
G258.E06   & -237.000& 200.137  & $0.771\pm0.005$&  98.64    &      $0.42$      & $8.0\pm0.4$ & $21\pm1$  &  $0.8^{+0.4}_{-0.1}$    & $10.1^{+0.4}_{-0.2}$    & $-$            & $-$        & $-$             &$86^{+3}_{-7}$ \\ 
G258.E07   & -231.065& 226.356  & $0.096\pm0.004$&  98.59    &      $0.39$      & $-$         & $-$       &  $-$                    & $-$                     & $-$     & $-$        & $-$             &$-$ \\ 
G258.E08   & -226.192& 178.898  & $1.052\pm0.004$&  99.30    &      $0.38$      & $5.6\pm0.5$ & $20\pm2$  &  $1.0^{+0.2}_{-0.2}$    & $9.8^{+0.4}_{-0.5}$     & $-$           & $-$        & $-$             &$84^{+6}_{-8}$ \\
G258.E09   & -222.287& 207.349  & $0.062\pm0.004$&  99.34    &      $0.28$      & $-$         & $-$       &  $-$                    & $-$                     & $-$     & $-$        & $-$             &$-$ \\ 
G258.E10   & -221.121& 132.805  & $0.049\pm0.004$&  99.34    &      $0.42$      & $-$         & $-$       &  $-$                    & $-$                     & $-$     & $-$        & $-$             &$-$ \\ 
G258.E11   & -181.135& -80.094  & $0.050\pm0.005$&  92.45    &      $0.28$      & $-$         & $-$       &  $-$                    & $-$                     & $-$     & $-$        & $-$             &$-$ \\ 
G258.E12   & -142.592& 213.335  & $0.081\pm0.004$&  92.71    &      $0.28$      & $-$         & $-$       &  $-$                    & $-$                     & $-$     & $-$        & $-$             &$-$ \\
G258.E13   & -128.106& 232.860  & $0.139\pm0.005$&  92.58    &      $0.24$      & $-$         & $-$       &  $-$                    & $-$                     & $-$     & $-$        & $-$             &$-$ \\ 
G258.E14   & -100.243& 433.103  & $0.286\pm0.007$&  92.23    &      $0.29$      & $-$         & $-$       &  $-$                    & $-$                     & $-$     & $-$        & $-$             &$-$ \\ 
G258.E15   & -91.938 & 268.910  & $0.063\pm0.004$&  93.19    &      $0.26$      & $-$         & $-$       &  $-$                    & $-$                     & $-$     & $-$        & $-$             &$-$ \\ 
G258.E16   & -83.742 & -38.296  & $0.069\pm0.004$&  95.52    &      $0.41$      & $3.6\pm0.2$ & $31\pm3$  &  $1.1^{+0.3}_{-0.2}$    & $9.5^{+0.7}_{-0.1}$     & $-$           & $-$        & $-$             &$87^{+3}_{-11}$ \\ 
G258.E17   & -80.913 & -29.622  & $2.606\pm0.006$&  97.06    &      $0.42$      & $2.5\pm0.3$ & $26\pm1$  &  $1.6^{+0.2}_{-0.4}$    & $9.3^{+0.8}_{-0.4}$     & $-$           & $-$        & $-$             &$82^{+9}_{-19}$ \\ 
G258.E18   & -76.567 & -0.937   & $0.128\pm0.006$&  97.01    &      $0.45$      & $-$         & $-$       &  $-$                    & $-$                     & $-$     & $-$        & $-$             &$-$ \\ 
G258.E19   & -75.785 & -18.445  & $0.257\pm0.004$&  95.43    &      $0.58$      & $-$         & $-$       &  $-$                    & $-$                     & $-$     & $-$        & $-$             &$-$ \\ 
G258.E20   & -75.071 & -2.984   & $0.105\pm0.004$&  97.10    &      $0.57$      & $-$         & $-$       &  $-$                    & $-$                     & $-$     & $-$        & $-$             &$-$ \\
G258.E21   & -74.637 & -20.985  & $0.289\pm0.004$&  95.43    &      $0.39$      & $-$         & $-$       &  $-$                    & $-$                     & $-$     & $-$        & $-$             &$-$ \\ 
G258.E22   & -70.991 & -91.331  & $0.046\pm0.004$&  95.52    &      $0.54$      & $-$         & $-$       &  $-$                    & $-$                     & $-$     & $-$        & $-$             &$-$ \\ 
G258.E23   & -38.632 & 45.489   & $5.683\pm0.030$&  91.61    &      $0.34$      & $-$         & $-$       &  $-$                    & $-$                     & $-$     & $-$        & $-$             &$-$ \\ 
G258.E24   & -25.660 & 25.568   & $0.551\pm0.010$&  90.74    &      $0.19$      & $-$         & $-$       &  $-$                    & $-$                     & $-$     & $-$        & $-$             &$-$ \\
G258.E25   & -23.076 & 15.581   & $0.081\pm0.004$&  92.76    &      $1.85$      & $-$         & $-$       &  $-$                    & $-$                     & $-$     & $-$        & $-$             &$-$ \\ 
G258.E26   & -18.779 & 2.611    & $0.096\pm0.004$&  92.97    &      $0.39$      & $-$         & $-$       &  $-$                    & $-$                     & $-$     & $-$        & $-$             &$-$ \\ 
G258.E27   & -14.944 & 19.765   & $0.577\pm0.009$&  91.22    &      $0.58$      & $-$         & $-$       &  $-$                    & $-$                     & $-$     & $-$        & $-$             &$-$ \\ 
G258.E28   & -12.03  & 23.851   & $0.327\pm0.004$&  89.95    &      $0.62$      & $-$         & $-$       &  $-$                    & $-$                     & $-$     & $-$        & $-$             &$-$ \\ 
G258.E29   & -9.272  & 14.484   & $0.361\pm0.009$&  91.00    &      $0.42$      & $-$         & $-$       &  $-$                    & $-$                     & $-$     & $-$        & $-$             &$-$ \\ 
G258.E30   & -6.817  & 4.279    & $6.123\pm0.027$&  91.75    &      $0.37$      & $-$         & $-$       &  $-$                    & $-$                     & $-$     & $-$                & $-$         &$-$ \\ 
G258.E31   & 0       & 0        & $8.280\pm0.010$&  90.87    &      $0.30$      & $2.3\pm0.1$ & $16\pm2$  &  $1.2^{+0.1}_{-0.3}$    & $9.3^{+0.7}_{-0.4}$     & $-$           & $-$        & $-$             &$81^{+9}_{-41}$ \\ 
G258.E32   & 0.011   & 10.109   & $0.574\pm0.005$&  90.38    &      $0.35$      & $-$         & $-$       &  $-$                    & $-$                     & $-$     & $-$       & $-$              &$-$ \\
G258.E33   & 1.358   & -85.382  & $8.525\pm0.014$&  93.90    &      $0.27$      & $5.8\pm0.5$ & $27\pm1$  &  $0.8^{+0.4}_{-0.1}$    & $9.8^{+0.3}_{-0.5}$     & $-$           & $-$       & $-$              &$82^{+7}_{-6}$ \\ 
G258.E34   &  5.587  & -14.959  & $0.874\pm0.004$&  89.95    &      $0.47$      & $-$         & $-$       &  $-$                    & $-$                     & $-$     & $-$        & $-$             &$-$ \\ 
G258.E35   & 5.605   & -57.646  & $0.290\pm0.013$&  93.85    &      $0.26$      & $-$         & $-$       &  $-$                    & $-$                     & $-$     & $-$        & $-$             &$-$ \\ 
G258.E36   & 12.613  & -95.870  & $2.008\pm0.008$&  94.07    &      $0.27$      & $4.8\pm0.4$ & $33\pm3$  &  $0.8^{+0.2}_{-0.1}$    & $9.7^{+0.3}_{-0.3}$     & $-$           & $-$        & $-$             &$84^{+6}_{-7}$ \\
G258.E37   & 15.097  & -140.760 & $0.268\pm0.011$&  93.98    &      $0.29$      & $-$         & $-$       &  $-$                    & $-$                     & $-$     & $-$        & $-$             &$-$ \\ 
G258.E38   & 16.132  & -23.134  & $1.324\pm0.011$&  91.31    &      $0.30$      & $2.6\pm0.5$ & $14\pm3$  &  $1.1^{+0.2}_{-0.3}$    & $9.3^{+1.0}_{-1.1}$     & $-$           & $-$        & $-$             &$82^{+8}_{-14}$ \\ 
G258.E39   & 16.185  & -75.846  & $0.533\pm0.004$&  93.28    &      $0.32$      & $3.9\pm0.2$ & $28\pm3$  &  $0.9^{+0.2}_{-0.2}$    & $9.6^{+0.4}_{-0.1}$     & $-$           & $-$        & $-$             &$87^{+3}_{-10}$ \\ 
G258.E40   & 25.962  & -28.428  & $0.267\pm0.006$&  90.43    &      $0.32$      & $-$         & $-$       &  $-$                    & $-$                     & $-$     & $-$        & $-$             &$-$ \\ 
G258.E41   & 29.263  & -52.185  & $0.156\pm0.005$&  92.58    &      $0.42$      & $-$         & $-$       &  $-$                    & $-$                     & $-$     & $-$        & $-$             &$-$ \\ 
G258.E42   & 40.974  & -36.043  & $2.573\pm0.010$&  91.26    &      $0.27$      & $3.2\pm0.6$ & $20\pm1$  &  $0.9^{+0.2}_{-0.2}$    & $9.4^{+0.9}_{-0.6}$     & $-$           & $-$       & $-$              &$78^{+10}_{-11}$ \\ 
G258.E43   & 52.169  & -71.208  & $0.059\pm0.006$&  92.05    &      $0.38$      & $6.0\pm0.5$ & $31\pm17$ &  $1.1^{+0.2}_{-0.4}$    & $9.9^{+0.6}_{-0.5}$     & $-$           & $-$       & $-$              &$83^{+6}_{-9}$ \\ 
G258.E44   & 57.102  & -64.080  & $14.482\pm0.028$& 91.57    &      $0.39$      & $4.9\pm0.3$ & $19\pm0.9$&  $1.3^{+0.1}_{-0.5}$    & $9.7^{+0.8}_{-0.3}$     & $-$           & $-$       & $-$           &$81^{+9}_{-14}$ \\ 
G258.E45   & 61.875  & -36.107  & $0.888\pm0.030$&  91.61    &      $0.27$      & $4.9\pm0.1$ & $43\pm3$  &  $0.8^{+0.3}_{-0.1}$    & $9.7^{+0.2}_{-0.4}$     & $-$           & $-$       & $-$              &$87^{+3}_{-9}$ \\ 
G258.E46   & 70.003  & -82.818  & $0.937\pm0.007$&  92.23    &      $0.34$      & $3.9\pm0.5$ & $21\pm3$  &  $1.0^{+0.2}_{-0.3}$    & $9.6^{+0.7}_{-0.8}$     & $-$           & $-$       & $-$              &$84^{+6}_{-11}$ \\
\hline
\end{tabular} \end{center}
\tablefoot{
\tablefoottext{a}{The reference position is $\alpha_{2000}=18^{\rm{h}}39^{\rm{m}}03^{\rm{s}}\!.630$ and 
$\delta_{2000}=-06^{\circ}24'11''\!\!.163$ (see Sect.~\ref{obssect}).}
\tablefoottext{b}{$P_{\rm{l}}$ and $\chi$ are the mean values of the linear polarization fraction and the linear polarization angle measured across the spectrum, respectively.}
\tablefoottext{c}{The best-fitting results obtained by using a model based on the radiative transfer theory of methanol masers 
for $\Gamma+\Gamma_{\nu}=1~\rm{s^{-1}}$ (Vlemmings et al. \cite{vle10}, Surcis et al. \cite{sur11a}). The errors were determined 
by analyzing the full probability distribution function.}
\tablefoottext{d}{The angle between the magnetic field and the maser propagation direction is determined by using the observed $P_{\rm{l}}$ 
and the fitted emerging brightness temperature. The errors were determined by analyzing the full probability distribution function.}
\tablefoottext{e}{In the fitting model we include the mean values for \tbo ~and \dvi.}
}
\label{G258_tab}
\end{table*}
\begin {table*}[h!]
\caption []{Parameters of the 6.7 GHz \meth ~maser features detected in G25.71-0.04.} 
\begin{center}
\scriptsize
\begin{tabular}{ l c c c c c c c c c c c c c}
\hline
\hline
\,\,\,\,\,(1)&(2)   & (3)      & (4)            & (5)       & (6)              & (7)         & (8)       & (9)                     & (10)                    & (11)                        & (12)         &(13)       & (14)                    \\
Maser     & RA\tablefootmark{a}&Dec\tablefootmark{a}& Peak flux & $V_{\rm{lsr}}$& $\Delta v\rm{_{L}}$ &$P_{\rm{l}}\tablefootmark{b}$ &  $\chi\tablefootmark{b}$   & $\Delta V_{\rm{i}}\tablefootmark{c}$ & $T_{\rm{b}}\Delta\Omega\tablefootmark{c}$& $P_{\rm{V}}$ & $\Delta V_{\rm{Z}}$ & $B_{\rm{||}}$  &$\theta\tablefootmark{d}$\\
          &  offset &  offset  & Density(I)     &           &                  &             &            &                         &                         &              &         &             &      \\ 
          &  (mas)  &  (mas)   & (Jy/beam)      &  (km/s)   &      (km/s)      & (\%)        &   (\d)    & (km/s)                  & (log K sr)              &   ($\%$)     & (m/s)  & (mG)               &(\d)       \\ 
\hline
G257.E01   & -164.687& 66.549  & $0.473\pm0.003$&  92.57    &      $0.46$      & $-$         & $-$       &  $-$                    & $-$                     & $-$     & $-$     & $-$                &$-$ \\ 
G257.E02   & -149.505& 58.381  & $1.495\pm0.004$&  92.09    &      $0.32$      & $-$         & $-$       &  $-$                    & $-$                     & $-$     & $-$     & $-$                &$-$ \\
G257.E03   & -144.266& 55.501  & $3.772\pm0.007$&  92.05    &      $0.28$      & $-$         & $-$       &  $-$                    & $-$                     & $-$     & $-$     & $-$                &$-$ \\
G257.E04   & -128.842& 46.229  & $0.269\pm0.005$&  91.34    &      $0.27$      & $-$         & $-$       &  $-$                    & $-$                     & $-$     & $-$     & $-$                &$-$ \\
G257.E05   & -126.208& 1.349   & $0.490\pm0.011$&  94.28    &      $0.35$      & $-$         & $-$       &  $-$                    & $-$                    & $-$      & $-$     & $-$                &$-$ \\
G257.E06   & -104.362& 46.667  & $1.727\pm0.005$&  91.39    &      $0.34$      & $-$         & $-$       &  $-$                    & $-$                     & $-$     & $-$     & $-$                &$-$ \\
G257.E07   & -102.795& 105.448 & $0.077\pm0.003$&  100.61   &      $0.40$      & $-$         & $-$       &  $-$                    & $-$                     & $-$     & $-$     & $-$                &$-$ \\
G257.E08   & -95.690 & 46.626  & $3.546\pm0.010$&  89.76    &      $0.43$      & $-$         & $-$       &  $-$                    & $-$                     & $-$     & $-$     & $-$                &$-$ \\
G257.E09   & -78.404 & 137.372 & $0.829\pm0.003$&  100.17   &      $0.33$      & $-$         & $-$       &  $-$                    & $-$                     & $-$     & $-$    & $-$                 &$-$ \\
G257.E10   & -77.185 & 81.546  & $1.139\pm0.006$&  93.67    &      $0.46$      & $-$         & $-$       &  $-$                    & $-$                     & $-$     & $-$    & $-$                 &$-$ \\
G257.E11   & -76.567 & 111.733 & $1.796\pm0.009$&  95.12    &      $0.39$      & $-$         & $-$       &  $-$                    & $-$                     & $-$     & $-$    & $-$                 &$-$ \\
G257.E12   & -76.083 & 46.719  & $3.271\pm0.008$&  89.98    &      $0.54$      & $1.4\pm0.5$ & $+22\pm14$&  $2.1^{+0.1}_{-0.3}$    & $9.0^{+0.3}_{-2.6}$     & $0.8$      & $-3.1\pm0.7$   & $>61$          &$90^{+47}_{-47}$ \\
G257.E13   & -71.456 & 125.005 &$22.384\pm0.039$&  95.08    &      $0.46$      & $1.6\pm1.5$ & $-15\pm12$&  $2.0^{+0.2}_{-0.4}$    & $9.1^{+0.8}_{-2.7}$     & $-$           & $-$    & $-$                 &$\bf{62^{+9}_{-51}}$ \\
G257.E14   & -70.166 & 180.090 & $0.086\pm0.003$&  102.54   &      $0.27$      & $-$         & $-$       &  $-$                    & $-$                     & $-$     & $-$    & $-$                 &$-$ \\
G257.E15   & -68.883 & 74.435  & $4.877\pm0.010$&  93.36    &      $0.29$      & $0.6\pm0.2$ & $-16\pm3$ &  $1.0^{+0.1}_{-0.3}$    & $9.5^{+0.9}_{-2.6}$     & $-$           & $-$    & $-$                 &$75^{+15}_{-39}$ \\
G257.E16   & -61.132 & 148.974 & $0.694\pm0.011$&  96.22    &      $0.33$      & $-$         & $-$       &  $-$                    & $-$                     & $-$     & $-$    & $-$                 &$-$ \\
G257.E17   & -47.165 & 175.220 & $0.153\pm0.003$&  96.83    &      $0.25$      & $-$         & $-$       &  $-$                    & $-$                     & $-$     & $-$    & $-$                 &$-$ \\
G257.E18   & -34.818 & 176.856 & $0.214\pm0.003$&  96.87    &      $0.23$      & $-$         & $-$       &  $-$                    & $-$                     & $-$     & $-$    & $-$                 &$-$ \\
G257.E19   &  0      &  0       &$50.172\pm0.068$&  95.56    &      $0.51$      & $0.4\pm0.2$ & $-21\pm48$&  $2.2^{+0.2}_{-0.2}$    & $8.3^{+1.0}_{-1.8}$     & $-$           & $-$    & $-$                 &$\bf{63^{+6}_{-52}}$ \\
G257.E20   &  8.210  & 14.284  & $2.095\pm0.009$&  94.20    &      $0.45$      & $0.6\pm0.2$ & $+26\pm13$&  $2.0^{+0.2}_{-0.2}$    & $8.6^{+1.0}_{-1.4}$     & $-$           & $-$    & $-$                 &$79^{+11}_{-38}$ \\
G257.E21   &  20.069 & 13.464  & $1.968\pm0.011$&  94.28    &      $0.32$      & $-$         & $-$       &  $-$                    & $-$                     & $-$     & $-$    & $-$                 &$-$ \\
G257.E22   &  23.950 & 95.390  & $0.391\pm0.003$&  92.92    &      $0.30$      & $-$         & $-$       &  $-$                    & $-$                     & $-$     & $-$    & $-$                 &$-$ \\
G257.E23   &  33.923 & 88.257  & $1.119\pm0.008$&  93.85    &      $0.27$      & $-$         & $-$       &  $-$                    & $-$                     & $-$     & $-$    & $-$                 &$-$ \\
G257.E24   &  62.063 & 258.951 & $0.896\pm0.003$&  98.06    &      $0.35$      & $-$         & $-$       &  $-$                    & $-$                     & $-$     & $-$    & $-$                 &$-$ \\
G257.E25   &  63.609 & 152.553 & $0.091\pm0.003$&  99.42    &      $0.47$      & $-$         & $-$       &  $-$                    & $-$                     & $-$     & $-$    & $-$                 &$-$ \\
G257.E26   &  159.790& 183.558 & $0.060\pm0.003$&  103.06   &      $0.54$      & $-$         & $-$       &  $-$                    & $-$                     & $-$     & $-$    & $-$                 &$-$ \\
\hline
\end{tabular} \end{center}
\tablefoot{
\tablefoottext{a}{The reference position is $\alpha_{2000}=18^{\rm{h}}38^{\rm{m}}03^{\rm{s}}\!.140$ and 
$\delta_{2000}=-06^{\circ}24'15''\!\!.453$ (see Sect.~\ref{obssect}).}
\tablefoottext{b}{$P_{\rm{l}}$ and $\chi$ are the mean values of the linear polarization fraction and the linear polarization angle measured across the spectrum, respectively.}
\tablefoottext{c}{The best-fitting results obtained by using a model based on the radiative transfer theory of methanol masers 
for $\Gamma+\Gamma_{\nu}=1~\rm{s^{-1}}$ (Vlemmings et al. \cite{vle10}, Surcis et al. \cite{sur11a}). The errors were determined 
by analyzing the full probability distribution function.}
\tablefoottext{d}{The angle between the magnetic field and the maser propagation direction is determined by using the observed $P_{\rm{l}}$ 
and the fitted emerging brightness temperature. The errors were determined by analyzing the full probability distribution function.}
}
\label{G257_tab}
\end{table*}
\begin {table*}[t!]
\caption []{Parameters of the 6.7 GHz \meth ~maser features detected in G28.31-0.39.} 
\begin{center}
\scriptsize
\begin{tabular}{ l c c c c c c c c c c c c c}
\hline
\hline
\,\,\,\,\,(1)&(2)   & (3)      & (4)            & (5)       & (6)              & (7)         & (8)       & (9)                     & (10)                    & (11)                        & (12)         &(13)     & (14)                      \\
Maser     & RA\tablefootmark{a}&Dec\tablefootmark{a}& Peak flux & $V_{\rm{lsr}}$& $\Delta v\rm{_{L}}$ &$P_{\rm{l}}\tablefootmark{b}$ &  $\chi\tablefootmark{b}$   & $\Delta V_{\rm{i}}\tablefootmark{c}$ & $T_{\rm{b}}\Delta\Omega\tablefootmark{c}$& $P_{\rm{V}}$ & $\Delta V_{\rm{Z}}$ & $B_{\rm{||}}$  &$\theta\tablefootmark{d}$\\
          &  offset &  offset  & Density(I)     &           &                  &             &            &                         &                         &              &                      &     &  \\ 
          &  (mas)  &  (mas)   & (Jy/beam)      &  (km/s)   &      (km/s)      & (\%)        &   (\d)    & (km/s)                  & (log K sr)              &   ($\%$)     &  (m/s)   & (mG)               &(\d)       \\ 
\hline
G283.E01   & -17.917 & -175.766 & $0.455\pm0.004$&   83.23   &      $0.30$      & $-$         & $-$       &  $-$                    & $-$                     & $-$     & $-$     & $-$                &$-$ \\ 
G283.E02   & -4.280  & -192.822 & $0.077\pm0.005$&   79.94   &      $0.23$      & $-$         & $-$       &  $-$                    & $-$                     & $-$     & $-$     & $-$                &$-$ \\ 
G283.E03   &   0     &    0     & $7.985\pm0.012$&   81.96   &      $0.27$      & $1.6\pm0.3$ & $37\pm2$  &  $0.9^{+0.1}_{-0.3}$    & $9.0^{+0.9}_{-1.1}$     & $-$           & $-$     & $-$                &$80^{+11}_{-19}$ \\ 
G283.E04   & 11.013  &  12.524  & $0.692\pm0.008$&   81.39   &      $0.47$      & $-$         & $-$       &  $-$                    & $-$                     & $-$     & $-$     & $-$                &$-$ \\ 
G283.E05   & 13.067  &  51.216  & $0.906\pm0.005$&   83.01   &      $0.32$      & $-$         & $-$       &  $-$                    & $-$                     & $-$     & $-$     & $-$                &$-$ \\ 
G283.E06   & 13.694  &  1.911   & $0.457\pm0.005$&   80.99   &      $0.27$      & $-$         & $-$       &  $-$                    & $-$                     & $-$     & $-$     & $-$                &$-$ \\
G283.E07   & 25.278  & -175.720 & $0.074\pm0.003$&   80.07   &      $0.18$      & $-$         & $-$       &  $-$                    & $-$                     & $-$     & $-$     & $-$                &$-$ \\
G283.E08   & 32.039  &  3.162   & $0.104\pm0.004$&   81.17   &      $0.26$      & $-$         & $-$       &  $-$                    & $-$                     & $-$     & $-$     & $-$                &$-$ \\
G283.E09   & 32.439  &  42.442  & $0.079\pm0.004$&   83.23   &      $0.46$      & $-$         & $-$       &  $-$                    & $-$                     & $-$     & $-$     & $-$                &$-$ \\ 
G283.E10   & 40.712  &  35.751  & $6.459\pm0.018$&   82.31   &      $2.09$      & $4.3\pm1.4$ & $-81\pm30$&  $1.0^{+0.2}_{-0.4}$    & $9.6^{+0.9}_{-2.1}$     & $-$           & $-$     & $-$                &$75^{+15}_{-27}$ \\
G283.E11   & 177.970 & -246.517 & $3.008\pm0.008$&   79.81   &      $0.21$      & $-$         & $-$       &  $-$                    & $-$                     & $-$     & $-$     & $-$            &$-$ \\ 
G283.E12   & 195.459 & -24.464  & $1.753\pm0.008$&   92.72   &      $0.47$      & $-$         & $-$       &  $-$                    & $-$                     & $-$     & $-$     & $-$            &$-$ \\
G283.E13   & 210.950 &  4.860   & $0.128\pm0.005$&   93.81   &      $0.34$      & $-$         & $-$       &  $-$                    & $-$                     & $-$     & $-$     & $-$            &$-$ \\ 
\hline
\end{tabular} \end{center}
\tablefoot{
\tablefoottext{a}{The reference position is $\alpha_{2000}=18^{\rm{h}}44^{\rm{m}}22^{\rm{s}}\!.030$ and 
$\delta_{2000}=-04^{\circ}17'38''\!\!.304$ (see Sect.~\ref{obssect}).}
\tablefoottext{b}{$P_{\rm{l}}$ and $\chi$ are the mean values of the linear polarization fraction and the linear polarization angle measured across the spectrum, respectively.}
\tablefoottext{c}{The best-fitting results obtained by using a model based on the radiative transfer theory of methanol masers 
for $\Gamma+\Gamma_{\nu}=1~\rm{s^{-1}}$ (Vlemmings et al. \cite{vle10}, Surcis et al. \cite{sur11a}). The errors were determined 
by analyzing the full probability distribution function.}
\tablefoottext{d}{The angle between the magnetic field and the maser propagation direction is determined by using the observed $P_{\rm{l}}$ 
and the fitted emerging brightness temperature. The errors were determined by analyzing the full probability distribution function.}
}
\label{G283_tab}
\end{table*}
\begin {table*}[h!]
\caption []{Parameters of the 6.7 GHz \meth ~maser features detected in G28.83-0.25.} 
\begin{center}
\scriptsize
\begin{tabular}{ l c c c c c c c c c c c c c}
\hline
\hline
\,\,\,\,\,(1)&(2)   & (3)      & (4)            & (5)       & (6)              & (7)         & (8)       & (9)                     & (10)                    & (11)                        & (12)         &(13)     & (14)                      \\
Maser     & RA\tablefootmark{a}&Dec\tablefootmark{a}& Peak flux & $V_{\rm{lsr}}$& $\Delta v\rm{_{L}}$ &$P_{\rm{l}}\tablefootmark{b}$ &  $\chi\tablefootmark{b}$   & $\Delta V_{\rm{i}}\tablefootmark{c}$ & $T_{\rm{b}}\Delta\Omega\tablefootmark{c}$& $P_{\rm{V}}$ & $\Delta V_{\rm{Z}}$ & $B_{\rm{||}}$  &$\theta\tablefootmark{d}$\\
          &  offset &  offset  & Density(I)     &           &                  &             &            &                         &                         &              &                      &   &   \\ 
          &  (mas)  &  (mas)   & (Jy/beam)      &  (km/s)   &      (km/s)      & (\%)        &   (\d)    & (km/s)                  & (log K sr)              &   ($\%$)     &  (m/s)   &  (mG)               &(\d)       \\ 
\hline
G288.E01   &-240.550 &  129.036 & $0.605\pm0.006$&  85.31    &      $0.26$      & $-$         & $-$       &  $-$                    & $-$                     & $-$     & $-$     & $-$                &$-$ \\ 
G288.E02   &-230.958 &  144.054 & $4.376\pm0.012$&  85.97    &      $0.25$      & $3.2\pm0.6$ & $-57\pm6$ &  $0.8^{+0.1}_{-0.2}$    & $9.5^{+0.7}_{-1.3}$     & $-$           & $-$     & $-$                &$80^{+10}_{-7}$ \\ 
G288.E03   &-117.049 &  30.556  & $0.371\pm0.007$&  86.14    &      $0.25$      & $-$         & $-$       &  $-$                    & $-$                     & $-$     & $-$     & $-$                &$-$ \\ 
G288.E04   &-102.832 &  25.745  & $3.167\pm0.010$&  85.04    &      $0.33$      & $1.3\pm0.2$ & $-5\pm4$  &  $1.3^{+0.1}_{-0.3}$    & $8.9^{+0.6}_{-0.6}$     & $-$           & $-$     & $-$                &$82^{+8}_{-21}$ \\ 
G288.E05   & -90.670 &  11.570  & $1.368\pm0.009$&  84.60    &      $0.41$      & $-$         & $-$       &  $-$                    & $-$                     & $-$     & $-$     & $-$                &$-$ \\ 
G288.E06   & -86.502 & -272.385 & $0.089\pm0.008$&  92.24    &      $1.58$      & $-$         & $-$       &  $-$                    & $-$                     & $-$     & $-$     & $-$                &$-$ \\ 
G288.E07   & -80.221 &  5.180   & $9.342\pm0.017$&  83.95    &      $0.33$      & $3.3\pm1.7$ & $-69\pm8$ &  $1.1^{+0.3}_{-0.3}$    & $9.4^{+0.2}_{-3.2}$     & $-$           & $-$     & $-$                &$79^{+11}_{-44}$ \\ 
G288.E08   & -74.569 & -30.281  & $0.848\pm0.004$&  80.30    &      $0.69$      & $-$         & $-$       &  $-$                    & $-$                     & $-$     & $-$     & $-$                &$-$ \\ 
G288.E09   & -64.006 & -36.369  & $0.240\pm0.004$&  79.73    &      $0.42$      & $-$         & $-$       &  $-$                    & $-$                     & $-$     & $-$     & $-$                &$-$ \\ 
G288.E10   & -46.249 & -98.236  & $0.944\pm0.032$&  83.55    &      $0.33$      & $-$         & $-$       &  $-$                    & $-$                     & $-$     & $-$     & $-$                &$-$ \\ 
G288.E11   & -35.058 & -28.030  & $3.598\pm0.009$&  81.40    &      $0.45$      & $-$         & $-$       &  $-$                    & $-$                     & $-$     & $-$     & $-$                &$-$ \\ 
G288.E12   & -31.004 & -16.800  & $3.275\pm0.010$&  81.88    &      $0.36$      & $-$         & $-$       &  $-$                    & $-$                     & $-$     & $-$     & $-$                &$-$ \\ 
G288.E13   & -9.535  & -215.977 & $0.045\pm0.004$&  93.91    &      $1.74$      & $-$         & $-$       &  $-$                    & $-$                     & $-$     & $-$     & $-$                &$-$ \\ 
G288.E14   & -8.336  &  0.401   & $0.555\pm0.007$&  84.25    &      $0.25$      & $-$         & $-$       &  $-$                    & $-$                     & $-$     & $-$     & $-$                &$-$ \\ 
G288.E15   & -8.108  &  1.835   & $0.462\pm0.004$&  82.54    &      $0.82$      & $-$         & $-$       &  $-$                    & $-$                     & $-$     & $-$     & $-$                &$-$ \\ 
G288.E16   & 0       &  0       &$25.018\pm0.032$&  83.55    &      $0.44$      & $0.5\pm0.1$ & $-6\pm7$  &  $1.8^{+0.1}_{-0.5}$    & $9.1^{+0.8}_{-1.2}$     & $-$           & $-$     & $-$                &$\bf{61^{+3}_{-47}}$ \\ 
G288.E17   & 70.344  & -279.736 & $1.592\pm0.008$&  92.24    &      $0.23$      & $-$         & $-$       &  $-$                    & $-$                     & $-$     & $-$     & $-$                &$-$ \\ 
G288.E18   & 82.962  & -283.184 &$24.891\pm0.043$&  91.89    &      $0.32$      & $1.3\pm0.3$ & $-87\pm8$ &  $1.2^{+0.2}_{-0.2}$    & $9.0^{+1.0}_{-0.9}$     & $-$           & $-$     & $-$                &$79^{+11}_{-35}$ \\ 
G288.E19   & 117.963 & -292.862 & $6.203\pm0.012$&  91.32    &      $0.25$      & $-$         & $-$       &  $-$                    & $-$                     & $0.6$       & $-1.1\pm0.3\tablefootmark{e}$    & $>21\tablefootmark{e}$ &$-$ \\ 
G288.E20   & 142.343 & -299.622 & $7.293\pm0.019$&  90.93    &      $0.31$      & $2.8\pm0.5$ & $-64\pm4$ &  $1.1^{+0.3}_{-0.2}$    & $9.4^{+0.9}_{-1.5}$     & $-$           & $-$     & $-$                &$83^{+5}_{-15}$ \\ 
G288.E21   & 164.325 & -303.360 & $0.545\pm0.007$&  90.44    &      $0.21$      & $-$         & $-$       &  $-$                    & $-$                     & $-$     & $-$     & $-$                &$-$ \\
\hline
\end{tabular} \end{center}
\tablefoot{
\tablefoottext{a}{The reference position is $\alpha_{2000}=18^{\rm{h}}44^{\rm{m}}51^{\rm{s}}\!.080$ and 
$\delta_{2000}=-03^{\circ}45'48''\!\!.494$ (see Sect.~\ref{obssect}).}
\tablefoottext{b}{$P_{\rm{l}}$ and $\chi$ are the mean values of the linear polarization fraction and the linear 
polarization angle measured across the spectrum, respectively.}
\tablefoottext{c}{The best-fitting results obtained by using a model based on the radiative transfer theory of 
methanol masers 
for $\Gamma+\Gamma_{\nu}=1~\rm{s^{-1}}$ (Vlemmings et al. \cite{vle10}, Surcis et al. \cite{sur11a}). The errors 
were determined 
by analyzing the full probability distribution function.}
\tablefoottext{d}{The angle between the magnetic field and the maser propagation direction is determined by using 
the observed $P_{\rm{l}}$ 
and the fitted emerging brightness temperature. The errors were determined by analyzing the full probability 
distribution function.}
\tablefoottext{e}{To model the circularly polarized emission we considered the error-weighted values of 
$\langle$\tbo$\rangle=9.4\cdot10^8$~K~sr and \dvi$=1.1$~\kms ~that best fit the total intensity emission.}
}
\label{G288_tab}
\end{table*}
\begin {table*}[t!]
\caption []{Parameters of the 6.7 GHz \meth ~maser features detected in G29.96-0.02.} 
\begin{center}
\scriptsize
\begin{tabular}{ l c c c c c c c c c c c c c }
\hline
\hline
\,\,\,\,\,(1)&(2)   & (3)      & (4)            & (5)       & (6)              & (7)         & (8)       & (9)                     & (10)                    & (11)                        & (12)         &(13)     & (14)                      \\
Maser     & RA\tablefootmark{a}&Dec\tablefootmark{a}& Peak flux & $V_{\rm{lsr}}$& $\Delta v\rm{_{L}}$ &$P_{\rm{l}}\tablefootmark{b}$ &  $\chi\tablefootmark{b}$   & $\Delta V_{\rm{i}}\tablefootmark{c}$ & $T_{\rm{b}}\Delta\Omega\tablefootmark{c}$& $P_{\rm{V}}$ & $\Delta V_{\rm{Z}}$ & $B_{\rm{||}}$  &$\theta\tablefootmark{d}$\\
          &  offset &  offset  & Density(I)     &           &                  &             &            &                         &                         &              &                      &    &   \\ 
          &  (mas)  &  (mas)   & (Jy/beam)      &  (km/s)   &      (km/s)      & (\%)        &   (\d)    & (km/s)                  & (log K sr)              &   ($\%$)     &  (m/s)  &  (mG)               &(\d)       \\ 
\hline
G29.E01    &-330.651 & 137.873  & $1.496\pm0.047$&  96.79    &      $0.32$      & $-$         & $-$       &  $-$                    & $-$                     & $-$     & $-$      & $-$               &$-$ \\ 
G29.E02    &-329.869 & 141.356  & $2.944\pm0.072$&  96.09    &      $0.22$      & $-$         & $-$       &  $-$                    & $-$                     & $-$     & $-$      & $-$               &$-$ \\ 
G29.E03    &-329.438 & 140.755  & $0.095\pm0.006$&  95.70    &      $1.14$      & $-$         & $-$       &  $-$                    & $-$                     & $-$     & $-$      & $-$               &$-$ \\ 
G29.E04    &-291.830 & 74.976   & $0.078\pm0.005$&  98.59    &      $0.32$      & $-$         & $-$       &  $-$                    & $-$                     & $-$     & $-$      & $-$               &$-$ \\ 
G29.E05    &-271.946 & 51.844   & $0.244\pm0.005$&  98.42    &      $0.38$      & $-$         & $-$       &  $-$                    & $-$                     & $-$     & $-$      & $-$               &$-$ \\ 
G29.E06    &-278.374 & 95.936   & $0.167\pm0.004$&  101.93   &      $0.26$      & $-$         & $-$       &  $-$                    & $-$                     & $-$     & $-$      & $-$               &$-$ \\ 
G29.E07    &-41.356  & 86.535   & $0.053\pm0.005$&  98.68    &      $0.26$      & $-$         & $-$       &  $-$                    & $-$                     & $-$     & $-$      & $-$               &$-$ \\ 
G29.E08    &-38.811  & 20.832   & $0.243\pm0.005$&  98.64    &      $0.23$      & $-$         & $-$       &  $-$                    & $-$                     & $-$     & $-$      & $-$               &$-$ \\ 
G29.E09    & 0       & 0        &$62.718\pm0.064$&  95.96    &      $0.37$      & $5.7\pm1.4$ & $75\pm8$  &  $1.1^{+0.2}_{-0.3}$    & $9.8^{+0.5}_{-0.3}$     & $-$           & $-$      & $-$               &$83^{+7}_{-10}$ \\ 
G29.E10    & 8.705   &-70.158   & $0.838\pm0.025$&  96.97    &      $0.33$      & $-$         & $-$       &  $-$                    & $-$                     & $-$     & $-$      & $-$               &$-$ \\ 
G29.E11    & 10.033  & 57.062   & $1.160\pm0.038$&  96.88    &      $0.25$      & $-$         & $-$       &  $-$                    & $-$                     & $-$     & $-$      & $-$               &$-$ \\ 
G29.E12    & 17.168  &-7.324    &$13.278\pm0.035$&  96.36    &      $0.33$      & $4.9\pm0.2$ & $77\pm5$  &  $1.1^{+0.3}_{-0.5}$    & $9.7^{+1.0}_{-2.1}$     & $-$           & $-$      & $-$               &$75^{+14}_{-8}$ \\ 
G29.E13    & 49.083  &-74.986   & $0.056\pm0.004$&  103.25   &      $0.36$      & $-$         & $-$       &  $-$                    & $-$                     & $-$     & $-$      & $-$               &$-$ \\ 
G29.E14    & 57.367  &-81.915   & $1.045\pm0.005$&  105.75   &      $0.50$      & $-$         & $-$       &  $-$                    & $-$                     & $-$     & $-$      & $-$               &$-$ \\ 
G29.E15    & 58.759  &-182.772  & $2.966\pm0.050$&  96.66    &      $0.25$      & $-$         & $-$       &  $-$                    & $-$                     & $-$     & $-$      & $-$               &$-$ \\ 
G29.E16    & 61.603  &-36.263   & $7.437\pm0.019$&  95.65    &      $0.45$      & $-$         & $-$       &  $-$                    & $-$                     & $-$     & $-$      & $-$               &$-$ \\ 
G29.E17    & 62.692  &-195.025  & $1.464\pm0.007$&  99.21    &      $0.31$      & $-$         & $-$       &  $-$                    & $-$                     & $-$     & $-$      & $-$               &$-$ \\ 
G29.E18    & 64.662  &-211.916  & $0.941\pm0.010$&  97.23    &      $0.40$      & $-$         & $-$       &  $-$                    & $-$                     & $-$     & $-$      & $-$               &$-$ \\ 
G29.E19    & 73.117  &-155.859  & $0.742\pm0.006$&  97.94    &      $0.28$      & $-$         & $-$       &  $-$                    & $-$                     & $-$     & $-$      & $-$               &$-$ \\ 
G29.E20    & 80.876  &-221.630  & $0.854\pm0.007$&  97.58    &      $0.86$      & $-$         & $-$       &  $-$                    & $-$                     & $-$     & $-$      & $-$               &$-$ \\ 
G29.E21    & 86.027  &-203.621  & $0.081\pm0.007$&  99.21    &      $3.77$      & $-$         & $-$       &  $-$                    & $-$                     & $-$     & $-$      & $-$               &$-$ \\ 
G29.E22    & 87.865  &-132.851  & $0.132\pm0.009$&  97.45    &      $0.25$      & $-$         & $-$       &  $-$                    & $-$                     & $-$     & $-$      & $-$               &$-$ \\ 
G29.E23    & 592.615 & -14.347  & $0.167\pm0.005$ & 92.05     &      $0.42$      & $-$         & $-$       &  $-$                    & $-$                     & $-$     & $-$      & $-$               &$-$ \\ 
G29.E24    & 593.021 & 68.5539  & $0.146\pm0.005$ & 100.88    &      $0.27$      & $-$         & $-$       &  $-$                    & $-$                     & $-$     & $-$      & $-$               &$-$ \\ 
G29.E25    & 685.252 & 137.028  & $3.400\pm0.054$ & 96.75     &      $0.45$      & $-$         & $-$       &  $-$                    & $-$                     & $-$     & $-$      & $-$               &$-$ \\ 
G29.E26    & 688.357 & 113.371  & $1.416\pm0.050$ & 96.75     &      $0.35$      & $1.4\pm0.1$ & $+46\pm4$ &  $1.2^{+0.2}_{-0.3}$       & $9.2^{+0.7}_{-1.6}$         & $-$       & $-$   & $-$          &$86^{+4}_{-15}$ \\ 
G29.E27    & 755.304 & 368.290  & $0.531\pm0.037$ & 100.26    &      $0.32$      & $-$         & $-$       &  $-$                    & $-$                     & $-$     & $-$      & $-$               &$-$ \\ 
G29.E28    & 761.407 & 150.597  & $0.471\pm0.009$ & 99.34     &      $0.47$      & $-$         & $-$       &  $-$                    & $-$                     & $-$     & $-$      & $-$               &$-$ \\ 
G29.E29    & 762.160 & 147.505  & $6.826\pm0.041$ & 100.31    &      $0.48$      & $-$         & $-$       &  $-$                    & $-$                     & $-$     & $-$      & $-$               &$-$ \\ 
G29.E30    & 764.094 & 137.281  & $1.961\pm0.041$ & 100.39    &      $0.43$      & $2.1\pm0.3$ & $+54\pm8$ &  $1.5^{+0.3}_{-0.3}$        & $9.0^{+0.8}_{-0.1}$         & $-$       & $-$  & $-$           &$86^{+4}_{-13}$ \\ 
G29.E31    & 766.061 & 71.0526  & $0.911\pm0.042$ & 100.35    &      $0.39$      & $-$         & $-$       &  $-$                    & $-$                     & $-$     & $-$     & $-$                &$-$ \\ 
G29.E32    & 776.240 & 140.797  & $0.230\pm0.006$ & 100.92    &      $0.26$      & $-$         & $-$       &  $-$                    & $-$                     & $-$     & $-$     & $-$                &$-$ \\ 
G29.E33    & 1145.428& 335.558  & $0.282\pm0.006$ & 101.36    &      $0.20$      & $-$         & $-$       &  $-$                    & $-$                     & $-$     & $-$     & $-$                &$-$ \\ 
G29.E34    & 1167.131& 330.568  & $0.120\pm0.006$ & 101.40    &      $0.15$      & $-$         & $-$       &  $-$                    & $-$                     & $-$     & $-$    & $-$                 &$-$ \\ 
\hline
\end{tabular} \end{center}
\tablefoot{
\tablefoottext{a}{The reference position is $\alpha_{2000}=18^{\rm{h}}46^{\rm{m}}03^{\rm{s}}\!.740$ and 
$\delta_{2000}=-02^{\circ}39'22''\!\!.299$ (see Sect.~\ref{obssect}).}
\tablefoottext{b}{$P_{\rm{l}}$ and $\chi$ are the mean values of the linear polarization fraction and the linear polarization angle measured across the spectrum, respectively.}
\tablefoottext{c}{The best-fitting results obtained by using a model based on the radiative transfer theory of methanol masers 
for $\Gamma+\Gamma_{\nu}=1~\rm{s^{-1}}$ (Vlemmings et al. \cite{vle10}, Surcis et al. \cite{sur11a}). The errors were determined 
by analyzing the full probability distribution function.}
\tablefoottext{d}{The angle between the magnetic field and the maser propagation direction is determined by using the observed $P_{\rm{l}}$ 
and the fitted emerging brightness temperature. The errors were determined by analyzing the full probability distribution function.}
}
\label{G29_tab}
\end{table*}
\begin {table*}[t!]
\caption []{Parameters of the 6.7 GHz \meth ~maser features detected in G43.80-0.13.} 
\begin{center}
\scriptsize
\begin{tabular}{ l c c c c c c c c c c c c c}
\hline
\hline
\,\,\,\,\,(1)&(2)   & (3)      & (4)            & (5)       & (6)              & (7)         & (8)        & (9)                     & (10)                    & (11)        & (12)                 &(13)   & (14)                        \\
Maser     & RA\tablefootmark{a}&Dec\tablefootmark{a}& Peak flux & $V_{\rm{lsr}}$& $\Delta v\rm{_{L}}$ &$P_{\rm{l}}\tablefootmark{b}$ &  $\chi\tablefootmark{b}$   & $\Delta V_{\rm{i}}\tablefootmark{c}$ & $T_{\rm{b}}\Delta\Omega\tablefootmark{c}$& $P_{\rm{V}}$ & $\Delta V_{\rm{Z}}$ & $B_{\rm{||}}$   &$\theta\tablefootmark{d}$\\
          &  offset &  offset  & Density(I)     &           &                  &             &             &                         &                         &             &                      &   &   \\ 
          &  (mas)  &  (mas)   & (Jy/beam)      &  (km/s)   &      (km/s)      & (\%)        &   (\d)     & (km/s)                  & (log K sr)              &   ($\%$)    &  (m/s)   &  (mG)               &(\d)       \\ 
\hline
G43.E01    & -183.872&  323.418 & $0.410\pm0.008$&  39.83    &      $0.18$      & $-$         & $-$        &  $-$                    & $-$                     & $-$            & $-$     & $-$                & $-$\\ 
G43.E02    & -23.358 &  65.174  & $0.180\pm0.006$&  40.80    &      $0.14$      & $-$         & $-$        &  $-$                    & $-$                     & $-$            & $-$     & $-$                & $-$\\ 
G43.E03    & -22.117 &  33.051  & $5.967\pm0.013$&  40.41    &      $0.20$      & $1.1\pm0.1$ & $-82\pm6$  &  $0.7^{+0.2}_{-0.1}$    & $8.9^{+0.5}_{-0.5}$     & $-$           & $-$     & $-$                & $79^{+11}_{-35}$ \\ 
G43.E04    & -19.239 &  19.505  & $1.554\pm0.013$&  40.41    &      $0.19$      & $4.4\pm0.9$ & $-75\pm5$  &  $<0.5$                 & $9.6^{+0.2}_{-2.5}$     & $-$           & $-$     & $-$                & $90^{+12}_{-12}$ \\ 
G43.E05    & -14.951 &  49.179  & $0.855\pm0.018$&  40.36    &      $0.19$      & $-$         & $-$        &  $-$                    & $-$                     & $-$            & $-$     & $-$                & $-$\\ 
G43.E06    & -12.977 &  15.278  & $2.861\pm0.014$&  40.10    &      $0.23$      & $1.6\pm0.1$ & $-85\pm4$  &  $0.8^{+0.1}_{-0.1}$    & $9.1^{+0.5}_{-0.1}$     & $-$           & $-$     & $-$                & $82^{+7}_{-18}$ \\ 
G43.E07    & -3.047  &  11.246  & $0.261\pm0.011$&  39.57    &      $0.67$      & $-$         & $-$        &  $-$                    & $-$                     & $-$            & $-$     & $-$                & $-$\\ 
G43.E08    &  0      &  0       & $5.960\pm0.017$&  39.70    &      $0.20$      & $-$         & $-$        &  $-$                    & $-$                     & $-$            & $-$     & $-$                & $-$\\ 
G43.E09    &  0.395  & -10.040  & $0.743\pm0.019$&  39.75    &      $0.13$      & $-$         & $-$        &  $-$                    & $-$                     & $-$            & $-$     & $-$                & $-$\\ 
G43.E10    &  1.918  &  3.857   & $1.432\pm0.010$&  39.57    &      $0.17$      & $-$         & $-$        &  $-$                    & $-$                     & $-$            & $-$     & $-$                & $-$\\ 
G43.E11    &  2.595  &  117.371 & $0.222\pm0.011$&  40.23    &      $0.11$      & $-$         & $-$        &  $-$                    & $-$                     & $-$            & $-$     & $-$                & $-$\\ 
G43.E12    &  43.500 &  20.298  & $0.339\pm0.009$&  39.92    &      $0.15$      & $-$         & $-$        &  $-$                    & $-$                     & $-$            & $-$     & $-$                & $-$\\ 
G43.E13    &  43.838 &  29.270  & $0.877\pm0.009$&  39.92    &      $0.18$      & $-$         & $-$        &  $-$                    & $-$                     & $-$            & $-$     & $-$                & $-$\\ 
G43.E14    &  90.328 &  139.080 & $0.120\pm0.008$&  40.58    &      $0.20$      & $-$         & $-$        &  $-$                    & $-$                     & $-$            & $-$     & $-$                & $-$\\ 
G43.E15    &  391.947& -301.041 & $3.417\pm0.020$&  43.17    &      $0.23$      & $-$         & $-$        &  $-$                    & $-$                     & $-$            & $-$     & $-$                & $-$\\ 
G43.E16    &  393.809& -314.392 & $1.392\pm0.014$&  43.04    &      $0.32$      & $-$         & $-$        &  $-$                    & $-$                     & $-$            & $-$     & $-$                & $-$\\ 
G43.E17    &  396.348& -324.905 & $0.531\pm0.009$&  42.86    &      $0.26$      & $-$         & $-$        &  $-$                    & $-$                     & $-$            & $-$     & $-$                & $-$\\ 
G43.E18    &  399.733& -288.101 & $0.849\pm0.014$&  43.04    &      $0.20$      & $-$         & $-$        &  $-$                    & $-$                     & $-$            & $-$     & $-$                & $-$\\ 
\hline       
\end{tabular} \end{center}
\tablefoot{
\tablefoottext{a}{The reference position is $\alpha_{2000}=19^{\rm{h}}11^{\rm{m}}53^{\rm{s}}\!.990$ and 
$\delta_{2000}=+09^{\circ}35'50''\!\!.300$ (see Sect.~\ref{obssect}).}
\tablefoottext{b}{$P_{\rm{l}}$ and $\chi$ are the mean values of the linear polarization fraction and the linear polarization angle measured across the spectrum, respectively.}
\tablefoottext{c}{The best-fitting results obtained by using a model based on the radiative transfer theory of methanol masers 
for $\Gamma+\Gamma_{\nu}=1~\rm{s^{-1}}$ (Vlemmings et al. \cite{vle10}, Surcis et al. \cite{sur11a}). The errors were determined 
by analyzing the full probability distribution function.}
\tablefoottext{d}{The angle between the magnetic field and the maser propagation direction is determined by using the observed $P_{\rm{l}}$ 
and the fitted emerging brightness temperature. The errors were determined by analyzing the full probability distribution function.}
}
\label{G43_tab}
\end{table*}

\end{appendix}

\end{document}